\documentclass{article}
\usepackage[utf8]{inputenc}
\usepackage{amsmath,amssymb}
\usepackage{oubraces}
\usepackage[utf8]{inputenc}
\usepackage{natbib}
\usepackage{graphicx}
\usepackage{float}
\usepackage{subcaption}
\usepackage{setspace}
\usepackage{color}
\usepackage[dvipsnames]{xcolor}
\usepackage{lineno}
\usepackage{url}
\usepackage[normalem]{ulem} %strikeout functionality
\usepackage{systeme}
\usepackage{array}

\usepackage{titling}
\providecommand{\keywords}[1]{\textbf{\textit{Keywords---}} #1}
\DeclareMathOperator*{\argmin}{arg\,min}

\thanksmarkseries{alph}

\newcolumntype{C}[1]{>{\centering\arraybackslash}p{#1}}
\newcolumntype{R}[1]{>{\raggedleft\arraybackslash}p{#1}}
\newcolumntype{L}[1]{>{\raggedright\arraybackslash}p{#1}}
\title{Nonstationary seasonal model for daily mean temperature distribution bridging bulk and tails}

\author{Mitchell Krock\thanks{Department of Statistics,
   Rutgers University, Piscataway, NJ, USA. e-mail:
  \texttt{mk1867@stat.rutgers.edu}} \and
  Julie Bessac\thanks{Mathematics and Computer Science Division, Argonne National Laboratory, Lemont, IL, USA}
  \and Michael L. Stein$^\text{a}$ \and Adam H. Monahan\thanks{School of Earth and Ocean Sciences, University of Victoria, Victoria, British Columbia, Canada}}

\date{October 2021}

\begin{document}

\maketitle

\begin{abstract}

In traditional extreme value analysis, the bulk of the data is ignored, and only the tails of the distribution are used for inference. Extreme observations are specified as values that exceed a threshold or as maximum values over distinct blocks of time, and subsequent estimation procedures are motivated by asymptotic theory for extremes of random processes. 
For environmental data, nonstationary behavior in the bulk of the distribution, such as seasonality or climate change, will also be observed in the tails. To accurately model such nonstationarity, it seems natural to use the entire dataset rather than just the most extreme values. It is also common to observe different types of nonstationarity in each tail of a distribution. Most work on extremes only focuses on one tail of a distribution, but for temperature, both tails are of interest. This paper builds on a recently proposed parametric model for the entire probability distribution that has flexible behavior in both tails. We apply an extension of this model to historical records of daily mean temperature at several locations across the United States with different climates and local conditions. We highlight the ability of the method to quantify changes in the bulk and tails across the year over the past decades and under different geographic and climatic conditions. 
The proposed model shows good performance when compared to several benchmark models that are typically used in extreme value analysis of temperature.

\end{abstract}


%%%%------------------------------------
\section{Introduction}\label{sec:intro}

\subsection{Changing distribution of surface air temperature across time} 

The probability distribution of surface air temperature (SAT) possesses nonstationary traits, such as seasonality and long-term trends, that can be difficult to capture with off-the-shelf models. This fact is particularly true for the tails of the distribution.
In this paper, we study SAT using a model for the entire probability distribution that has versatile behavior in both tails. Some areas where such a model is useful include finance, insurance, and environmental science \citep{finkenstadt2003,reiss2007}.  The tails of the distribution may be more important than the bulk, as unlikely events in the stock market or environment can have more severe consequences than an ordinary event. With temperature data, both tails of the model distribution are important and of particular public interest due to growing impacts from climate change on human health, the environment, and the economy.  Early in 2021, Texas experienced record low temperatures in February from winter storms, which led to massive power grid failures \citep{texas}, while in the Pacific Northwest of America, record high temperatures in June led to increased hospitalizations and fatalities \citep{portland,wwa}. On longer time scales (months to years), both tails of precipitation distributions matter for the same reasons as temperature. Even when the focus is on the extremes, an understanding of the full distribution is often still of interest. 
Temporal changes may be different in the upper and lower tails, and nonstationary patterns such as seasonality and long-term trends will also be observed in the bulk of the distribution. To our knowledge, this study is the first attempt to model nonstationary temperature data on the daily scale with a concentration on behavior in both tails.

Temporal changes in temperature distribution are influenced by various phenomena.  
Seasonal variability is largely driven by the seasonal cycle in solar radiation, which depends strongly on latitude.
However, large-scale oceanic and atmospheric circulation patterns \citep{mckinnon2013} and local geography (e.g., elevation, distance to the ocean) also have a large impact on the seasonal patterns at specific locations. %\citep{vandendool1981,brooks1917,brooks1918}. 
For monthly means, temperature seasonality has been traditionally studied through circular harmonics, which have been shown to capture a large  amount of the variability (up to $99\%$) \citep{legates1990}. However, at shorter time resolution, temperature exhibits more complex seasonal patterns. 

Rising global mean temperature has drawn attention for many decades,  %\citep{hansen1987global,hansen2010,ipcc2018}
 and regional temperatures exhibit various rates and patterns of change, including in their extremes.   
 A large literature has been generated on the topic; an extensive consolidation can be found in assessment reports of the Intergovernmental Panel on Climate Change \citep{IPCC}. 
 In the following, we discuss literature focusing on statistical aspects of SAT extremes. 
 
Most previous works have focused on statistics of a single tail of a temperature distribution but rarely on the entire distribution of temperature changing over time \citep{IPCCch11}. 
For instance, among the large quantity of work on global mean temperature, \citet{rahmstorf2011} and \citet{poppick2017} examined trends, and \citet{mckinnon2016} studied long-term changes of temperature quantiles.  
\cite{meehl2009} showed that ratios of frequencies of daily hot maxima over frequencies of daily cold minima under nonstationary climate conditions in the United States exhibit strong asymmetry leading to more frequent hot extremes. 
 Other studies have considered the evaluation and quantification of changing temperature in historical datasets (measurements and/or model output) \citep{legates1990,tarleton1995, hansen2010,rhines2017}, while recent studies focus on projections under various greenhouse gas concentration pathways by leveraging the use of multi-model ensembles or large single-model ensembles \citep{huang2016,haugen2018,wehner2020}. 
 Several works have investigated the statistical aspects of extreme quantification under nonstationarity \citep{cheng2014,gilleland2017, IPCC}.  
\citet{katz1992} proposed  a  statistical hypothesis test  to analyse the sensitivity of  extreme events %(threshold exceedances or block maxima)
to changes in the location and  shape of climate  distributions, emphasizing the impact of the  scale parameter on extreme occurrences.
\citet{robin2020} proposed a nonstationary framework for extremes including model data and observations.
\cite{grotjahn2016} reviewed statistical methodologies, dynamics, modeling efforts, and trends dedicated to temperature extremes. 

Very few works have focused on jointly learning from the bulk and tails of SAT. 
However, pivotal insights about changing climate have been obtained by quantifying the link  between changing extremes and other statistics of the bulk for the quantity of interest. 
For instance, \citet{huybers2014} and \citet{huang2016} respectively linked  changing extremes to changing mean temperature using reanalysis data and climate model projections. 
\citet{huybers2014} proposed  a metric to link  the mean to an extreme quantile. 
These works motivate the use of statistical models that can simultaneously assess changes in the tails and bulk of a distribution.%, where despite a strong interest on extremes, the understanding of the full distribution is needed and of interest. 

\subsection{Statistical models for bulk and tails}
A desirable property in a statistical model for a probability distribution is the ability to simultaneously make inferences about the bulk of the distribution and its upper and lower tails. A bulk-and-tails model has the appealing ability to produce  simulations from an entire distribution with flexible behaviors in both tails, which could be useful in the context of stochastic weather generators for extremes \citep{semenov2008, furrer2008}.

Classical theory for extreme value statistics deals with a small fraction of the most extreme outcomes (large and/or small depending on context) of a sequence of independent, identically distributed (or stationary) random variables. Two common models, the generalized Pareto distribution (GPD) for threshold exceedances and the generalized extreme value distribution (GEV) for block-maxima, are motivated by asymptotic theory and can be readily fit by practitioners with pre-existing software. 
% First, a generalized pareto distribution (GPD) is fit to the exceedances of some user-specified threshold. Second, a generalized extreme value distribution (GEV) is fit to a collection of ``block maxima'', which corresponds to the set of maxima from blocks of time for which the independence (stationary) assumption is reasonable. Both distributions have a positive scale parameter and a shape parameter. The GEV distribution also has a location parameter, while the threshold parameter in the GPD distribution is fixed before model fitting. The sign of the shape parameter controls the behavior of the tail, with a positive shape producing a heavy tail and a negative shape producing a light tail. A limitation of both models is that they only fit a single tail of the distribution with only a small portion of the dataset. The restrictive assumption of stationarity can be addressed by allowing the mean (for the GEV) and shape  parameters to vary with time and estimating the corresponding parameters with maximum likelihood. With the GPD, the threshold can also be chosen to vary with time.
The restrictive assumption of stationarity can be addressed by treating parameters of the GPD or GEV distribution as functions of time-dependent covariates \citep{davison1990}.
See \citet{coles} for a more extensive introduction to extreme value theory.

An essential limitation of GPD and GEV models is that they only fit a single tail of the distribution with only a small portion of the dataset.
This practice of ignoring the majority of the available data is particularly concerning when trying to accurately infer any nonstationarities, as these patterns are strongly tied to the behavior in non-extreme observations. \citet{nogaj2007}, \citet{eastoe2009}, and \citet{hess-20-3527-2016} acknowledged this issue and proposed preprocessing of the data with mean and variance functions to capture the nonstationarity. The processed data are assumed to be stationary, and standard GEV/GPD methods can then be used for inference. Analyses using this ``transformed-stationary'' approach demonstrate the value of modeling  nonstationary extremes without ignoring the bulk of the data, but this methodology is limited to a single tail at a time, and is accompanied by standard issues with GEV/GPD fitting, such as selection of a block-size or threshold \citep{coles, scarrott2012}.

A recent model  proposed by \citet{stein2020} fits the entire distribution and has flexible behavior in both tails.
Other works have also attempted to bridge the tails and bulk of a distribution; see \citet{scarrott2012} for a review. These proposals largely rely on mixture models \citep{frigessi2002, carreau2009,bopp2017,yadav2021} or combining function composition with cumulative distribution functions \citep{naveau2016, tencaliec2020, steinextremes}. \citet{huang2019} suggest a semiparametric approach incorporating log-histosplines. A number of the limitations of these approaches, such as flexible behavior in only one tail, restrictions to positive or heavy-tailed variables, or the need to numerically compute a normalizing constant, are obviated by the approach of \citet{stein2020}, which provides a comprehensive approach to handle the bulk and both tails of a distribution.

\subsection{Proposed contributions}
In this paper, we extend the recent work of \citet{stein2020} and demonstrate a flexible model for a random variable whose distribution changes over time. 
In particular, we extend the distribution from \citet{stein2020} to a seasonal model accounting for seasonality, long-term trends, and the interaction between these two characteristics. 
We illustrate the flexibility and ability of the model to capture changing distributions of ground measurements of daily surface temperature. 
In particular, the bulk, lower tail, and upper tail of temperature vary differently across seasons and over the long-scale study period. 
A number of stations representing the diversity of seasonality and long-term changes in temperature are selected from across the United States. 
We highlight the capability of the proposed model to  quantify these changing temperature patterns and extremes. 

We outline the paper as follows: in Section \ref{sec:data}, we describe the daily temperature data used in this study. In Section \ref{sec:model}, we describe the model from \citet{stein2020} and its extension.  Finally, in Section \ref{sec:results}, we present results from applying our methodology to daily temperature data and compare it to alternatives in both the bulk and tails.

%%------------------------------------------------------
\section{Surface air temperature data}\label{sec:data}

%%---------------
%\subsection{Global summary of the day}

\begin{figure}
    \centering
    \includegraphics[scale=.5]{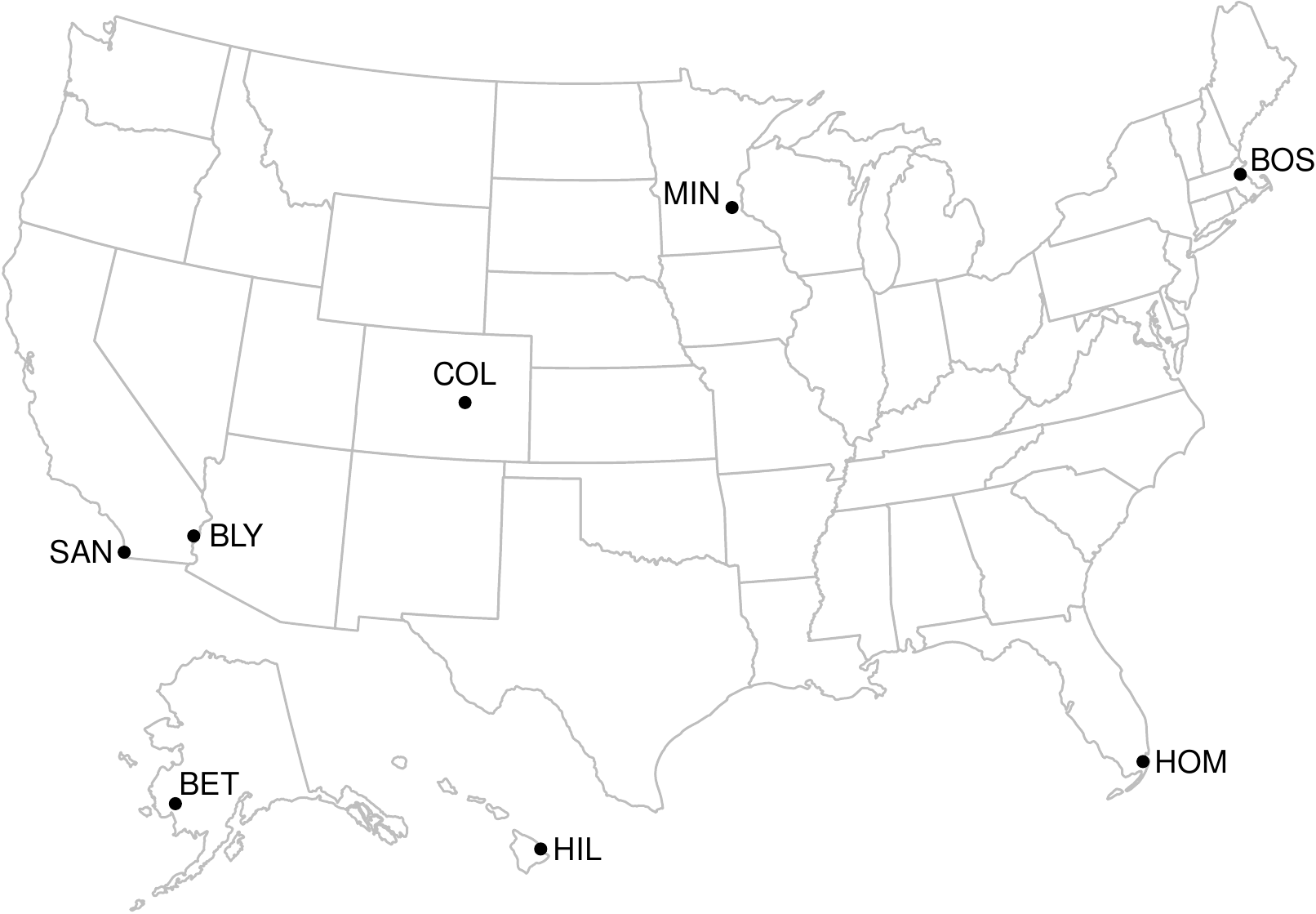}
     \caption{The eight locations from the GSOD database that are studied.}
    \label{fig:locationplots}
\end{figure}

Daily SAT used in this paper are provided by the National Climatic Data Center's Global Surface Summary of the Day (GSOD) \citep{GSOD}. 
The GSOD database contains meteorological measurements from weather stations across the world. 
The \texttt{R} package \texttt{GSODR} \citep{GSODR} offers a helpful interface to work with this dataset. 
We selected several U.S.\ cities from varying locations with different geographies, climates, and local conditions: Bethel, Alaska; Colorado Springs, Colorado; Minneapolis, Minnesota; Boston, Massachusetts; Hilo, Hawaii; San Diego, California; Blythe, California; and Homestead, Florida. Under the K\"{o}ppen climate classification \citep{koppen}, Bethel is a subarctic climate (Dfc), Colorado Springs is a dry semi arid climate (Bsk) at high elevation, Minneapolis is a humid continental region (Dfa) with large seasonal variation, Boston is a mixture between humid continental climate (Dfa) and a humid subtropical climate (Cfa), Hilo is a tropical rainforest which receives significant rainfall (Af), San Diego is a mixture between a Mediterranean climate (Csa) and a semi-arid climate (Bsh), Blythe is a hot desert climate (Bwh), and Homestead is a tropical savannah climate (Aw) that nears a tropical monsoon climate (Am). These eight locations are shown in Figure \ref{fig:locationplots}. Time periods of the SAT observations range from the 1940's to near present (2020). The records present missing data that are not imputed in this study.  Data are recorded on an hourly basis and subsequently averaged to create daily summary values.  Table \ref{tab:stationinfo} in Appendix \ref{app:data} provides additional information about the SAT measurements at these eight stations. 

\begin{figure}
    \centering
    \includegraphics[scale=.22]{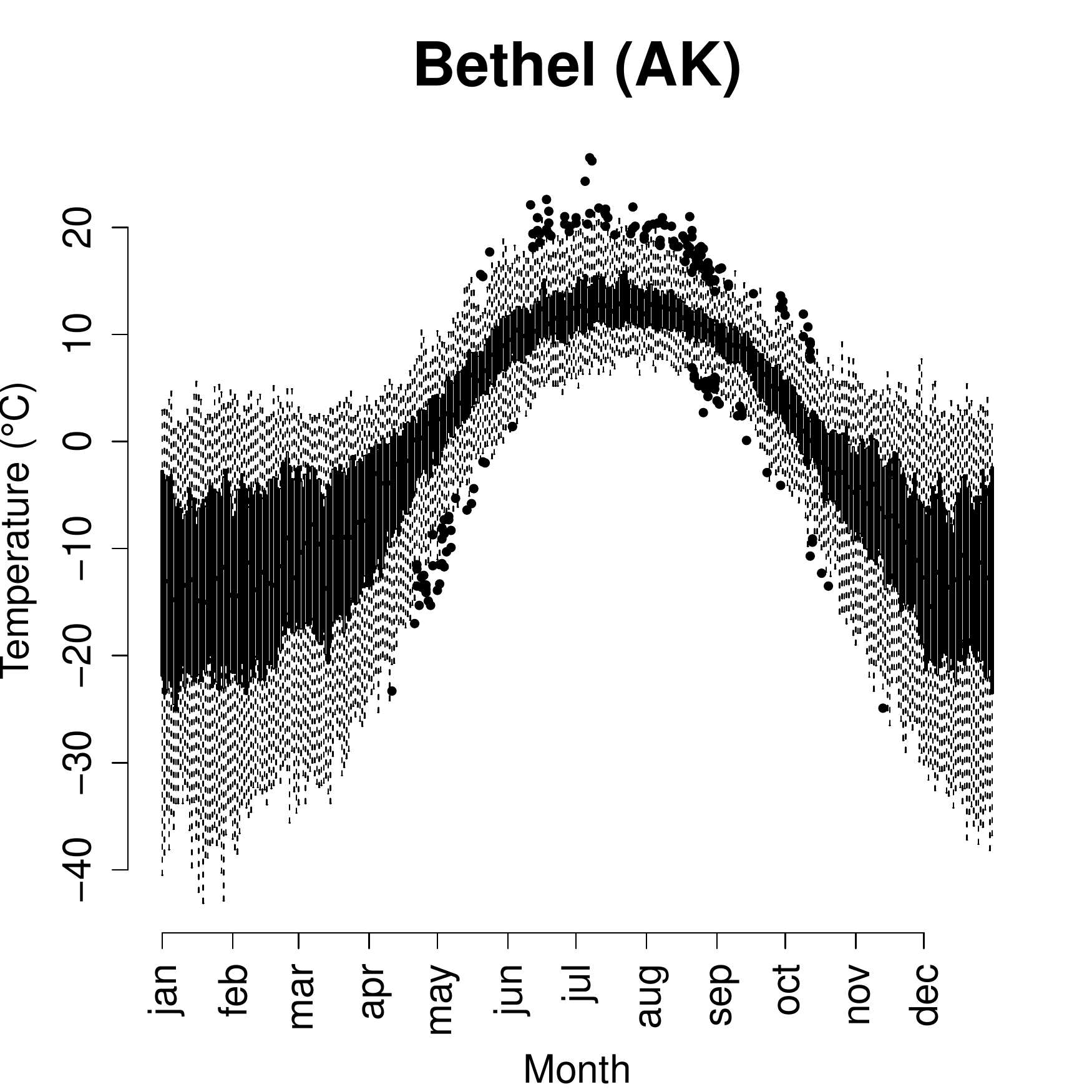} 
    \includegraphics[scale=.22]{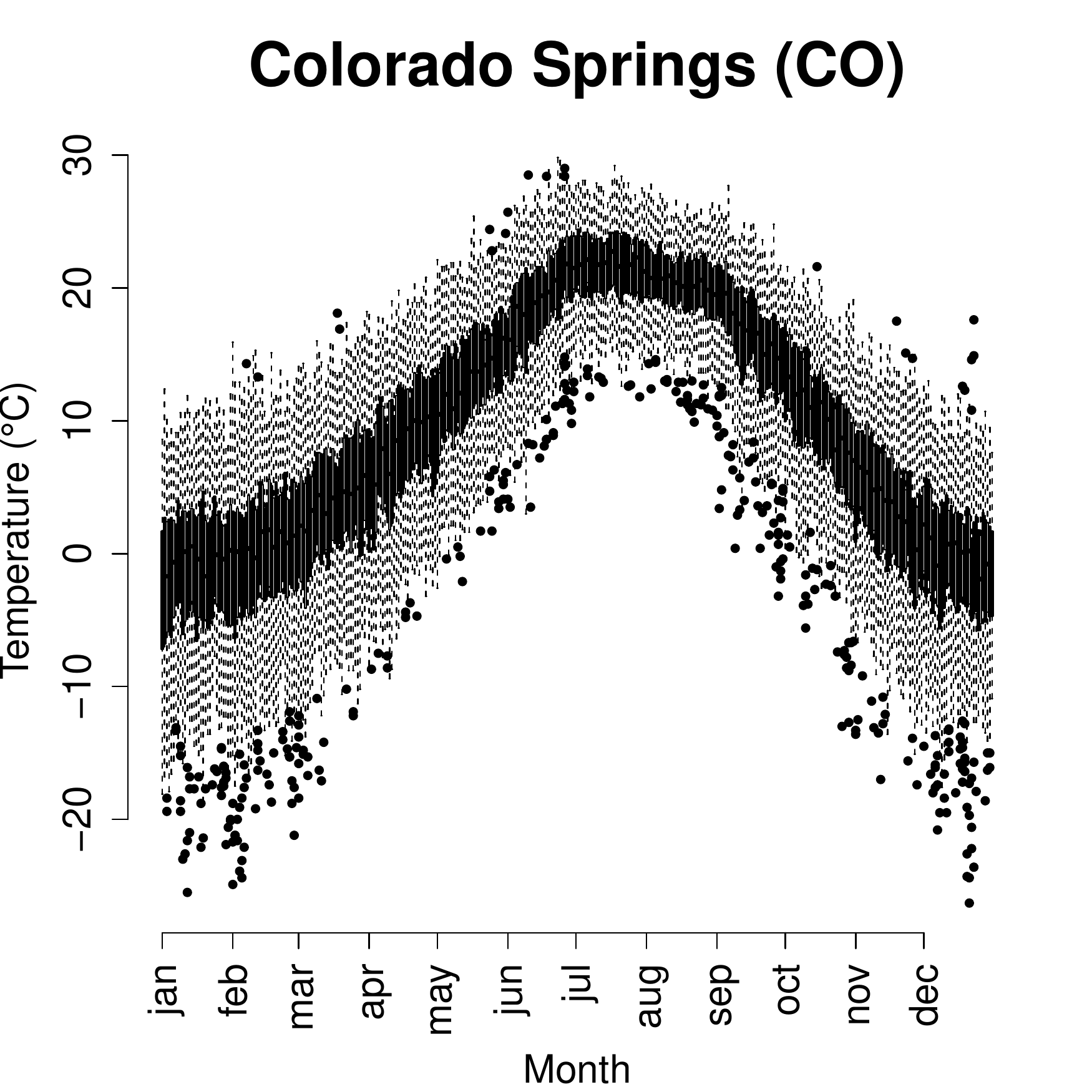}
    \includegraphics[scale=.22]{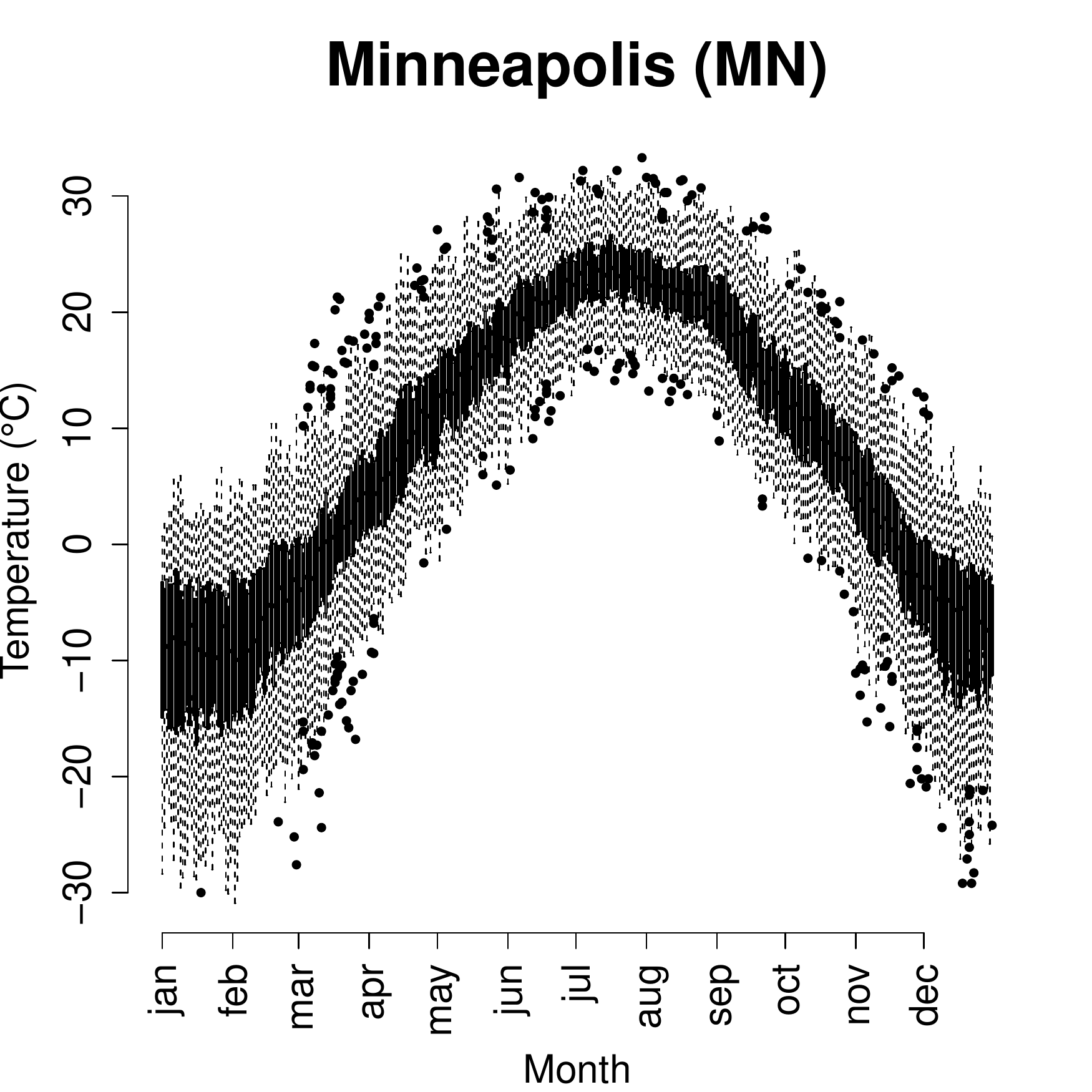}
     \includegraphics[scale=.22]{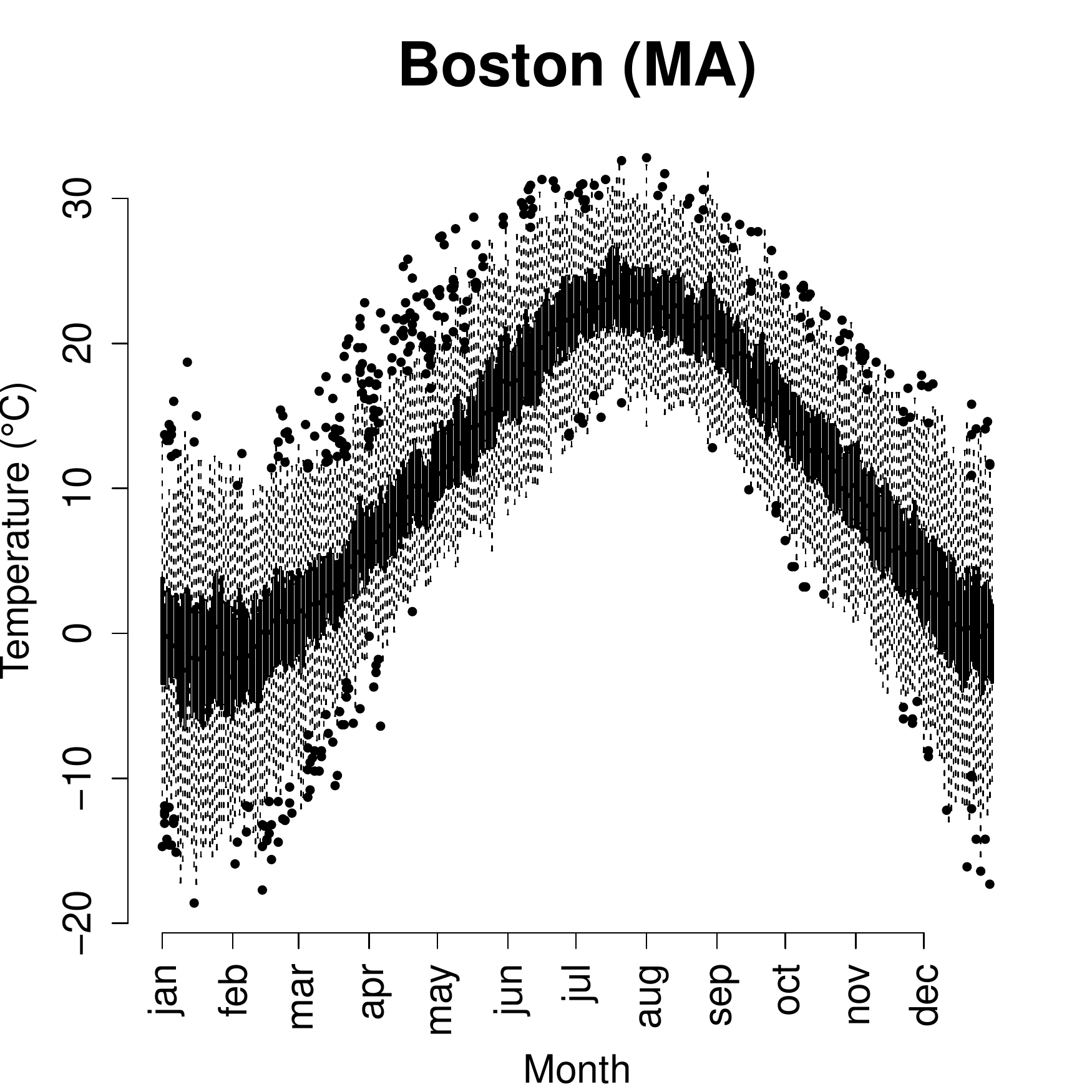}
    \includegraphics[scale=.22]{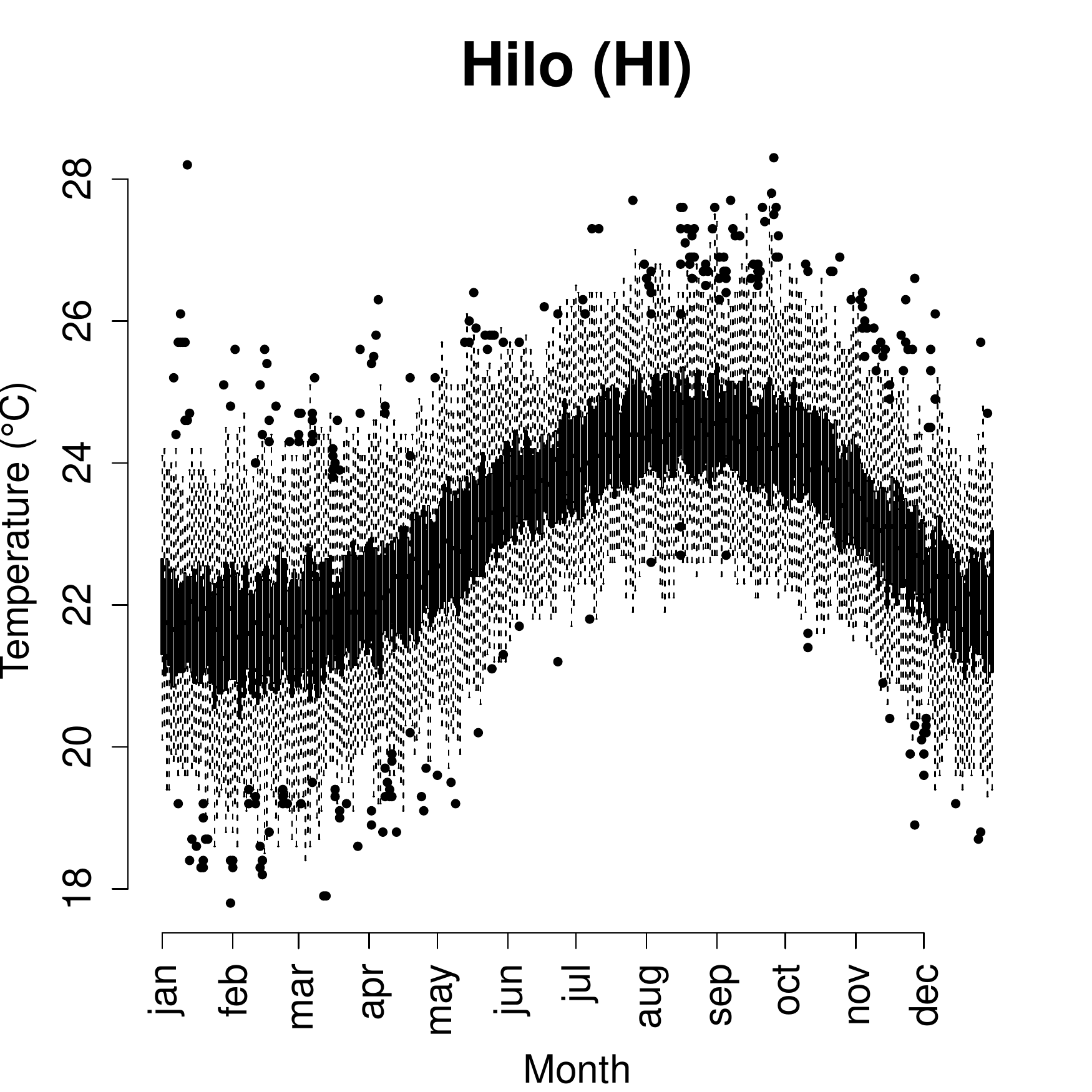} 
     \includegraphics[scale=.22]{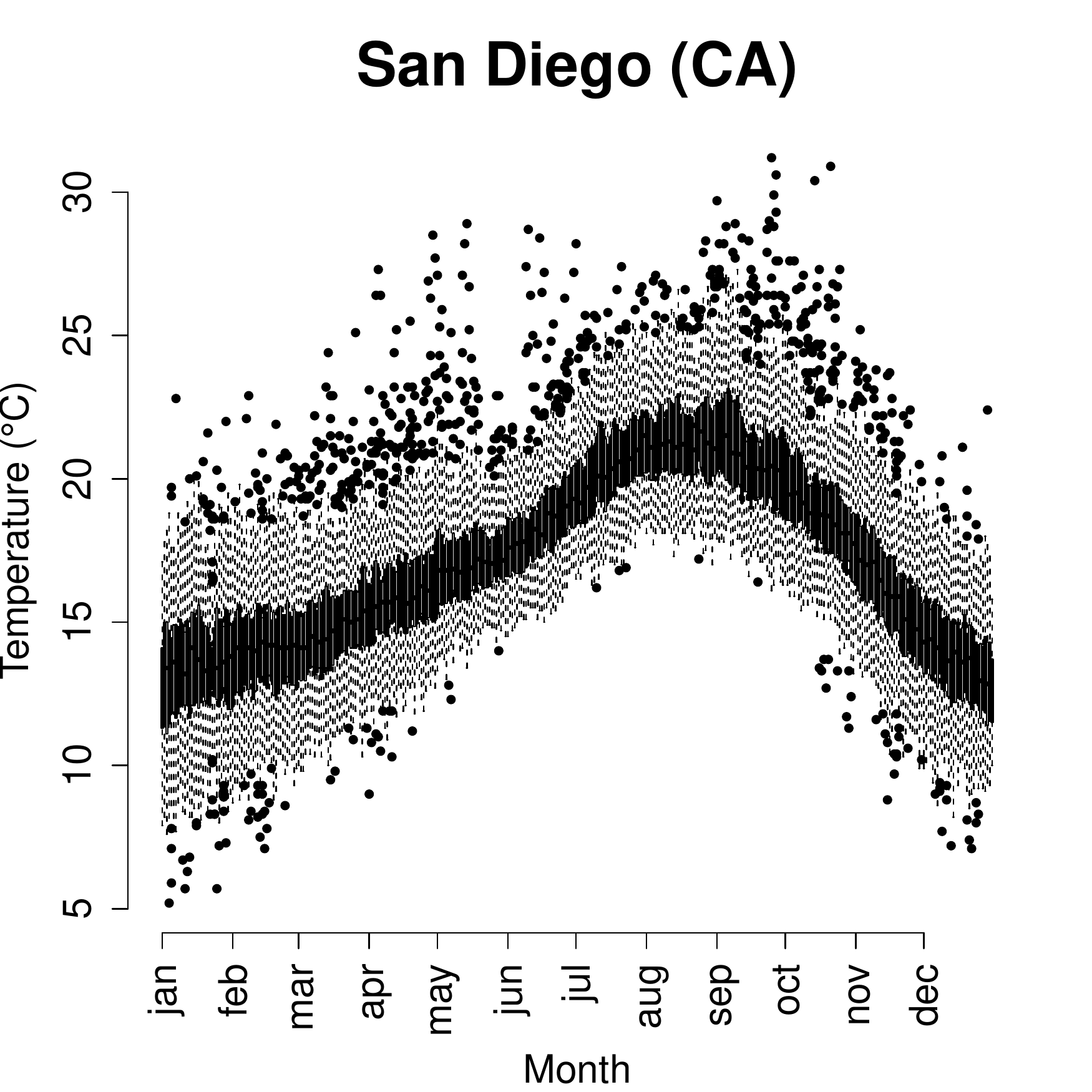} 
     \includegraphics[scale=.22]{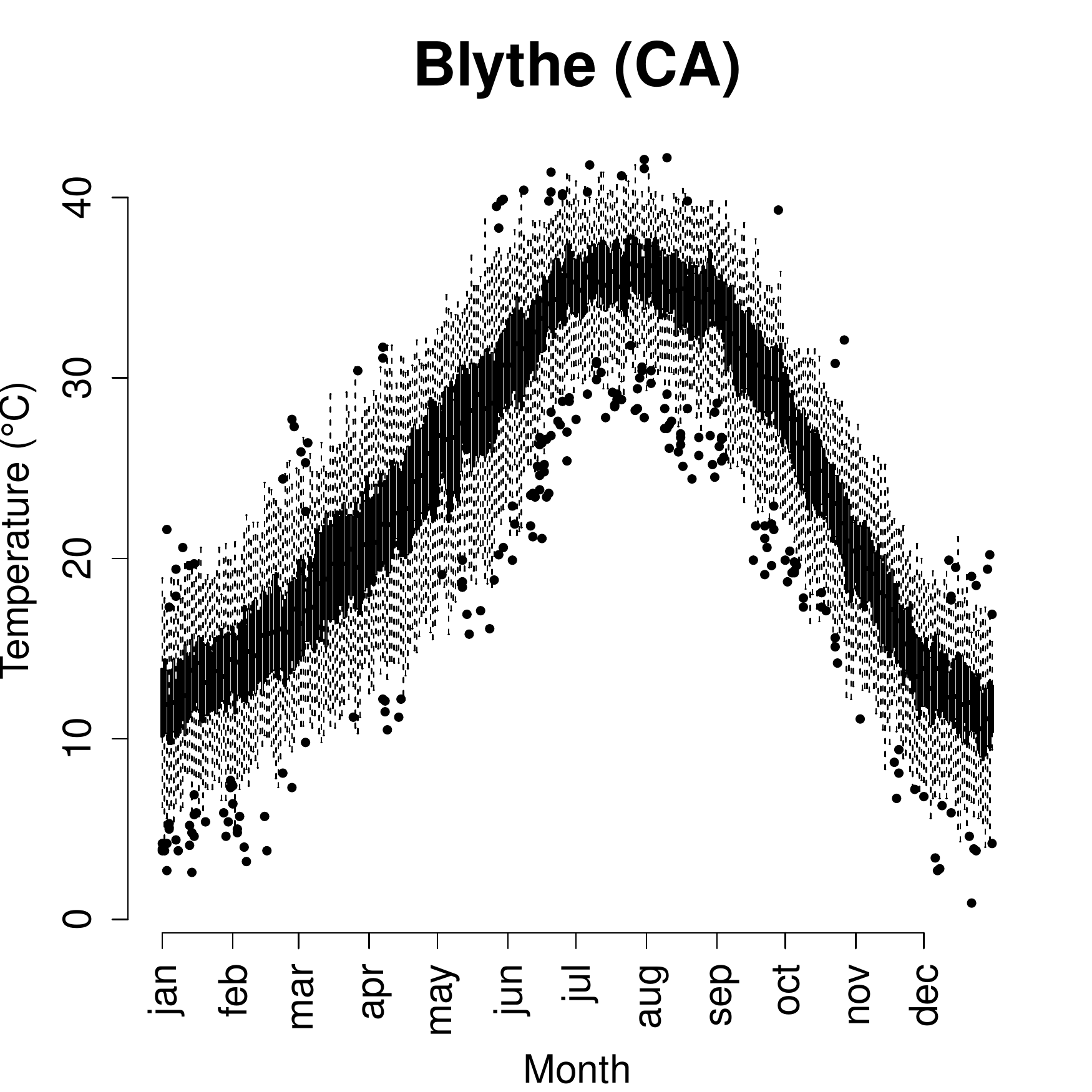}
     \includegraphics[scale=.22]{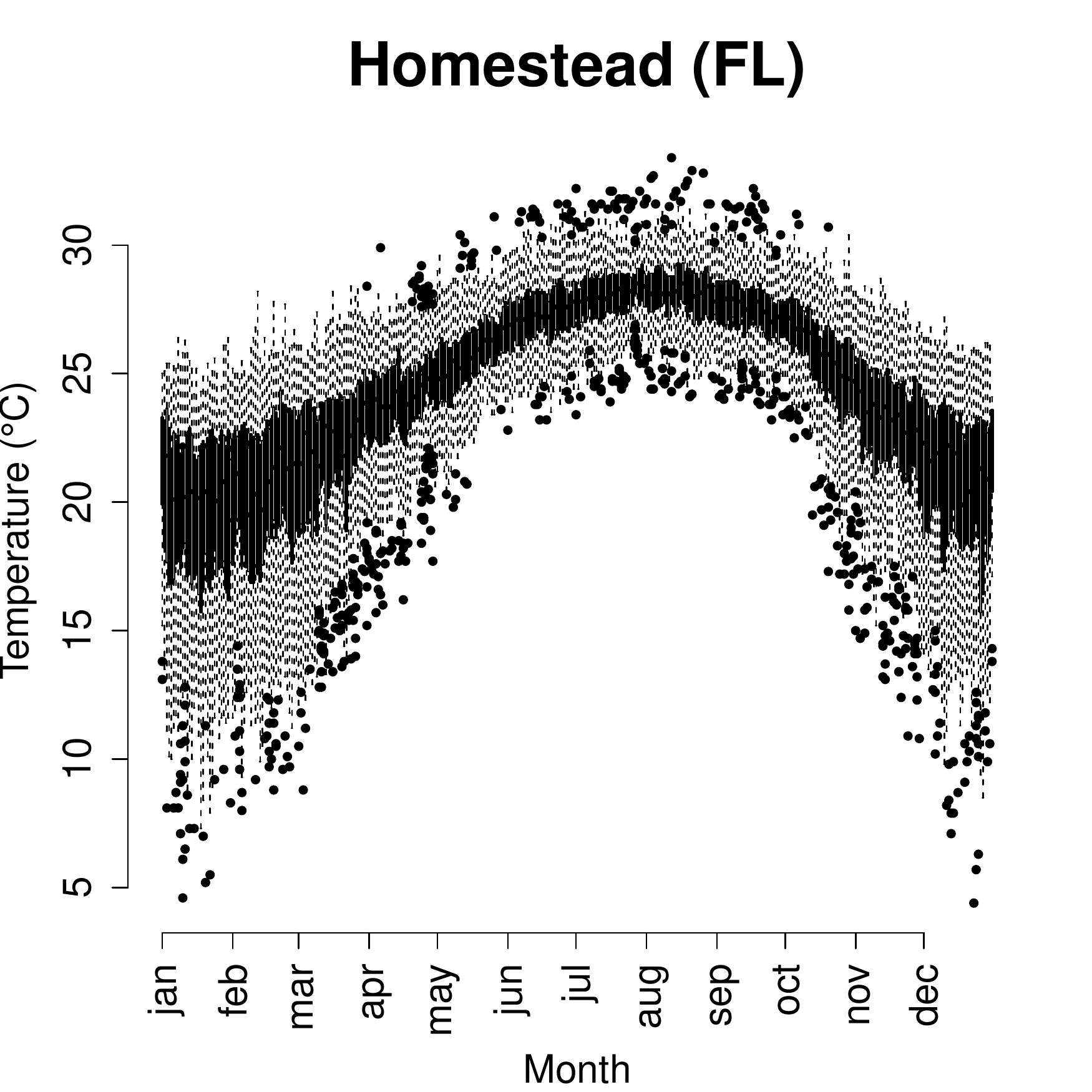}
     
     \caption{Boxplots of daily mean SAT at the stations considered across every day of the year for the entire studied time-period. 
     Boxes of the  boxplot are delimited by  the 0.25 and 0.75  quantiles  $q_{0.25}$ and $q_{0.75}$,  and whiskers are $q_{0.25} - 1.5 \times $IQR  and  $q_{0.75} + 1.5 \times $IQR,  with IQR $= q_{0.75} - q_{0.25}$ denoting the interquartile range. }
    \label{fig:boxplot_season}
\end{figure}

The selected stations across the U.S.\  exhibit  different seasonal patterns of SAT, as shown in boxplots in Figure \ref{fig:boxplot_season} (arranged very roughly in geographic order). 
Most stations experience a slow increase of temperature at the end of winter, followed by a faster temperature decrease at the end of the summer, indicating complex seasonality patterns not captured by a single annual harmonic. Bethel in particular sees long winters and short summers.
The most northern and continental locations (i.e.,\ all cities in the top row of Figure \ref{fig:boxplot_season} and Blythe) exhibit the largest amplitude of seasonality. Coastal stations at Hilo, San Diego, and Homestead experience relatively warm temperatures year round; the climate at Hilo is especially uniform. San Diego has distinctive seasonality during the first part of the year (January to June), showing an almost linear increase in the temperature  with an inflection around June-July.  
Figure \ref{fig:boxplot_season} also illustrates year-to-year variability of SAT on each day of the year.  
Seasonal variability is evident in the bulk of the distributions, represented by the box heights, as well as in the tails. Variability tends to be larger in winter and is particularly strong in the most northern locations such as Bethel. 
Lower and upper tails often exhibit different seasonal variability, which is evident in the various whisker\footnote{See caption of Figure \ref{fig:boxplot_season} for a  description of boxplot features.} lengths and the number of observations beyond the whiskers. 
In Bethel, many more observations lie outside the boxplot whiskers in the summer than in the winter. In Colorado Springs, values beyond the whiskers are more abundant and spread out in the lower tail than in the upper tail. Minneapolis and Homestead also present a lack of values beyond the whiskers in the upper tail during winter months, while Boston and Hilo show this behavior to a lesser extent in the lower tails during summer. The upper tail of San Diego sees the greatest number of observations beyond the whiskers, as well as the most significant spread in extremes. It is  noteworthy that tail heaviness does not imply greater variability, as the SAT range in San Diego is the second smallest of the locations considered (larger only than that of Hilo).
%San Diego also shows the greatest number observations beyond the whiskers of the daily box plots, although the number of occurrences and spread is much more significant in the upper tail. 
The observed asymmetries in tail behavior indicate a departure from Gaussianity and the need for non-standard distribution models.  
%These outliers indicate departure from Gaussianity of these distributions, as well as skewness, and suggest the  need for non-standard distribution models.  
%

%%------------------------------------------------------
\section{Seasonal model for bulk and tails of temperature distribution}\label{sec:model}

In this section, we detail the statistical model used to fit the temperature data described in Section \ref{sec:data}. The proposed model relies on a univariate model for bulk and tails (Section \ref{sec:bat_model}) and is extended to a seasonal model (Section \ref{sec:seasonal_model}) with parameters which are estimated by maximum likelihood (Section \ref{sec:mle}).

%%---------------
\subsection{Model for bulk and tails}  \label{sec:bat_model}
\citet{stein2020} proposed a flexible parametric model which we refer to as the bulk-and-tails (BATs) distribution. The ``s'' is included in the acronym to distinguish from ``bulk-and-tail'' models that tend to have limited flexibility in the lower tail. Consider the random variable $X$ whose cumulative distribution function (cdf) $P(X \le x)$ is given by $T_\nu(H_\theta(x))$, where $T_\nu$ is the cdf of a student-$t$ random variable with $\nu$ degrees of freedom and $H_\theta$ is a monotone-increasing function with six parameters that control the upper and lower tails. To be precise, define $\Psi(x) = \log(1+e^x)$ %, which tends to $0$ as $x \to -\infty$ and behaves linearly as $x \to \infty$, and let
and
\begin{equation} \label{hfunction}
 H_\theta(x) = \left\{ 1 + \kappa_1 \Psi \left( \frac{x-\phi_1}{\tau_1} \right) \right\}^{1/\kappa_1} - \left\{ 1 + \kappa_0 \Psi \left( \frac{\phi_0-x}{\tau_0} \right) \right\}^{1/\kappa_0}  
\end{equation}
where $(\phi_i,\tau_i,\kappa_i)$ are the location, scale, and shape parameters of the upper $(i=1)$ and lower $(i=0)$ tails. 
Like in the GPD and GEV distributions, the parameters $\kappa_i$ control the shape of the tails, with positive values producing a heavy tailed distribution and negative values producing a thin tailed distribution with bounded support in that tail. That is, if $\kappa_1 \ge 0$, the support $[L,U]$ has upper bound $U= \infty$, while if $\kappa_1 <0$, then $U = \phi_1 + \tau_1 \Psi^{-1} (-1/\kappa_1)$. Similarly, $L=-\infty$ if $\kappa_0 \ge 0$ and $L = \phi_0 - \tau_0 \Psi^{-1}(-1/\kappa_0)$ if $\kappa_0 < 0$. The cases $\kappa_i=0$ are defined by continuity. For example, when both shape parameters equal zero, $H_\theta$ reads
\begin{equation*} \label{hfunction0}
 H_\theta(x) = \exp \left( \frac{x- \phi_1}{\tau_1}  \right) - \exp \left( \frac{\phi_0 - x}{\tau_0}  \right).
\end{equation*}
Taking the derivative of the BATs cdf with respect to $x$, we obtain the probability distribution function (pdf) $t_\nu(H_\theta(x)) H'_\theta(x)$, where $t_\nu$ is the pdf of the student-$t$ distribution with $\nu$ degrees of freedom.

\citet{stein2020} showed that this distribution can behave like any three-parameter GPD in either tail. Derivatives of the BATs pdf with respect to model parameters can be calculated analytically, aiding maximum likelihood estimation. Moreover, there is no need to numerically compute any normalizing constant when writing the density, an obstacle often handled with Markov chain Monte Carlo methods \citep{gelman1998,moller2006}. Although the BATs distribution does not directly follow from any first-order limit theorems like in traditional extremes methodology, these properties produce a versatile and practical density for modeling purposes.

In this work, we attempt to model nonstationary data by allowing the scale parameters $\tau_i$ and location parameters $\phi_i$ to change with time, both seasonally and with long-term trends. In some scenarios, it may be appropriate to allow $\nu$ and $\kappa_i$ to also vary with time, but estimation for these parameters is already difficult when they are held constant. 
%As described in the following section, we use the logarithm of CO$_2$  equivalent as a covariate in the location parameters to account for climate change. 
With a relatively simple parameterization for $\tau_i$ and $\phi_i$, we fit our model to daily SAT records at the eight U.S. cities discussed in Section 2.

%%---------------
\subsection{Seasonal extension with long-term trend}\label{sec:seasonal_model}
Here we describe the seasonal model used to fit the daily SAT data to seasonal variations with a long-term trend. 
To capture nonstationary behaviors in SAT, namely seasonality, climate change trends, and their interaction, we introduce covariates which allow parameters of the $H_\theta$ function \eqref{hfunction} to depend upon time. Reasonable choices for seasonal covariates include harmonics or a periodic spline basis; we choose the latter for more flexibility in the main seasonal term, as Figure \ref{fig:boxplot_season} shows evidence of complex seasonal patterns. For a climate change covariate, we use the logarithm of CO$_2$ equivalent obtained from the PRIMAP emission time series \citep{co2data} and freely available for download at \url{https://www.pik-potsdam.de/paris-reality-check/primap-hist/}. The PRIMAP time series is available on a yearly basis through 2018, and we regress PRIMAP values on the historic Mauna Loa CO$_2$ dataset to predict the values at 2019 and 2020. In this study, the log CO$_2$ equivalent serves as a proxy for climate change induced by greenhouse gases. 
Finally, as seasonal patterns can also be affected by the changing climate, an interaction term between seasonality and long-term trend is added, where the seasonality is modeled with annual harmonics. 

Let $y$ represent the year and $d$ represent the Julian day since the first observation, modulo 365.25. Write $S_j(d)$ to denote the $j$\textsuperscript{th} spline basis function at day $d$ and $C(y)$ to denote the value of the log CO$_2$ equivalent at year $y$. Suppressing the subscripts $i=0,1$ for $\tau_i$ and $\phi_i$, we use the following parameterizations:
\begin{align} 
\phi(d,y) &= \alpha_0 + \sum_{j=1}^8 \alpha_{j} S_j(d) + C(y) \left(\beta_1  + \beta_2 \cos \left( \frac{2 \pi d}{365.25} \right)
+ \beta_3 \sin \left( \frac{2 \pi d}{365.25} \right)  \right) \nonumber\\ 
\log(\tau(d)) &= \gamma_0 + \sum_{j=1}^8 \gamma_j S_j(d). \label{steinparameterization}
\end{align}
The long-term trend and its interaction with seasonality are only considered in the location parameters; we represent seasonality in the interaction term with circular harmonics for parsimony purposes. For simplicity, we use the same seasonal basis functions (periodic cubic splines from the \texttt{R} package \texttt{pbs} \citep{pbs}) for the location and scale parameters. We fix the number of seasonal basis functions at eight. %, which in theory is able to capture different behavior at the beginning and end of each of the four seasons. 
Using fewer basis functions at cities with more complicated seasonal temperature variations presented problems with the optimization convergence and also a worsened fit based on model diagnostics. We found that the choices made here provided a good overall compromise between capturing the nuances of the seasonal distributions at some sites and limiting the total number of parameters for reasons of computational stability and controlling model complexity.

%%---------------
\subsection{Maximum likelihood estimation and its uncertainty}\label{sec:mle}
Suppose there are independent observations $x_1,\dots,x_m$ of temperature at a location. The corresponding likelihood is
\begin{equation} \label{likelihood}
\prod_{i=1}^m t_\nu(H_\theta(x_i)) H_\theta'(x_i)
\end{equation}
where $H_\theta$ is defined as in \eqref{hfunction} and $t_\nu$ is the student-$t$ pdf with $\nu$ degrees of freedom. Note that the likelihood \eqref{likelihood} does not explicitly include the temporal dependence between daily observations, although the uncertainty quantification described in Section \ref{sec:mle} does take this dependence into consideration. Ignoring the temporal dependence does not bias point estimates of marginal distributions of individual days \citep{varin2011}.
In total, there are 45 parameters which are estimated via maximum likelihood: 13 parameters each for the upper and lower locations, 8 parameters each for the upper and lower scales, two shape parameters, and the degrees of freedom $\nu$. We enforced two constraints $\kappa_i/\nu > -0.5$ so that the likelihood is twice-differentiable at its endpoints when $\kappa_i$ is negative. Optimization was performed in \texttt{Julia} with the \texttt{IPOPT} solver \citep{ipopt} and automatic differentiation from \texttt{ForwardDiff.jl} \citep{forwarddiff} to efficiently obtain the gradient and Hessian. 
%In this high-dimensional parameter space, different parameter starting values can lead to different local optima. 
For our initial guess, $\nu$ was set to 10, and all trend-related parameters and shape parameters were initially set to zero. Initial seasonal coefficients for the location parameters were obtained from linear regression. Coefficients for the log scale parameters were initially set to zero, except for the intercept term, which was profiled over with a small grid of values. 

Obtaining uncertainties of estimated parameters is not straightforward.
We perform uncertainty quantification with a bootstrapping procedure accounting for temporal dependence. Using stratified block bootstrap resamples, 
it is possible to obtain confidence intervals for any desired function of the parameters with the percentile bootstrap method \citep{efron1993}. Although classical bootstrapping operates under the assumption of independent and identically distributed data, the block bootstrap is commonly used in the setting of temporal dependence \citep{lahiri2003}. 
Under the assumption that sub-annual temporal dependence is stronger than interannual dependence, selecting a block size of a year is natural choice, and we resample years based on a decadal stratification to preserve aspects of  climate change seen over the observation period. That is, to create a bootstrapped dataset, we sample years (with replacement) from each decade, preserving the number of years in each decade from the original dataset. Maximum likelihood estimation is performed on each bootstrapped dataset, and the desired functions of parameters  (e.g.,\ quantiles, changes in quantiles over years, or parameters themselves) are computed for each fit. After ordering the bootstrapped quantities of interest from smallest to largest, pointwise 95\% confidence intervals are obtained as the 2.5 and 97.5 percentiles of the ordered values. Bootstrap parameter estimates were obtained using the local optima from the fit on the full dataset as the initial guess. Example confidence intervals from 200 bootstrap samples are shown later in Figure \ref{fig:bootstraps} in Section \ref{sec:climatechange} and in Table \ref{tab:kappaboottable} in Appendix \ref{app:param}. 
We emphasize that these uncertainty estimates are  computed for the daily marginal distribution of SAT.

% In Figure \ref{fig:bootstraps}, we display bootstrap uncertainties for quantiles at three of the stations with high quality records which go back to the 1940s. Pointwise 99\% confidence intervals are obtained from the percentile bootstrap method with 200 stratified block bootstrap resamples where the block length is a year and stratified resamples are taken over each decade. 
% Different levels of uncertainty are observed for the various quantiles; more precisely, these uncertainty levels also vary with the spatial location. 
% The median quantile presents the least estimation uncertainty over the different locations, whereas the higher/lower quantiles exhibit the most estimation variability. 
% More specifically, Bethel (AK) shows a higher uncertainty in the cold quantiles compared with the other quantiles. As we will see later, the lower quantiles in Bethel are experiencing the most rapid warming among our eight observation sites, so this large uncertainty is unsurprising. Boston (MA) present similar uncertainties for both cold and hot high quantiles, whereas San  Diego (CA) shows the most estimation uncertainty in its warmer quantiles. In other stations with more missing data, the choice of block size and corresponding stratification may not be so clear. Nonetheless, the purpose of this short study was to show that bootstrapping provides an avenue for uncertainty quantification under our framework with a high-dimensional parameter space and various temporal patterns in the observations.

%%------------------------------------------------------
\section{Results on changing daily mean temperature distributions} \label{sec:results}
%%---------------

In this section, we present results of the proposed model fit to daily mean SAT. 
Section \ref{sec:quantiles} provides a visual evaluation of the quality of the fit models.
Section \ref{sec:CVcomparison} shows a quantitative comparison of the proposed model to several benchmark models. In particular, we compare with skew-normal and generalized Pareto distributions to respectively assess the bulk and each tail of the fit distribution. 
Finally, in Section \ref{sec:climatechange}, we highlight how the proposed model is able to capture the changing seasonal patterns due to long-term trends. 

To begin, we describe the benchmark distributions which are compared to the BATs distribution. 
First, with less of an emphasis on the tails of the distribution, we compare to a skew-normal distribution with time-varying location, scale, and skewness parameters. The skew-normal distribution was found to provide reasonable fits for temperatures in \citet{steinextremes}. 
With $\phi$ and $\Phi$ respectively denoting the pdf and cdf of a standard normal random variable, the skew-normal pdf is 
\begin{equation} \label{skewnormal}
    f(x) = \frac{2}{\sigma} \phi\left( \frac{x - \mu}{\sigma} \right) \Phi\left( \alpha \left( \frac{x - \mu}{\sigma} \right) \right)
\end{equation}
where $\alpha$ is a skewness parameter, $\mu$ is a location parameter, and $\sigma>0$ is a scale parameter. We use the same spline-basis parameterization for the location and scale parameters as in \eqref{steinparameterization} and also let the skewness vary in time using the same parameterization as the location parameter but without a climate-change covariate. Maximum likelihood estimation for the skew-normal model is performed in Julia.

To focus on the tails of the distribution, we compare with nonstationary GPD models. The GPD pdf is
\begin{equation} \label{GPDdensity}
 f(x)=   \frac {1}{\sigma } \left(1+\frac {\xi (x-\mu )}{\sigma }\right)^{-1-1/\xi}
\end{equation}
where $\mu$ is a user-specified threshold, $\xi$ is a shape parameter, and $\sigma>0$ is a location parameter. If the shape parameter is negative, the tail is bounded; if the shape is nonnegative, the tail is unbounded. Specifically, the GPD support is $[\mu,U]$ where $U = \mu - \sigma / \xi$ if $\xi < 0$ and $U =  \infty$ if $\xi \ge 0.$
We construct the GPD threshold $\mu$ as a quantile regression\footnote{See Appendix \ref{app:qr} for a description of  quantile regression.} at the $p_\mu=0.95$ quantile, using eight periodic spline basis functions as covariates. When considering the lower tail, we can multiply the data by $-1$ and work in the typical peaks-over-threshold setting. For GPD parameters, we consider a constant shape parameter and a temporal scale parameter whose logarithm varies in time like the location parameter in \eqref{steinparameterization}.  All GPD models were fit by maximum  likelihood with the \texttt{R} package \texttt{extRemes} \citep{extRemes}.

%%---------------
\subsection{Seasonal quantile evaluation}\label{sec:quantiles}
\begin{figure}
    \centering
    \includegraphics[scale=.22]{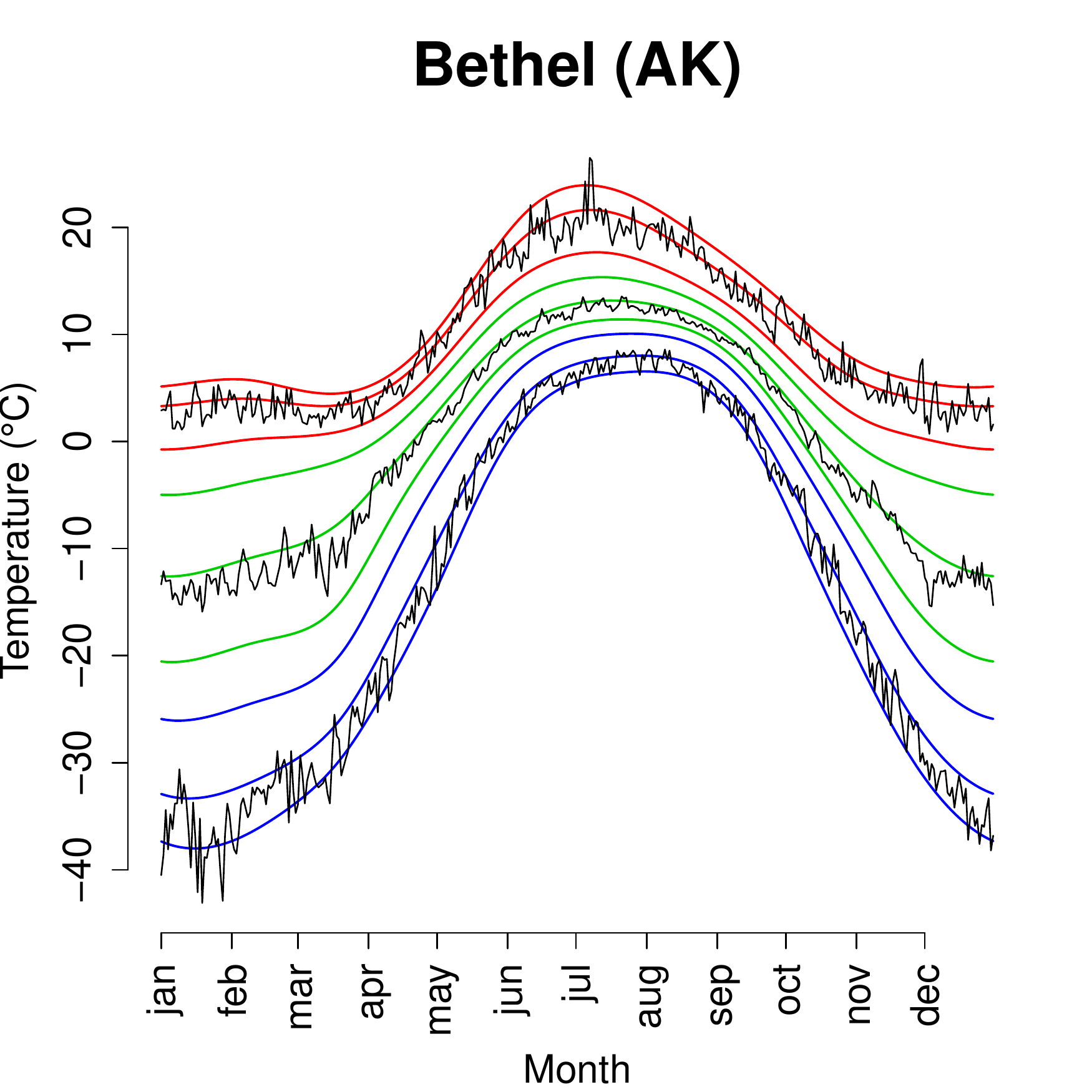}
    \includegraphics[scale=.22]{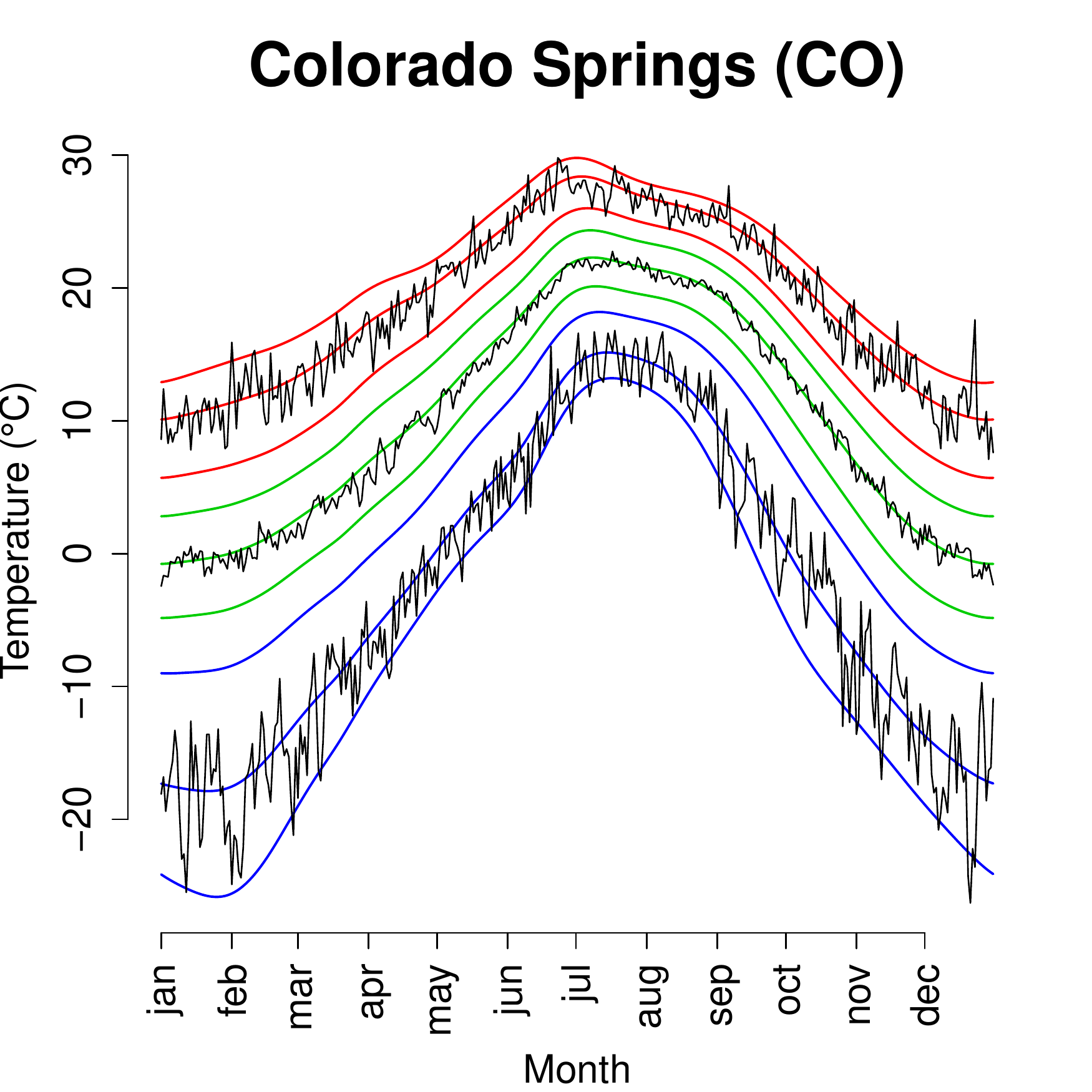}  
    \includegraphics[scale=.22]{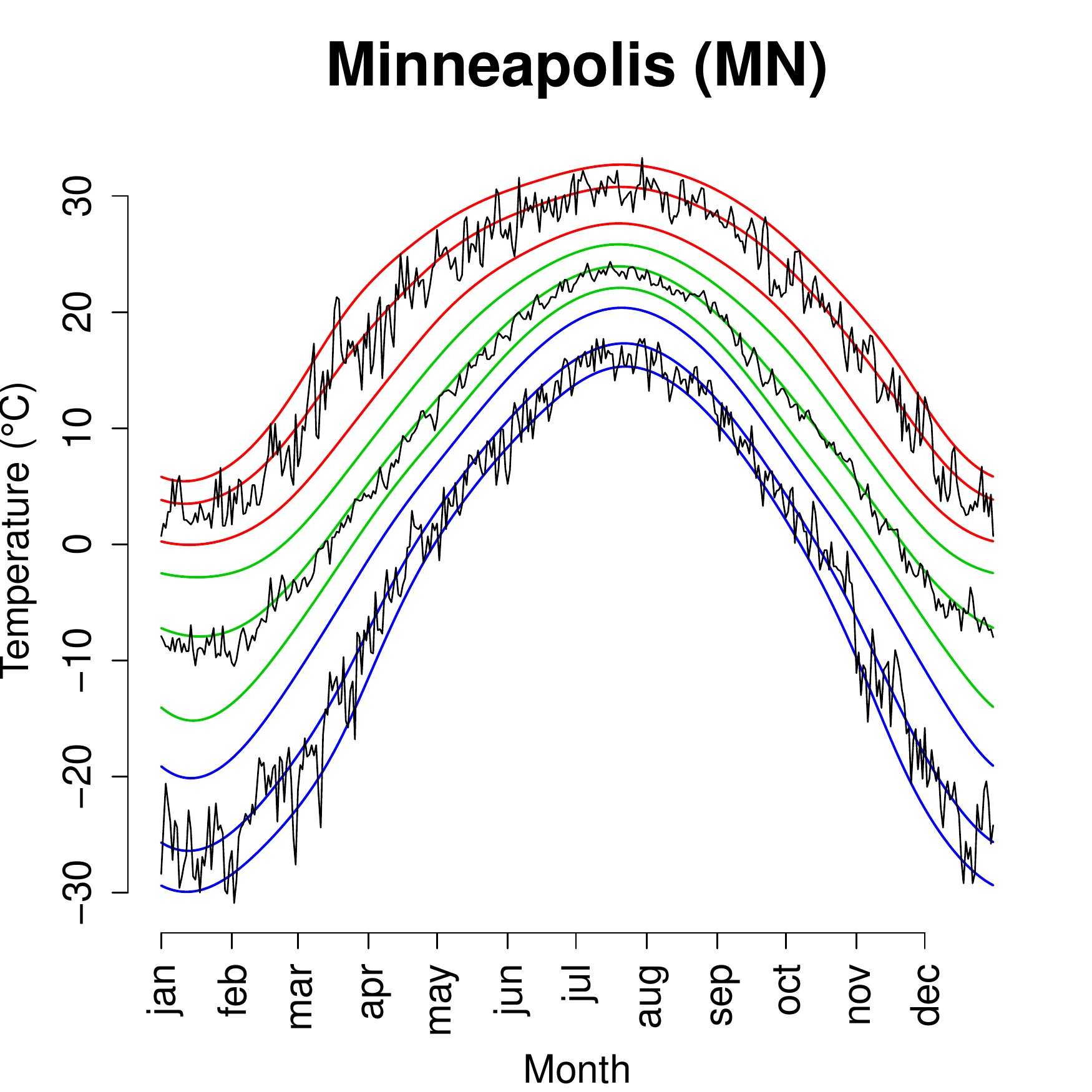} 
    \includegraphics[scale=.22]{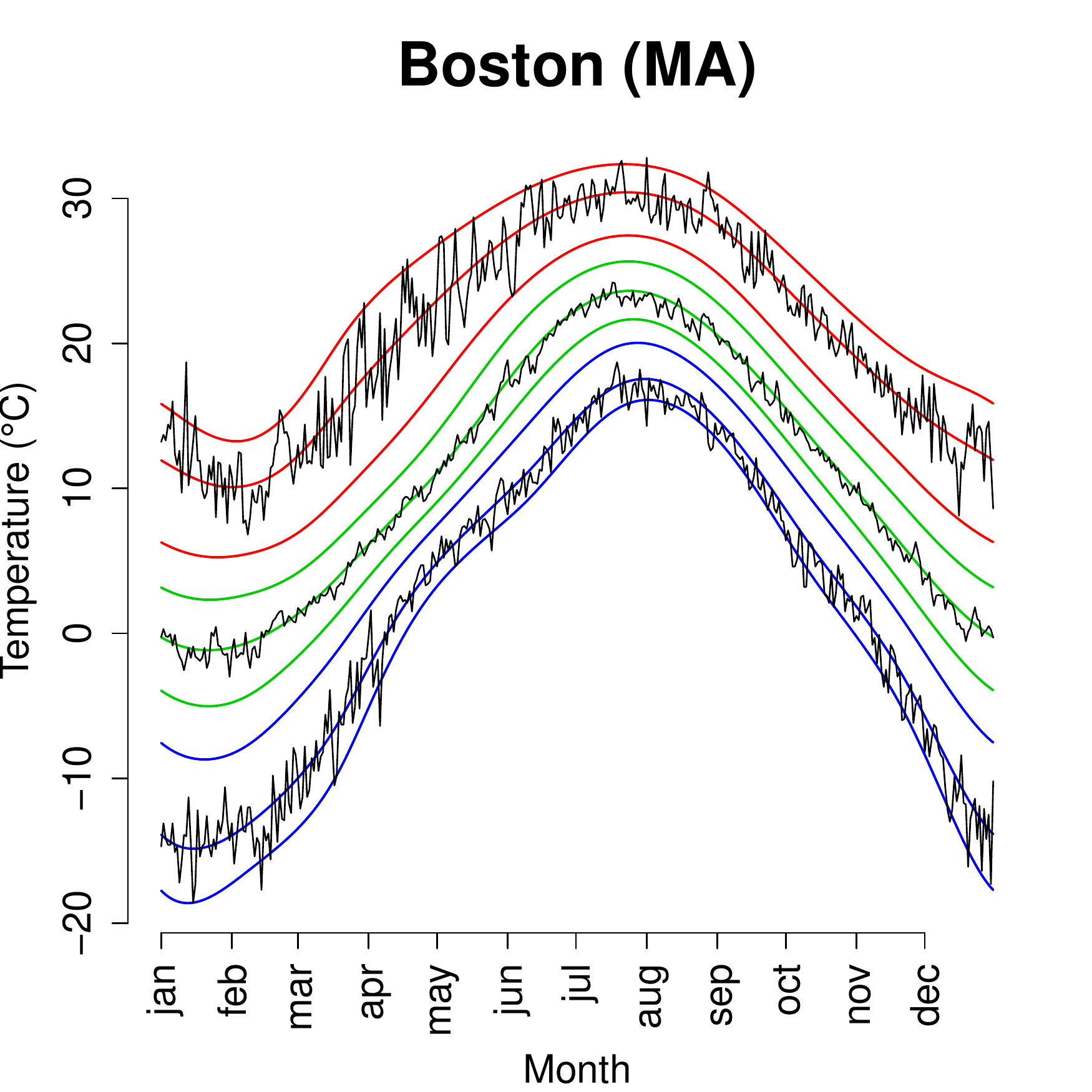}
    \includegraphics[scale=.22]{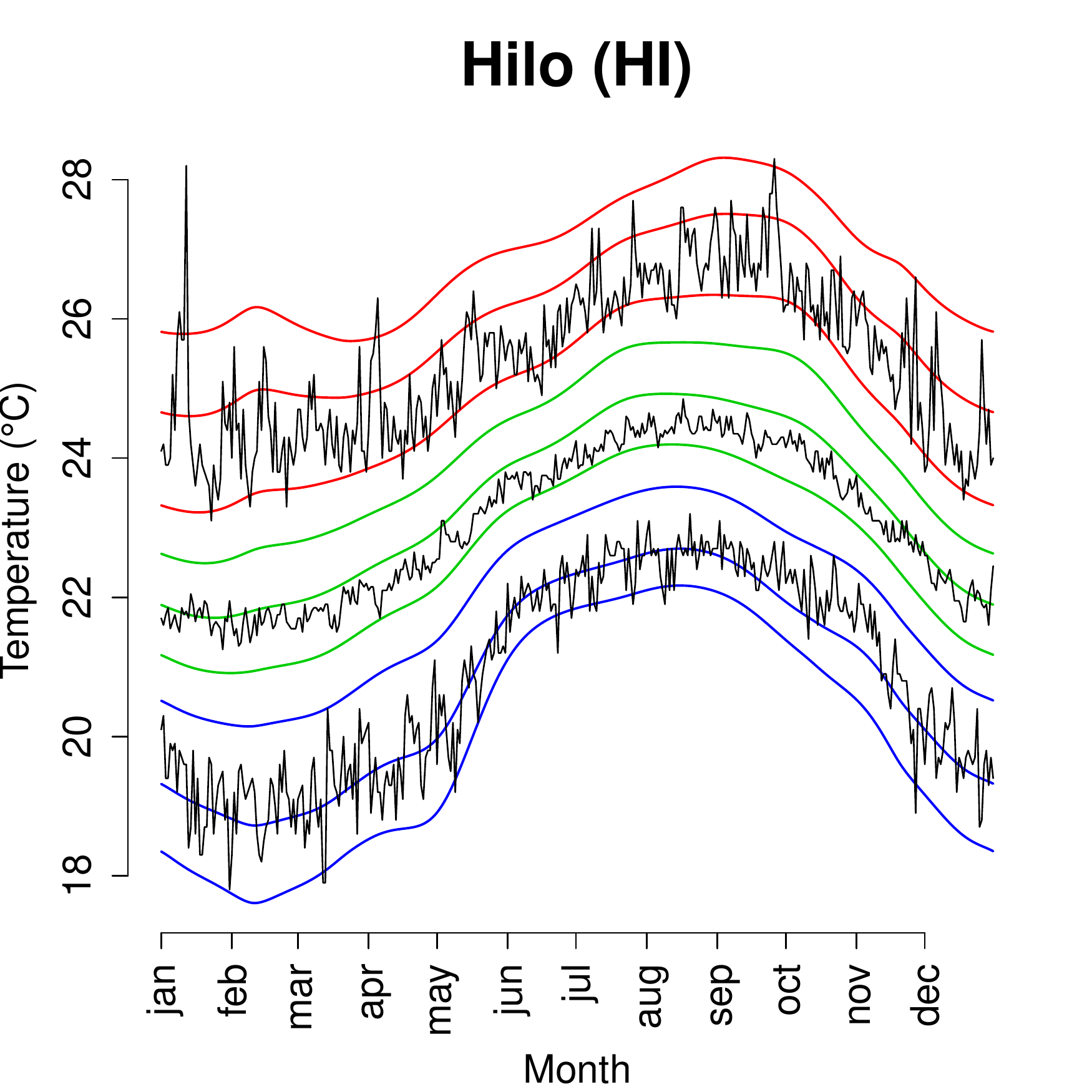} 
     \includegraphics[scale=.22]{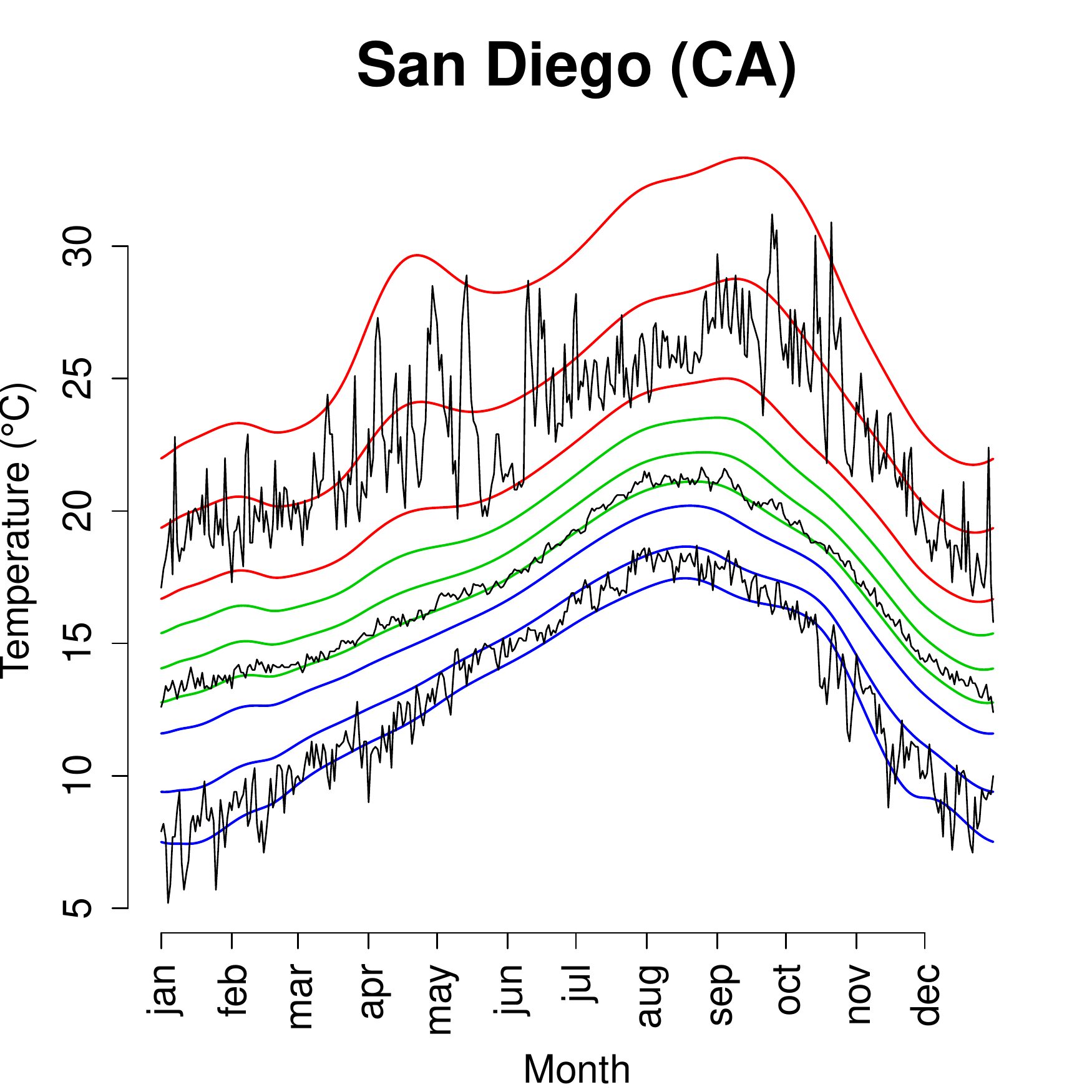} 
    \includegraphics[scale=.22]{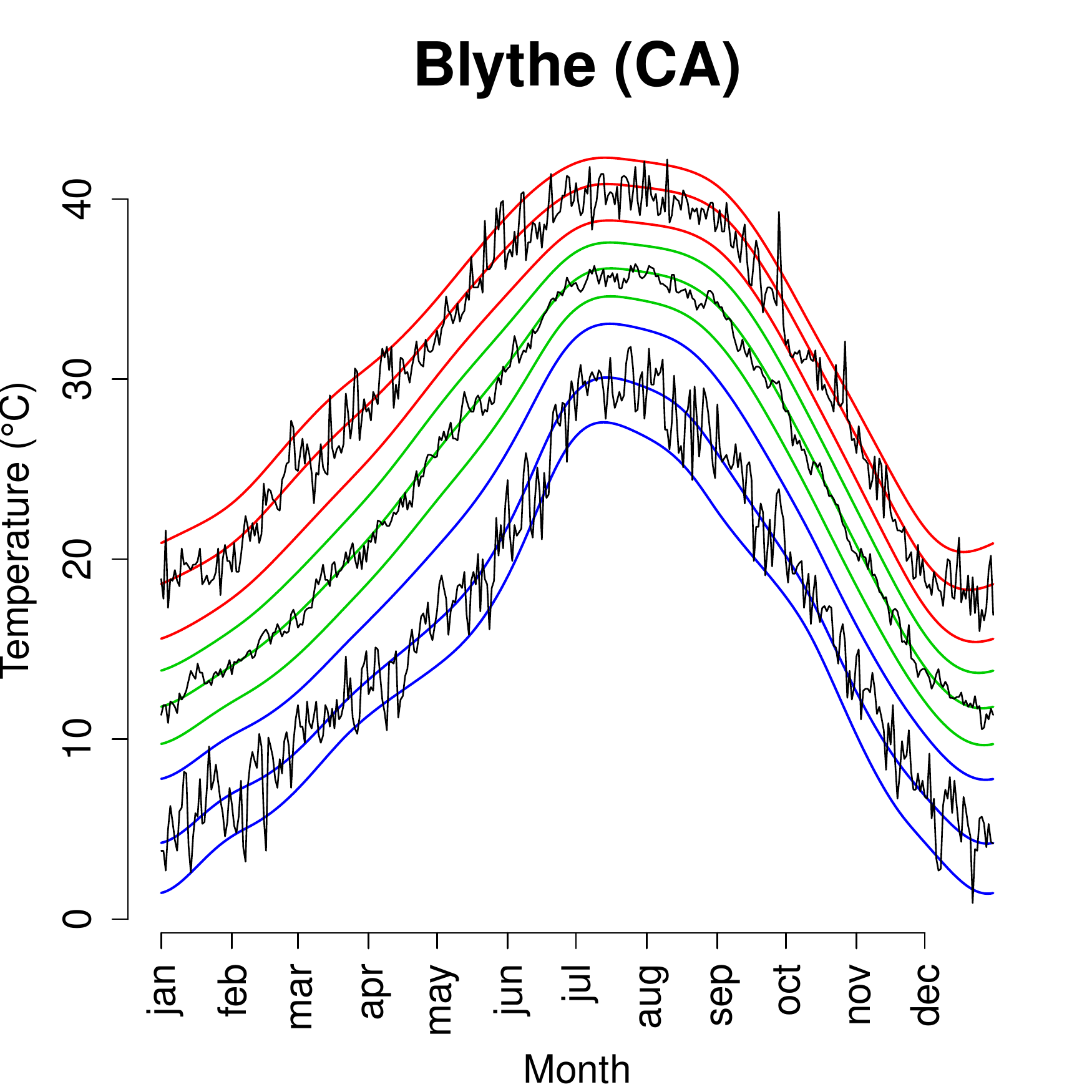}
    \includegraphics[scale=.22]{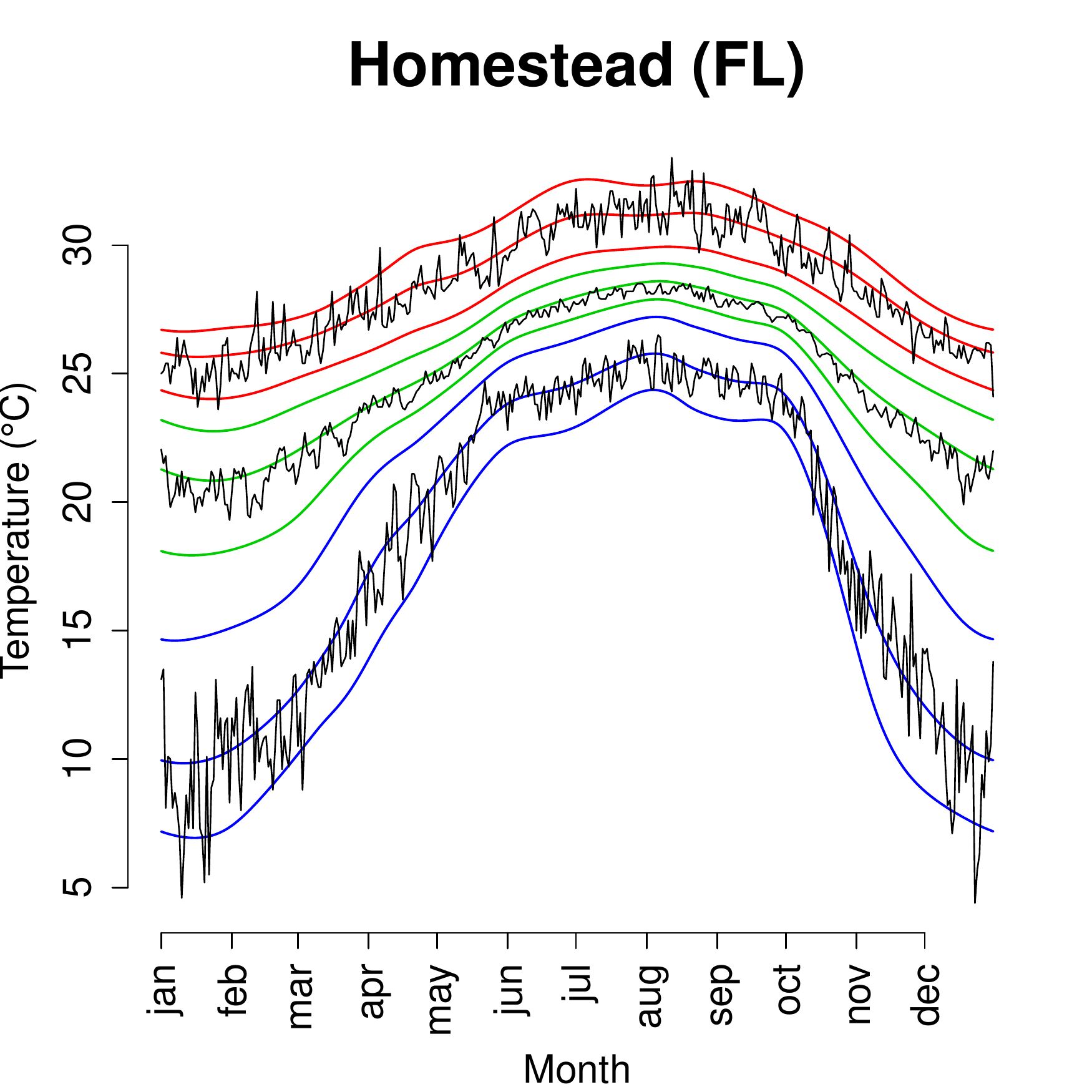}

    \caption{BATs quantile estimates for the year 2020: 0.001, 0.01, 0.1 (blue), 0.25, 0.5, 0.75 (green), and 0.9, 0.99, 0.999 (red). %Solid line is the BATs estimate, dashed line is a quantile regression. 
    Black lines show the minimum/median/maximum observation for that day of year, taken over all years.}
    \label{fig:quantiles}
\end{figure}

%This section contains results from our model fit. 
Given the model parameter estimates, it is possible to express the distribution of daily mean SAT at any day and year, provided that the log CO$_2$ equivalent is available for that year. 
First, in Figure \ref{fig:quantiles}, we examine estimated quantiles during the year 2020. Quantiles from the BATs model are shown for values 0.001, 0.01, 0.1 (blue), 0.25, 0.5, 0.75 (green), and 0.9, 0.99, 0.999 (red). 
%For comparison, the dashed line shows the fit from a quantile regression with the same parameterization as the location parameters in \eqref{steinparameterization}. 
The black lines in each panel are observed daily maximum, minimum, and medians taken across each day of the year over all years. These maximum and minimum values serve as a proxy for upper and lower tail descriptions which would require a large amount of data to quantify precisely. We also emphasize that these maximum and minimum values are taken from the entire time series, unlike the BATs curves that are shown for the year 2020.
Overall, the fitted BATs model provides a very flexible representation across the year for the bulk and both tails of the distributions, capturing: a) larger spread in both tails and sometimes in the interquartile range during winter (e.g.,\ Colorado Springs and Homestead); b) different spread in the lower and upper tails (e.g.,\ San Diego); c) different seasonality patterns in each tail and in the interquartile range, such as in Bethel, where the upper and lower tails exhibit significantly different seasonal patterns; d) and asymmetric seasonal patterns across the year for the eight stations where the temperature increase at the end of the winter is slower than the decrease in early fall.   
% At San Diego, there is a visible discrepancy in the estimated median of the data with the observed median, which is perhaps the  most obvious indicator of climate change from just these plots. 
%The difference between the BATs and quantile regression curves is larger in the more extreme quantiles, as the more extreme quantile regression curves must envelop more of the maximum/minimum observations.
Quantile regression can be used to study quantiles and tail behaviors in similar fashion to Figure \ref{fig:quantiles}, but this approach typically involves separate analyses of the desired quantiles and comes with the possibility of producing crossing quantile curves.
%Although none of the quantile regression curves cross, it is a potential issue when using this technique to study extremes.
%, especially when looking further in the tails. 
To avoid crossing quantiles in quantile regression, practitioners will need to turn to specialized methods \citep{hequantile, quantilecross1, cannon}.
The use of the BATs model for the entire distribution eliminates the risk of crossing quantiles.
%\julie{do we want to highlight these features as things we learn about local climates and discuss them further?}
%Stations present similar quality fits, further supporting the flexibility of the seasonal BAT model. 

An interesting point in Figure \ref{fig:quantiles} is the large spike in January in Hilo, which corresponds to heavy rainfall and flash flooding on January 12, 2020. In terms of the BATs model, this event corresponds to the $1-(8 \times 10^{-6})$ quantile, which is extremely far in the estimated tail of the distribution given the approximately 50-year duration of the Hilo data. After refitting with this value removed, the January 12 value unsurprisingly lies even further in the tail at the $1-(2 \times 10^{-7})$ quantile, and the upper tail behavior $\kappa_1/\nu$ (see Table \ref{tab:kappaboottable}) changes from $-0.008$ to $-0.05$. While this example illustrates that a single outlying observation can have a drastic impact on tail behavior, it is noteworthy that the BATs lower tail behavior $\kappa_0/\nu$ is unaffected by the January 12 2020 observation and remains around $-0.05$ even after removal. %\julie{i am not sure about keeping the following sentence: }Another dataset in a nearby location did not possess such an outlier, raising questions about the quality of this observation, but we proceed without removing this point from consideration. 
Overall, the estimates appear to provide good fits to the seasonal evolution of SAT in the different locations considered.

\begin{figure}[H]
    \centering
    \includegraphics[scale=.22]{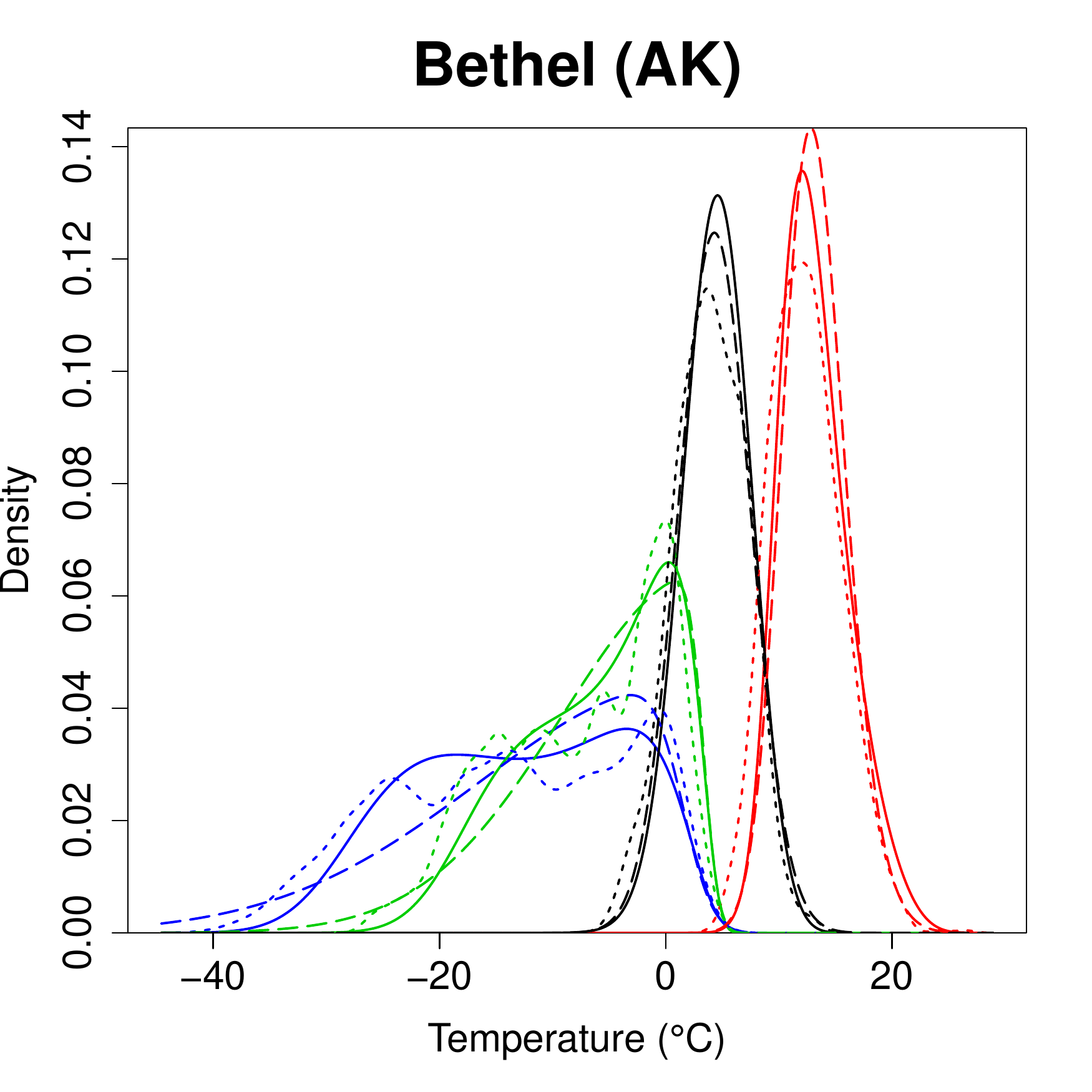} 
    \includegraphics[scale=.22]{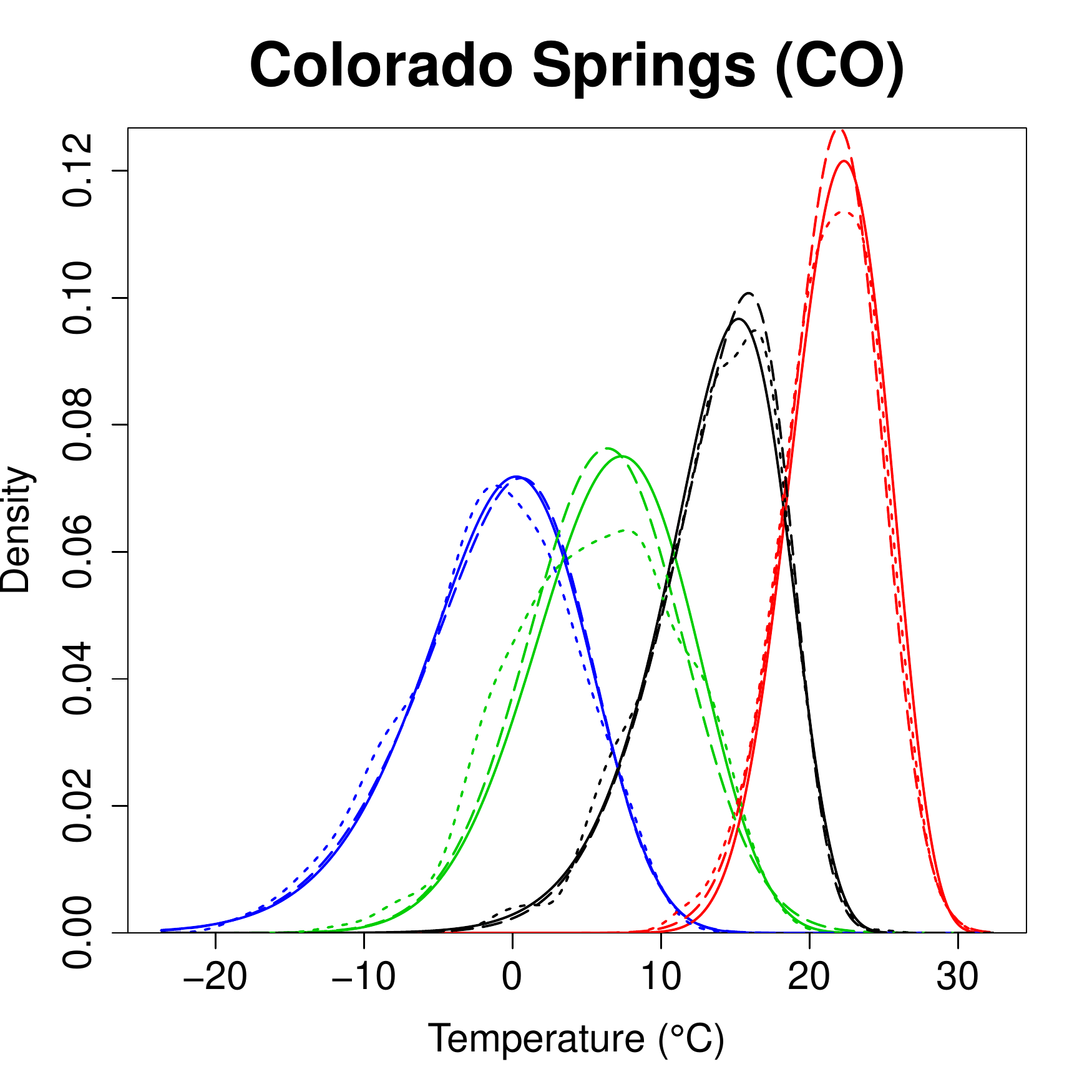}
    \includegraphics[scale=.22]{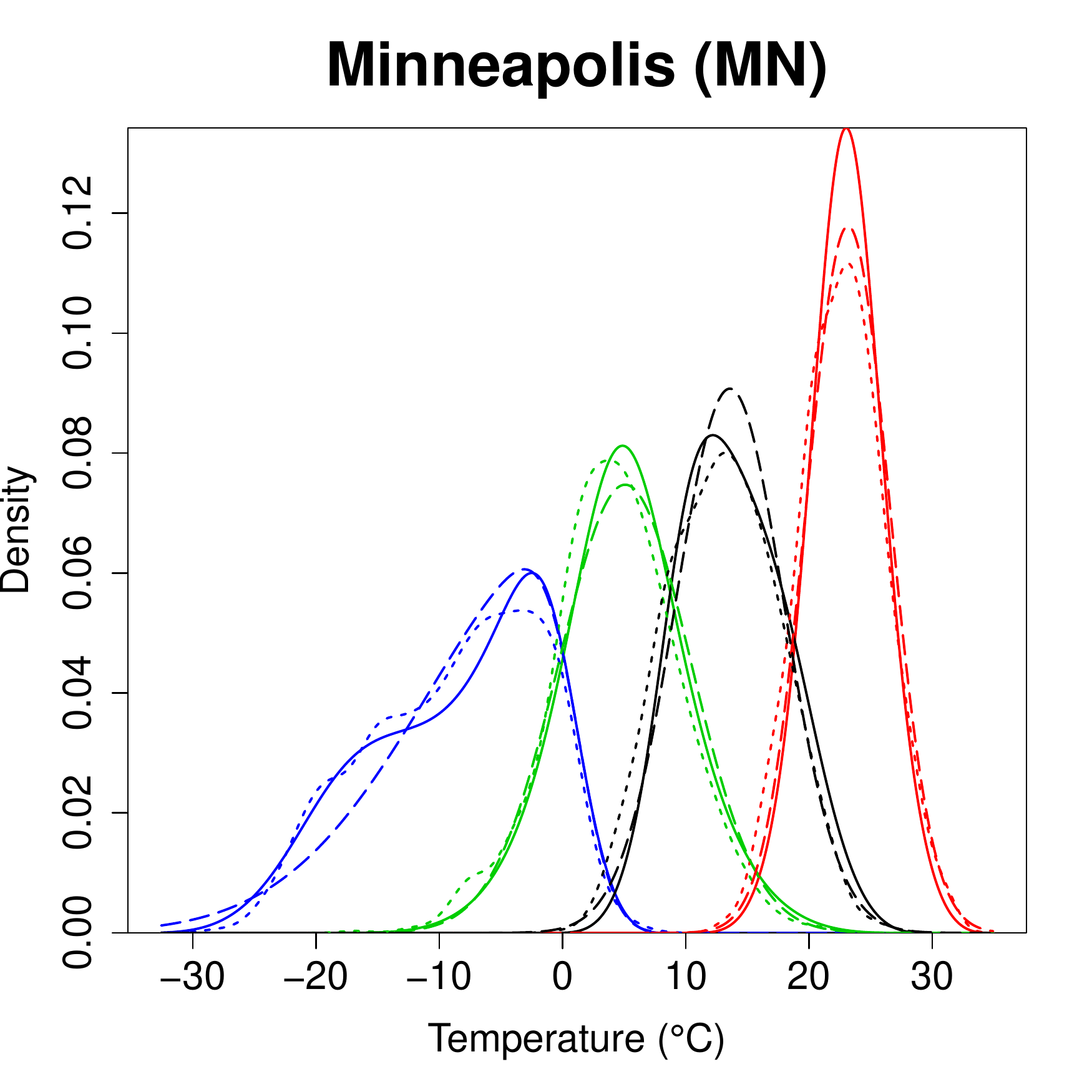}  
    \includegraphics[scale=.22]{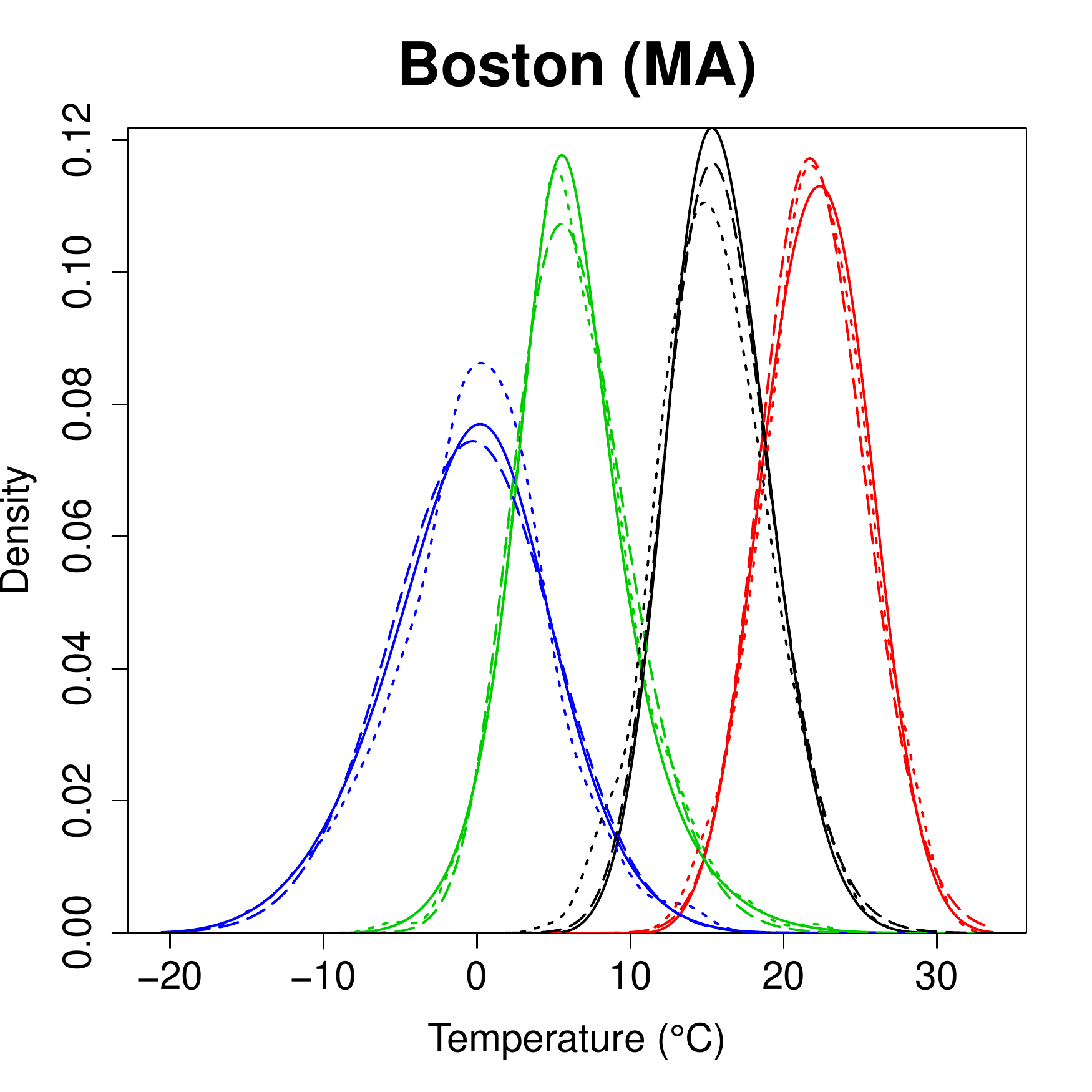}
    \includegraphics[scale=.22]{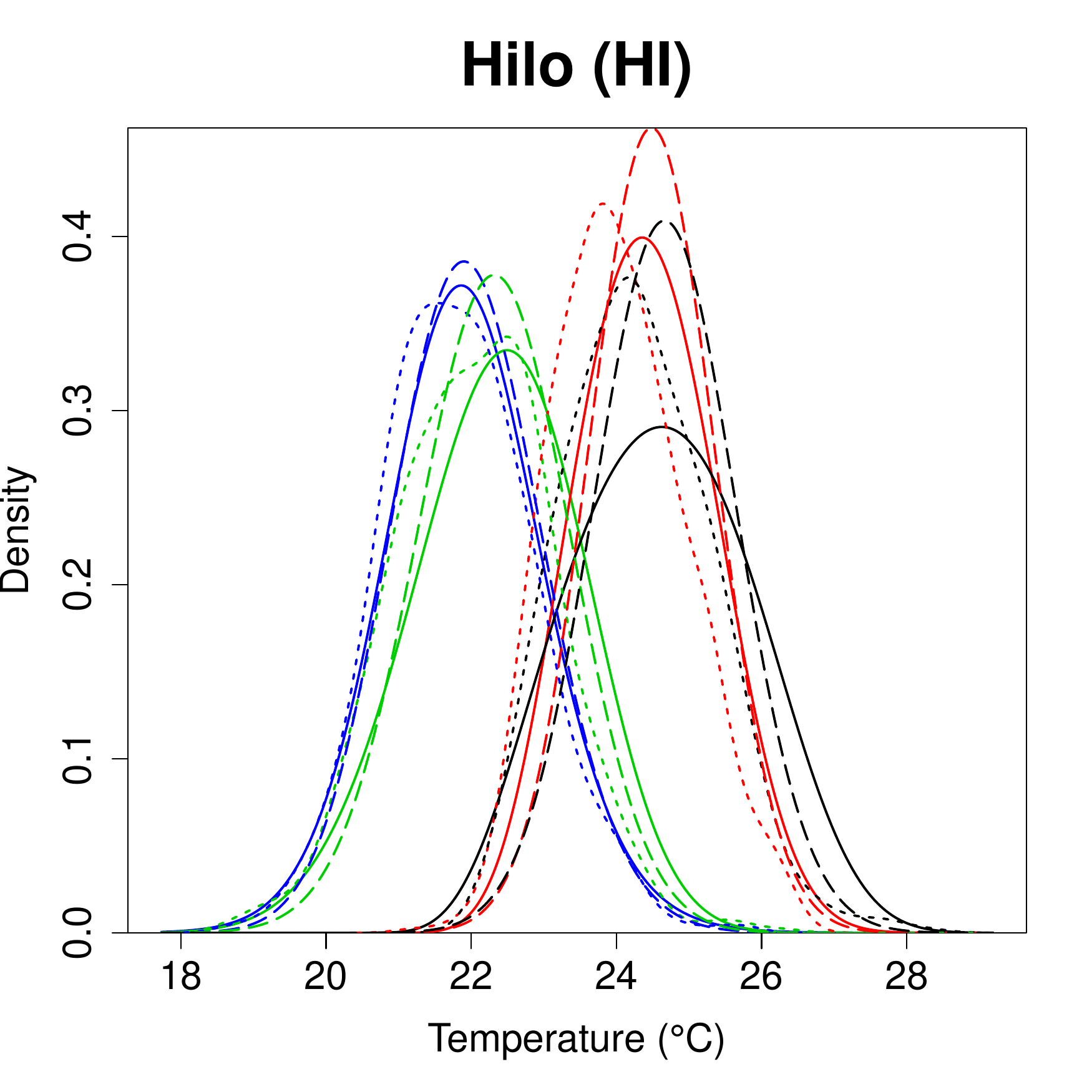} 
    \includegraphics[scale=.22]{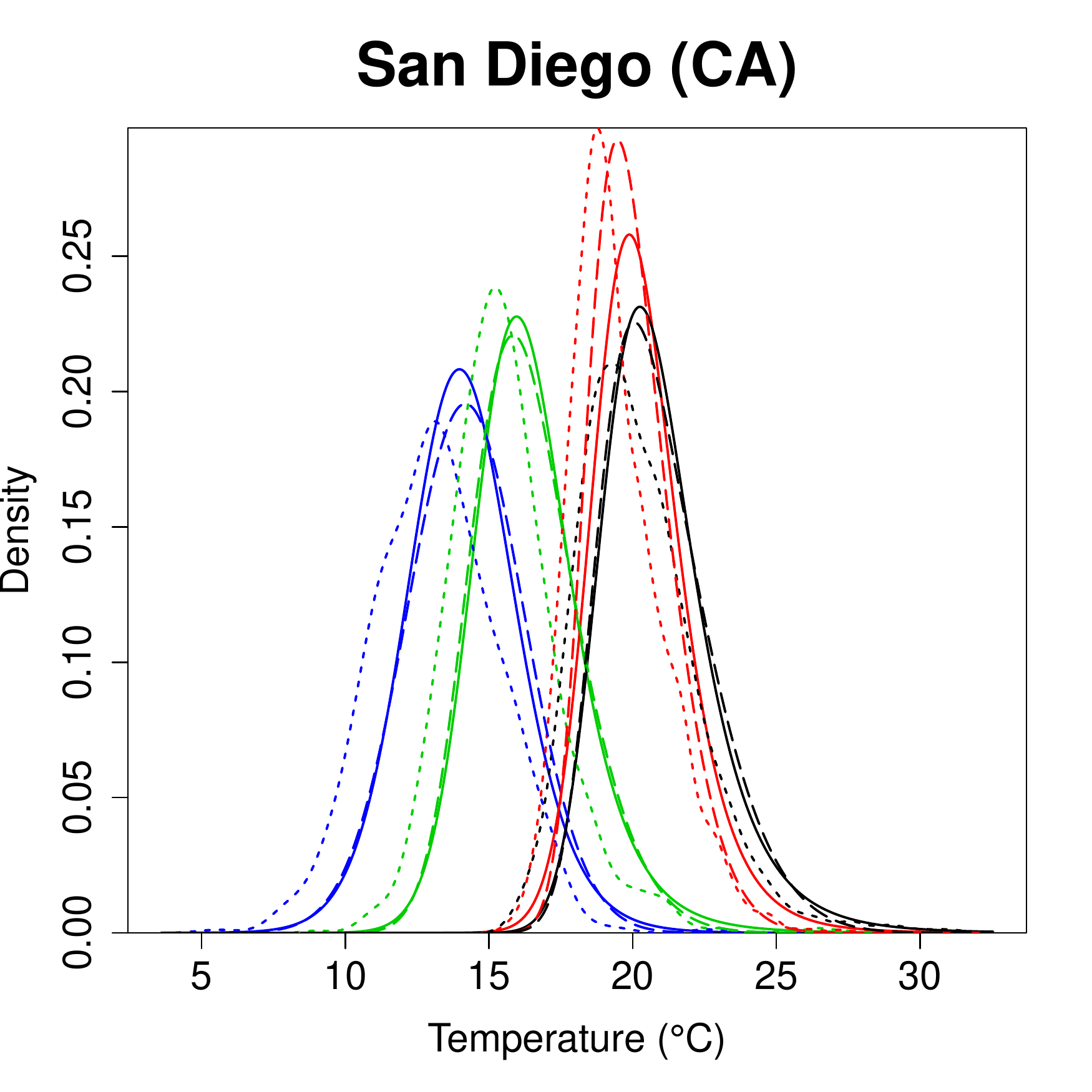} 
    \includegraphics[scale=.22]{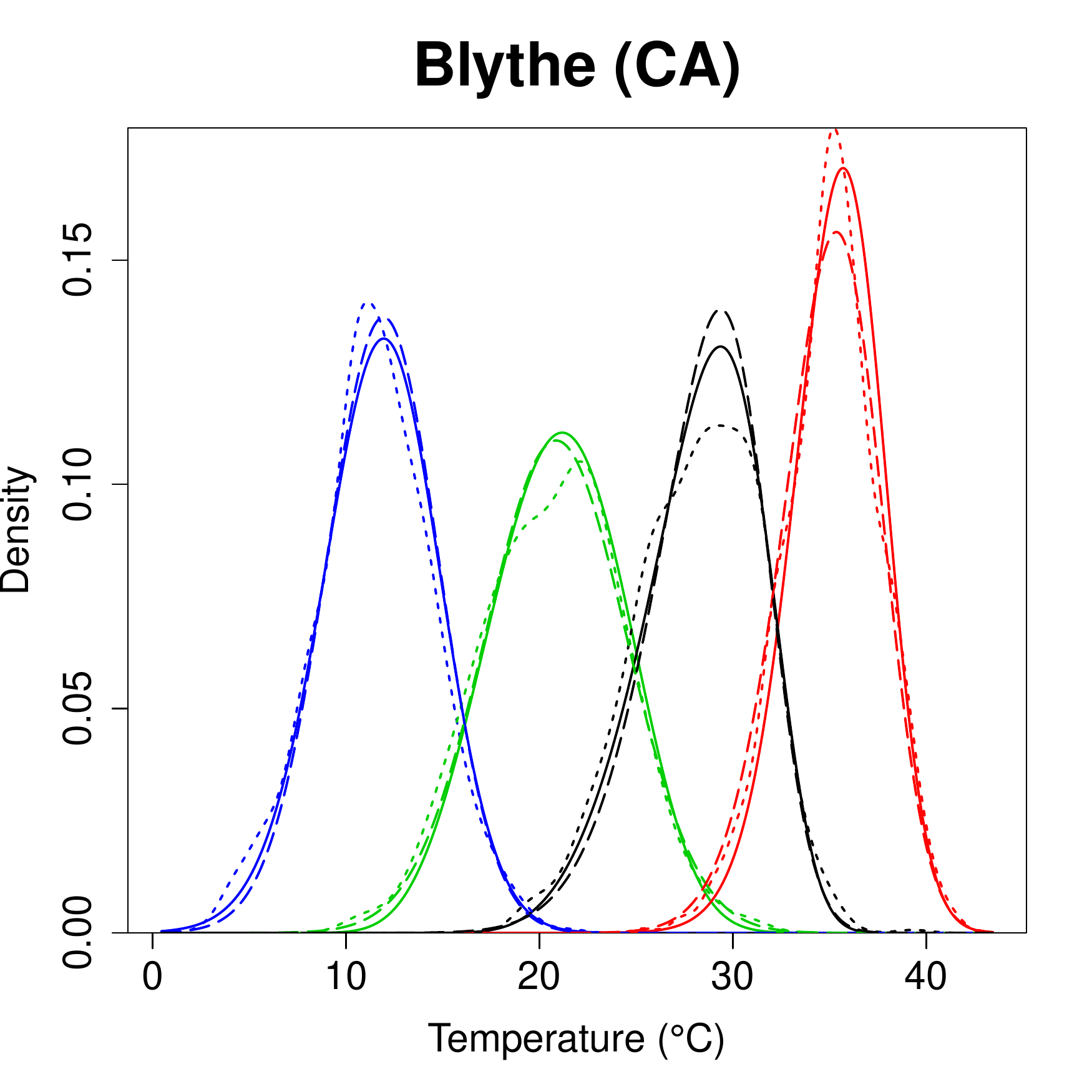}
    \includegraphics[scale=.22]{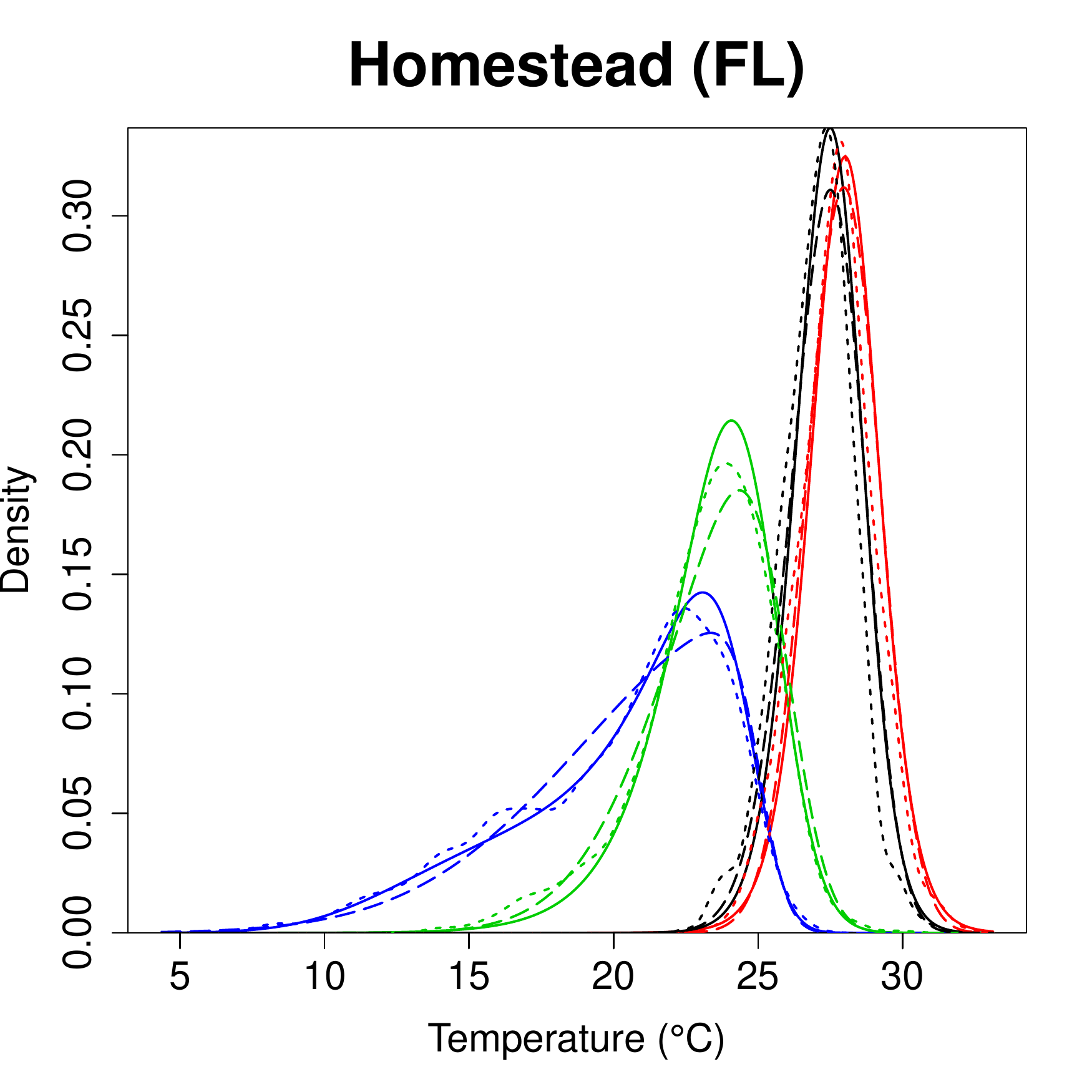}
    \caption{Density estimates for SAT distributions from the seasonal BATs model (solid), skew-normal model (long dash), and a kernel density estimate (short dash). Blue, green, red, and black curves correspond to the first day of January, April, July, and October, respectively. Parametric curves are for 2020, while the KDE curves are for the whole record.}
    \label{fig:density}
\end{figure}

To further explore the distributions across the year, we plot individual estimated monthly densities for the first day of each of January, April, July, and October (all in 2020) from parametric models (BATs and skew-normal) along with empirical kernel density estimates (KDE) in Figure \ref{fig:density}. %Blue colors correspond to January, green corresponds to April, red corresponds to July, and black corresponds October. The solid line shows the seasonal BATs estimate, the short dashed line shows the KDE, and the long dashed line shows a skew-normal estimate, which will be described more in Section \ref{sec:CVcomparison}. Each curve represents the first day of a month in the year. 
Seasonal BATs and skew-normal curves are shown for the year 2020, while the KDE curves come from a Gaussian kernel density estimate using observations from a fifteen-day window over all available years.\footnote{See Appendix \ref{app:kde} for a more precise definition of the kernel density estimate.} 
The three density curves at each  location seem to largely agree with each other for all seasons and locations, but there are some notable differences. For example, in the winter and spring at Bethel, the KDE suggests a multimodal density which cannot be captured by the skew-normal estimate. In contrast, the BATs distribution has the ability to model bimodal distributions (as shown in \citet{stein2020}). The presence of pronounced shoulders to the distributions which the BATs distribution can model but the skew-normal cannot is also found in the winters of Minneapolis and Homestead.
These results suggest that the seasonal BATs model can capture daily mean temperature distributions more reliably than a skew-normal distribution and provide flexible densities matching observed ones. 
A more quantitative comparison of the skew-normal and BATs models is continued in the  next section, along with a closer examination of tail properties. 
%This visual evaluation indicates a reliable fit, the quality of which we now qualify quantitatively. %  that can be further used to learn patterns of the temperature data as in the following section. 

%

\subsection{Cross-validation comparison} \label{sec:CVcomparison}

We further compare our methodology to two relevant models through a quantitative analysis. 
To compare entire distributions, we calculate the cross-validated continuous ranked probability score (CRPS) \citep{CRPS} of the BATs and skew-normal models over the entire year and for each of the four seasons individually. 
CRPS  is typically used in the evaluation of probabilistic forecasts as  a tool to assess calibration  (statistical  consistency between forecast and verification data) and sharpness (dispersion of the distribution to be evaluated). By definition, CRPS values are nonnegative, and smaller values indicate a better match. 
The CRPS between a single observation $x$ and cdf $F$ is defined as
\begin{equation} \label{CRPS}
CRPS(F,x) = \int_{-\infty}^{\infty}    (F(y) - \mathbf{1}[y \ge x])^2 \; dy. %= \int_{-\infty}^x (F(y))^2 \; dy + \int_{x}^\infty (1-  F(y))^2 \; dy
\end{equation}
The CRPS is typically reported as the average of instantaneous CRPS as in \eqref{CRPS} over a set of observations. 
Scores here are calculated in a cross-validation procedure where folds are obtained from blocks of four or five (consecutive) years, leading to  between 10 and 20 folds depending upon the station. A scaled version of the average difference (BATs minus skew-normal) for CRPS values within a given fold is shown as a percentage in Table \ref{tab:skewcomparisontable}. In all cities, the CRPS difference value is negative, indicating that the BATs model fits better over the whole year, but there are some combinations of seasons and cities where the skew-normal performs better. The most noticeable improvement provided by BATs is seen in Bethel, particularly during the wintertime, consistent with the visual evidence of skew-normal misfitting the winter months shown in Figure \ref{fig:density}.
\begin{table}[t] 
\tabcolsep=0.1cm
\centering
\begin{tabular}{p{1cm}|C{1.7cm}C{1.7cm}C{1.7cm}C{1.7cm}C{1.7cm}C{1.7cm}C{1.7cm}C{1.7cm}}
     & BET (AK) & BLY (CA) & BOS (MA) & COL (CO) & HIL (HI)  & HOM (FL) & MIN (MN) & SAN (CA) \\ \hline
Year &  -0.93 &  -0.14 &   -0.10 &       -0.10 & -0.23 &     -0.18 &       -0.01 &     -0.44 \\
 DJF &  -3.05 &  -0.27 &  -0.05 &      -0.07 & -0.13 &     -0.12 &       -0.18 &     -0.22 \\
 MAM &  -0.56 &   0.09 &   -0.10 &      -0.11 & -0.05 &     -0.05 &       -0.08 &      -0.70 \\
 JJA &   0.02 &  -0.48 &  -0.15 &      -0.21 & -0.24 &     -0.12 &        0.08 &     -0.42 \\
 SON &  -0.17 &   0.09 &   -0.10 &       0.01 & -0.49 &     -0.39 &        0.12 &      -0.40 \\
\end{tabular}
\caption{Cross-validated CRPS comparison between the entire distribution for BATs and skew-normal. Table values equal $100  \sum_{k=1}^K \texttt{mean}(\texttt{BATs}(k) - \texttt{Skew}(k))/ \sum_{k=1}^K \texttt{mean}(\texttt{BATs}(k))$ where $k$ is a cross-validation fold. Negative values indicate better performance in the BATs fit. Results are tabulated for the entire year (`Year' row) and by season (DJF: December-January-February, MAM: March-April-May, JJA: June-July-August, SON: September-October-November).}
\label{tab:skewcomparisontable}
\end{table}

 To compare the GPD and the non-thresholded models (BATs and skew-normal), we use a variant of the CRPS. A direct CRPS comparison is not sensible because the GPD only fits a single tail of a distribution. One option to deal with this is to change the cdf in the CRPS definition \eqref{CRPS} to the cdf conditional on being larger than the threshold. Conditioning puts the cdf of the BATs model on the same range, $[0,1]$, as the GPD model when looking above the threshold. However, this approach does not take into account the model's ability to estimate the probability of being larger than the chosen threshold. Instead, we consider a new random variable 
\begin{equation*}
Z = \begin{cases}
\mu & X \le \mu \\
X & X > \mu,
\end{cases}
\end{equation*} 
which is censored to equal the threshold when an observation falls below the threshold. For the BATs and skew-normal models, which fit the entire temperature distribution, this censored cdf equals
\begin{equation*}
    F_Z(z) = \begin{cases}
0 & z < \mu \\
F_X(z) & z \ge \mu,
\end{cases}
\end{equation*} 
and for the GPD models, the censored cdf equals
\begin{equation*} \label{censor}
    F_Z(z) = \begin{cases}
0 & z < \mu \\
p_\mu + (1-p_\mu) F_X(z) & z \ge \mu.
\end{cases}
\end{equation*}
As indicated by the common notation, the censoring value $\mu$ and the GPD threshold $\mu$ are identical, but in theory they could be different. Here, $p_\mu=0.95$ corresponds to the quantile regression level that creates the threshold $\mu$.  

To assess only tail behavior, we consider a weighted CRPS  \citep{gneiting2011, taillardat2019extreme} where the weight function in the integral equals one if the observation lies in the tail and zero otherwise. To be precise, the indicator-weighted score expresses as
\begin{equation*} \label{wCRPS}
wCRPS(F,x; q) =  \int_{-\infty}^{\infty}    (F(y) - \mathbf{1}[y \ge x])^2 \mathbf{1}[y \ge q] \; dy 
\end{equation*}
where $F$ is a censored cdf described above. The role of $q$ is to examine behavior of the distribution far in the tail by ignoring values smaller than $q$. Both the GPD threshold $\mu$ and the wCRPS threshold $q$ are obtained from a cross-validated quantile regression with eight periodic splines as covariates, with $\mu$ taken as the $p_\mu = 0.95$ quantile and $q$ as the $p_q$ quantile, where $p_q \ge 0.95$. A similar procedure can be conducted to assess the lower tails of the distributions. 
Results from this comparison are shown in Table \ref{tab:tailtable1}. For the most part, the BATs model and skew-normal appear to provide a slightly better fit than the GPD in this tail analysis. %, supporting the idea that studying nonstationary extremes with all possible data can improve the quality of fit in the tails. 
In Bethel, the skew-normal distribution fits the lower tail poorly, due primarily to its inability to reproduce behavior in DJF, as seen earlier in Figure \ref{fig:density} and Table \ref{tab:skewcomparisontable}. Besides the lower tail in Bethel, the BATs model and skew-normal are relatively comparable with a slight advantage to BATs.  %neither distribution systematically outperforming the other. 
Boxplots for Colorado Springs in Figure \ref{fig:boxplot_season}  showed asymmetric behavior in the upper and lower tails, with the lack of observations beyond the upper boxplot whiskers indicating a non-Gaussian upper tail. Accordingly, the BATs and skew-normal wCRPS values in the lower tail are quite similar, yet the BATs model provides a better fit in the upper tail. In cases where the performance of BATs is particularly poor---namely, the lower tails at Blythe and Homestead---the trouble largely comes from misfit in the first cross validation fold, where fitting was done on the hold-out data which did not include initial observation years due to the configuration of the cross validation folds. The temperature records at these two locations had significant gaps after their earliest years, with Homestead missing over a decade, and Blythe missing over 30 years (see Appendix \ref{app:data}).

\begin{table}[t] 
\centering
\resizebox{\linewidth}{!}{%
\tabcolsep=0.05cm
\begin{tabular}
{r|R{1cm}R{1cm}|R{1cm}R{1cm}|R{1cm}R{1cm}|R{1cm}R{1cm}|R{1cm}R{1cm}|R{1cm}R{1cm}|R{1cm}R{1cm}|R{1cm}R{1cm}}  
      & \multicolumn{2}{c}{BET (AK)} & \multicolumn{2}{c}{BLY (CA)} & \multicolumn{2}{c}{BOS (MA)} & \multicolumn{2}{c}{COL (CO)} & \multicolumn{2}{c}{HIL (HI)} & \multicolumn{2}{c}{HOM (FL)} & \multicolumn{2}{c}{MIN (MN)} & \multicolumn{2}{c}{SAN (CA)} \\
    $p_q$  & BATs         & Skew        & BATs        & Skew        & BATs        & Skew         & BATs          & Skew           & BATs        & Skew       & BATs         & Skew          & BATs          & Skew           & BATs          & Skew          \\ \hline
 0.95 &      -0.53 &      -0.42 &      -0.51 &      -0.87 &      -0.13 &      -0.03 &      -0.46 &      -0.30 &      -4.08 &      -2.89 &      -2.41 &      -2.38 &      -0.28 &      -0.25 &       -1.20 &      -1.12 \\
 0.99 &      -0.17 &      -0.19 &      -0.71 &      -1.05 &       -0.10 &       0.03 &      -0.35 &       0.10 &      -2.06 &      -1.12 &      -2.26 &      -2.17 &      -0.09 &       0.00 &      -0.04 &      -0.04 \\
 0.995 &      -0.13 &      -0.17 &      -0.84 &      -1.19 &      -0.06 &       0.06 &      -0.36 &      0.18 &      -1.28 &      -0.63 &      -2.06 &      -1.89 &      -0.09 &      -0.01 &       0.18 &        0.00 \\
 \hline
 0.05 &      -0.35 &       3.34 &      0.27 &      -0.27 &      -0.13 &      -0.06 &      -0.23 &      -0.21 &      -0.81 &      -0.46 &      -0.24 &       -0.30 &      -0.37 &      0.15 &       -1.50 &      -0.10 \\
 0.01 &      -0.06 &      13.38 &      0.91 &      -0.26 &      -0.05 &       0.02 &      -0.09 &      -0.09 &       -0.50 &       0.03 &       0.18 &      -0.04 &       -0.30 &       0.90 &      -0.82 &      0.15 \\
0.005 &      -0.09 &      20.44 &      0.94 &      -0.37 &      -0.06 &        0.00 &      -0.06 &      -0.06 &      -0.42 &       0.07 &        0.30 &       0.06 &      -0.25 &      1.18 &      -0.45 &      0.13 \\
\end{tabular}
}
\caption{Cross-validated wCRPS comparison of the tail behavior beyond quantile $p_q$. Estimates of $\mu$ (GPD threshold where  observations are censored; corresponds to $p_\mu =0.95$ quantile (upper tail) or $p_\mu=0.05$ quantile (lower tail)) and $q$ (wCRPS threshold; corresponds to $p_q$ quantile in left column) are obtained from cross-validated quantile regressions with eight periodic splines as covariates. 
Table values equal $100  \sum_{k=1}^K \texttt{mean}(\texttt{BATs}(k) - \texttt{GPD}(k))/\sum_{k=1}^K \texttt{mean}(\texttt{GPD}(k))$ and $100  \sum_{k=1}^K \texttt{mean}(\texttt{Skew}(k) - \texttt{GPD}(k))/ \sum_{k=1}^K \texttt{mean}(\texttt{GPD}(k))$ where $k$ is a cross-validation fold. Negative values indicate better performance than the GPD fit.}
\label{tab:tailtable1}
\end{table}

\subsection{Changing distributions over the years} \label{sec:climatechange}
\begin{figure}
    \centering
    \includegraphics[scale=.22]{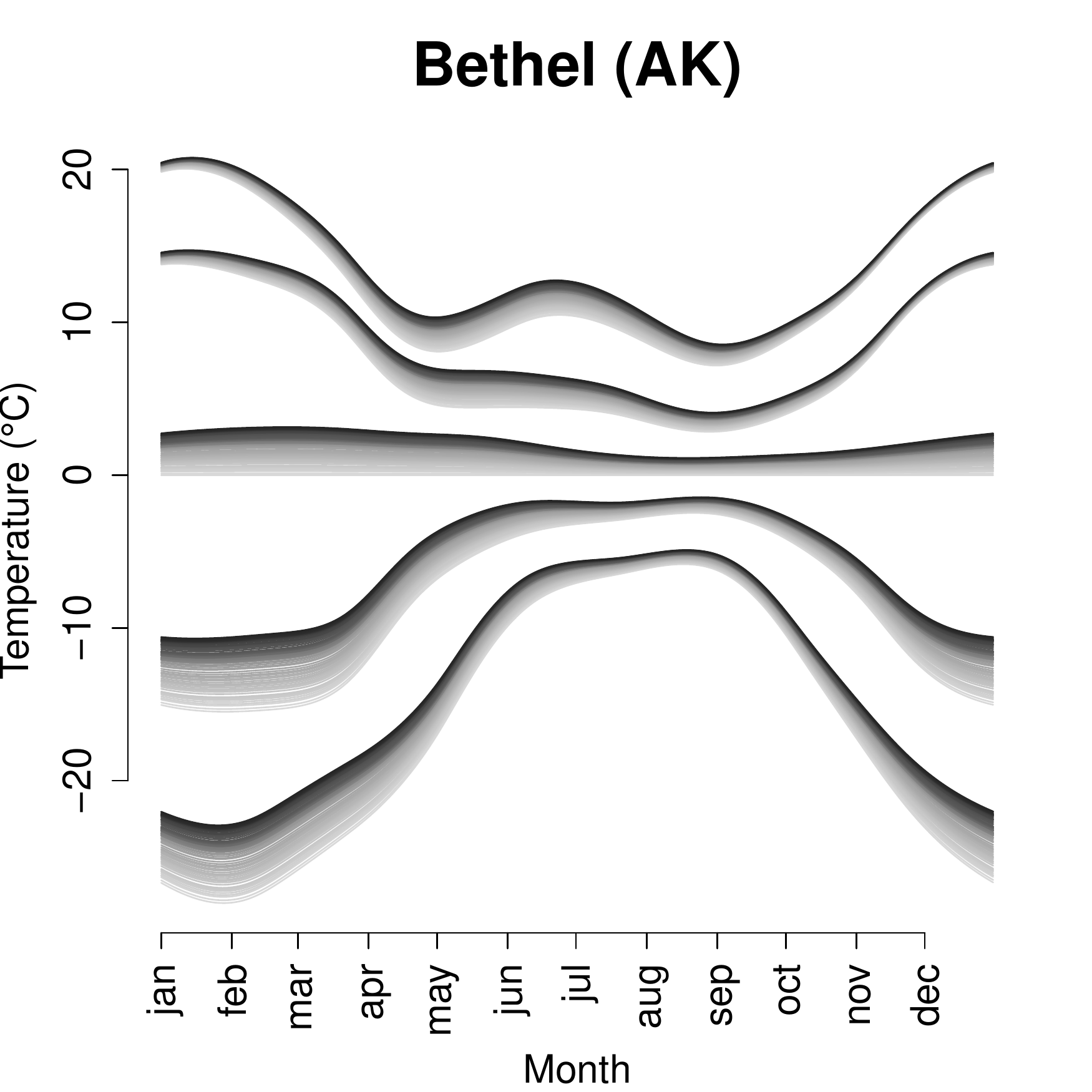}
    \includegraphics[scale=.22]{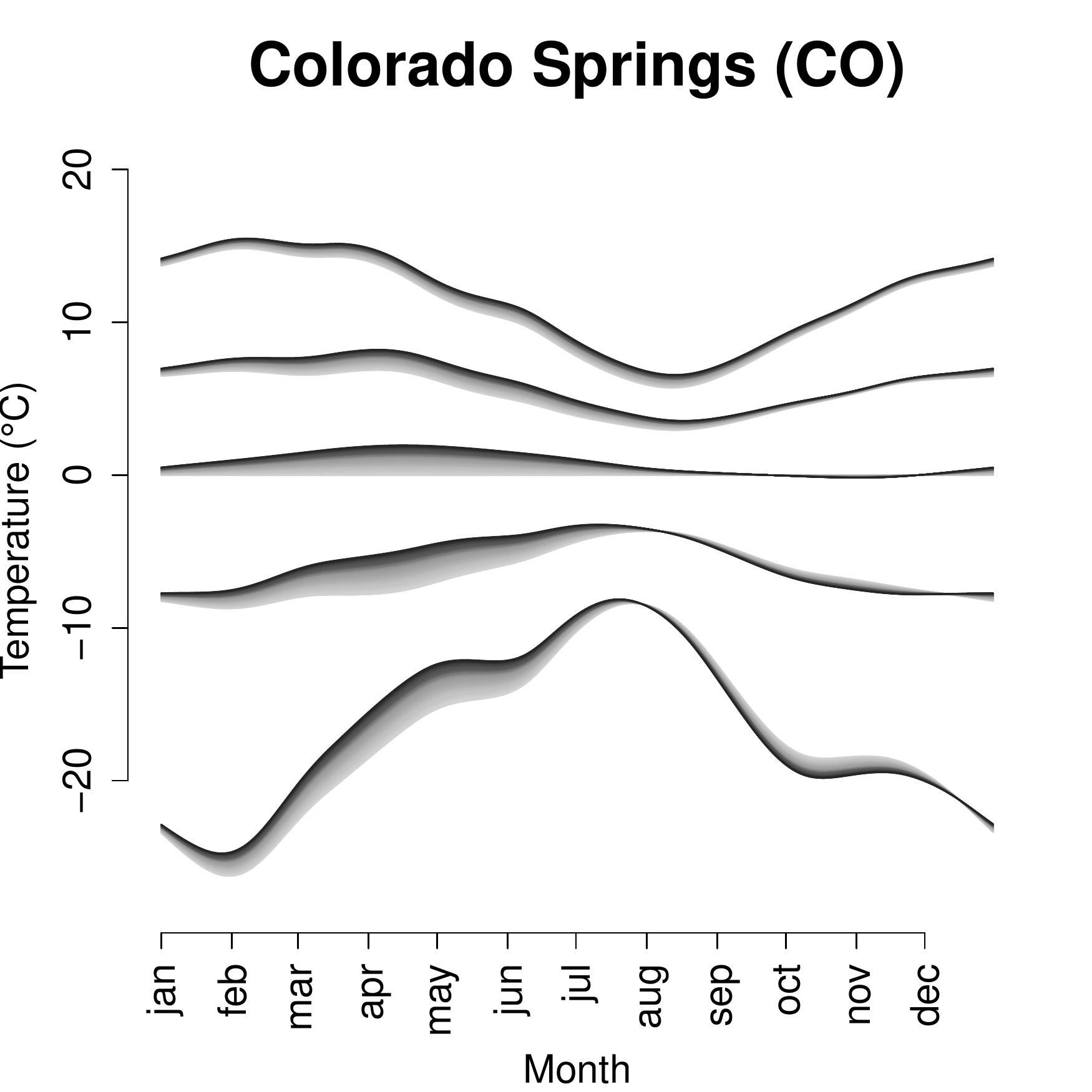}
    \includegraphics[scale=.22]{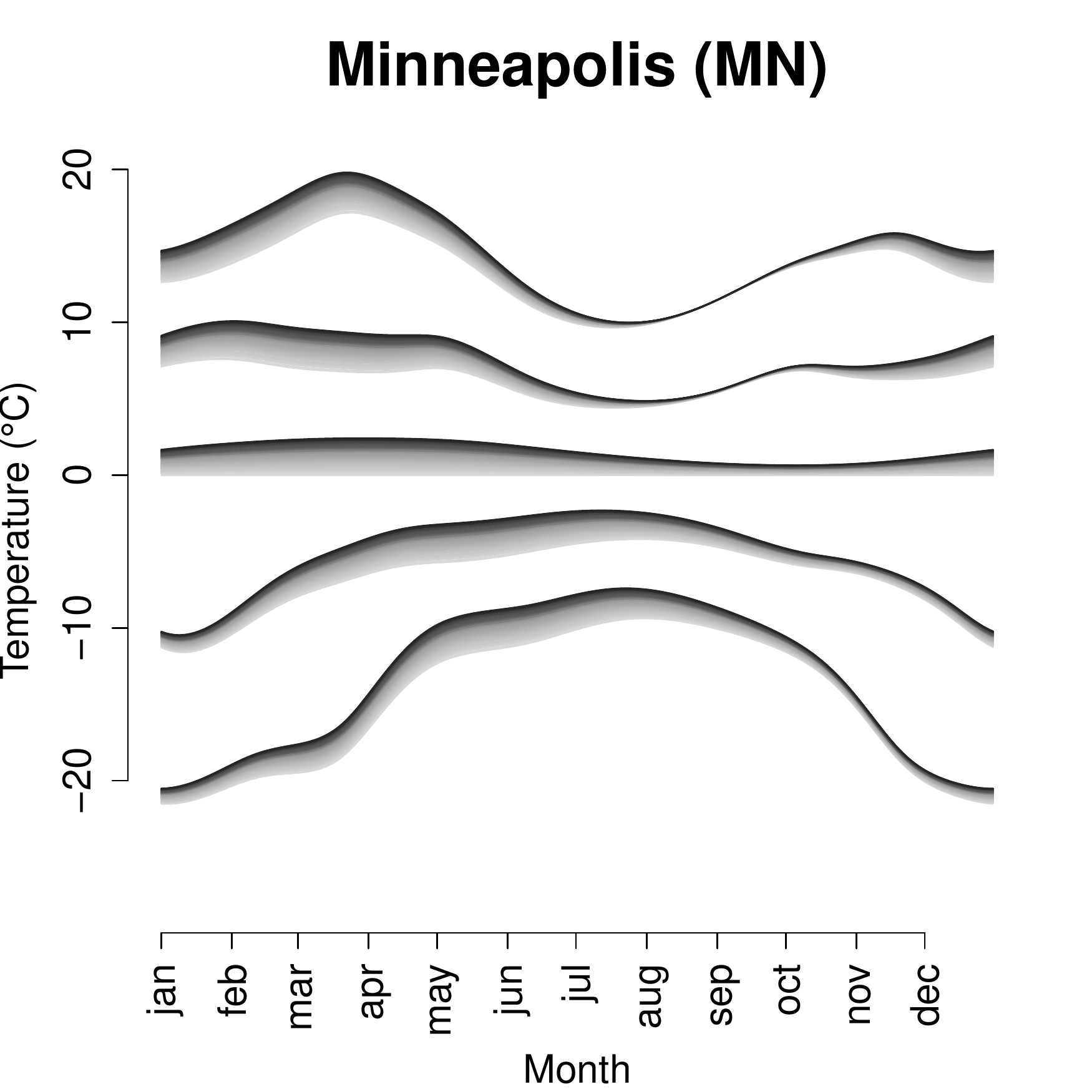}
    \includegraphics[scale=.22]{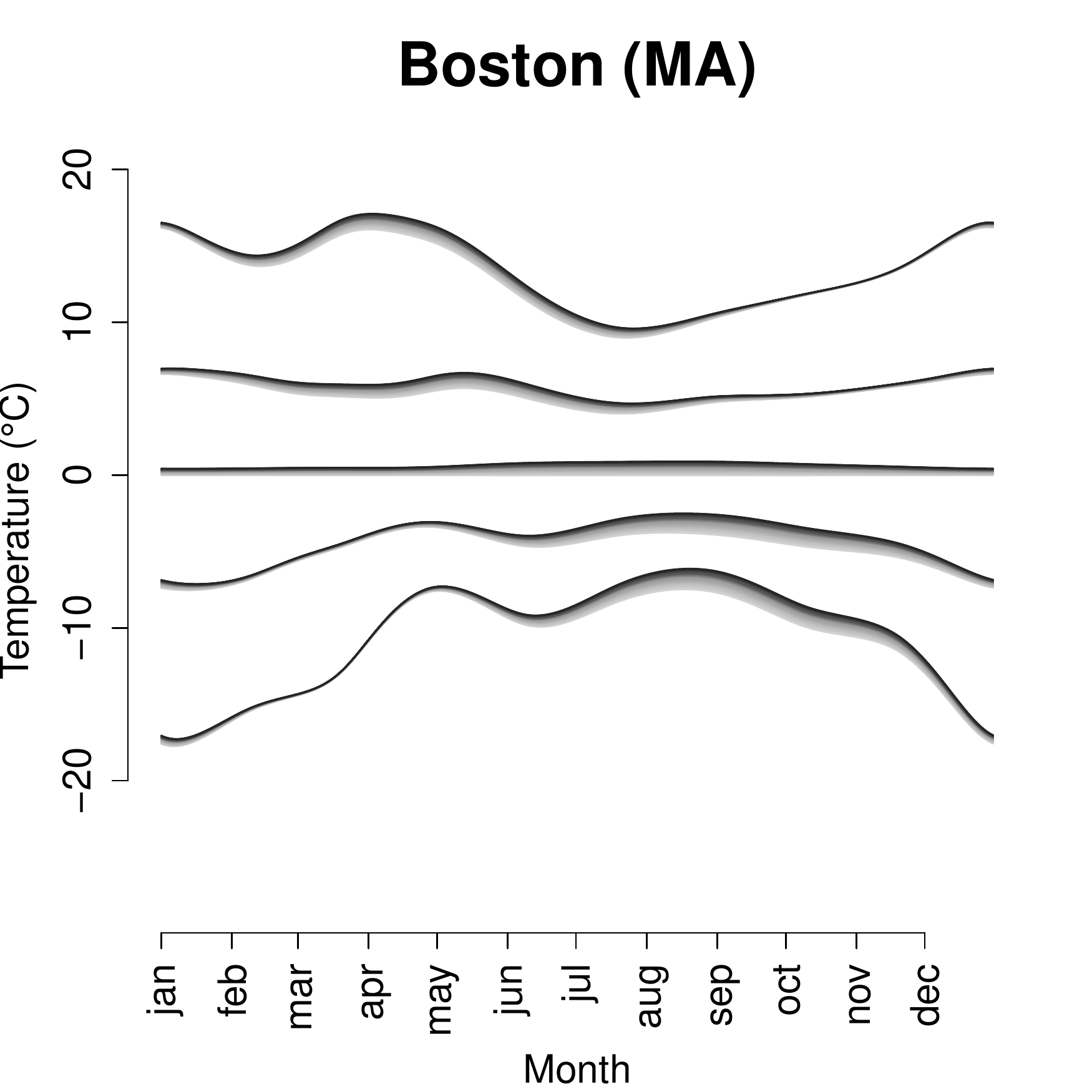}
    \includegraphics[scale=.22]{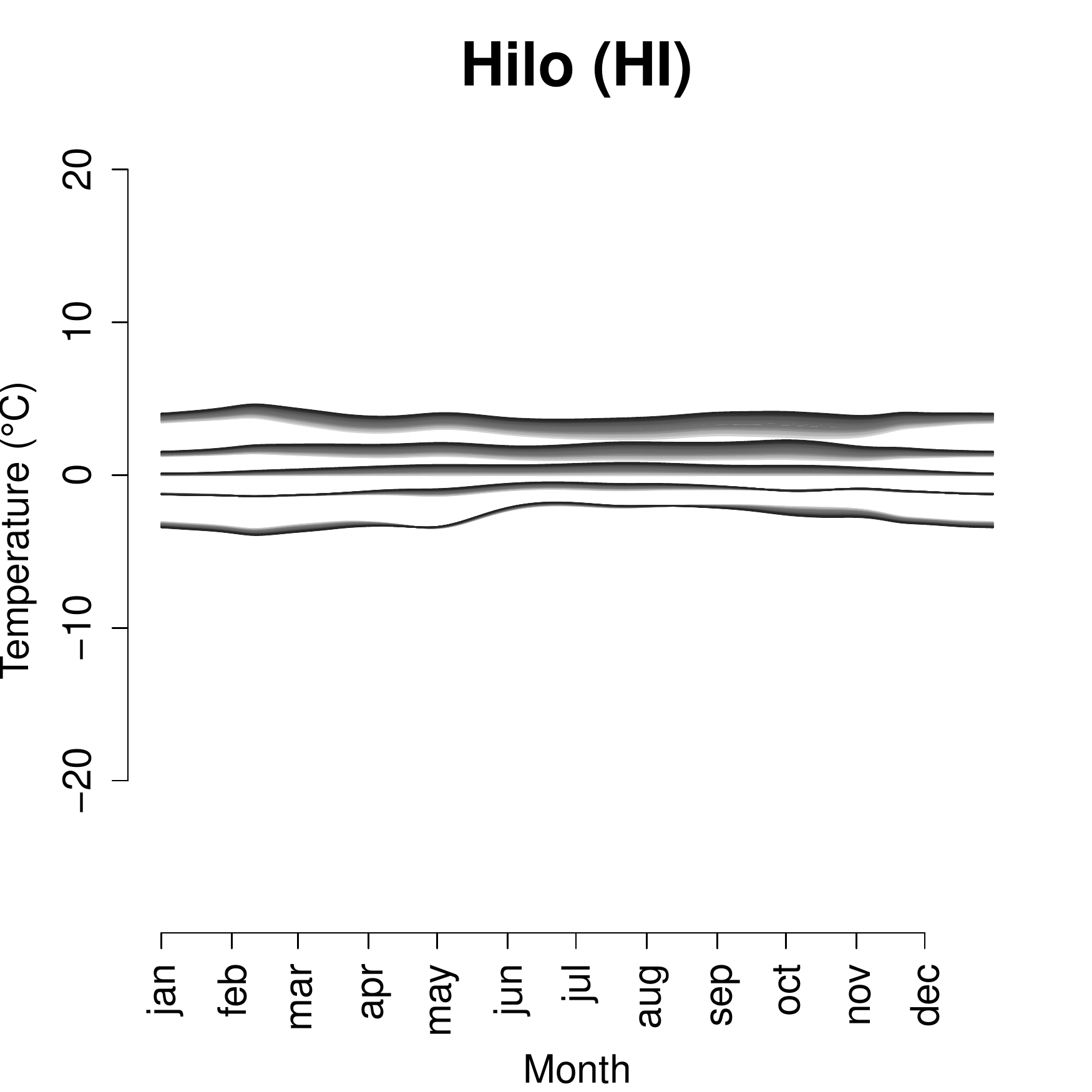}
    \includegraphics[scale=.22]{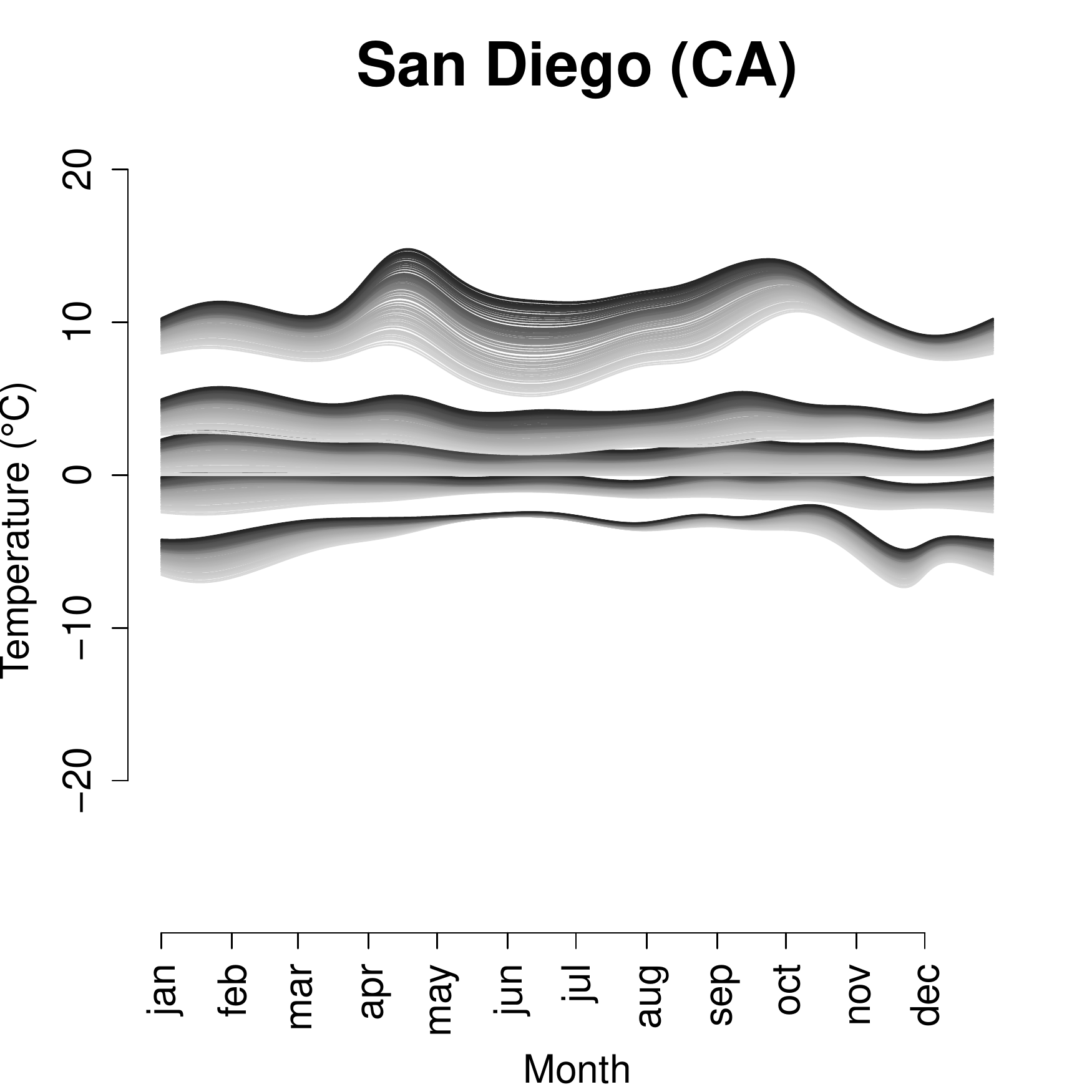}
    \includegraphics[scale=.22]{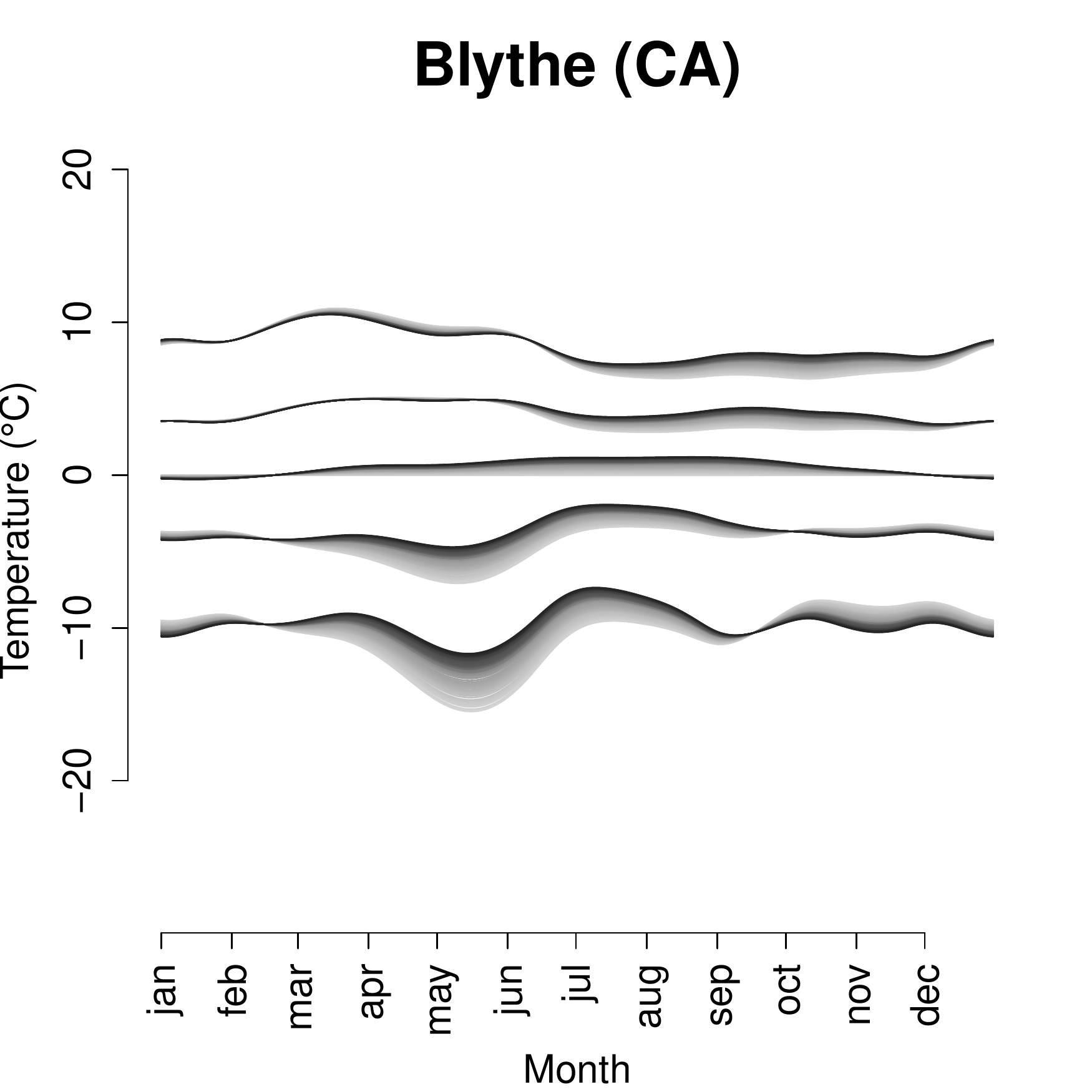}
    \includegraphics[scale=.22]{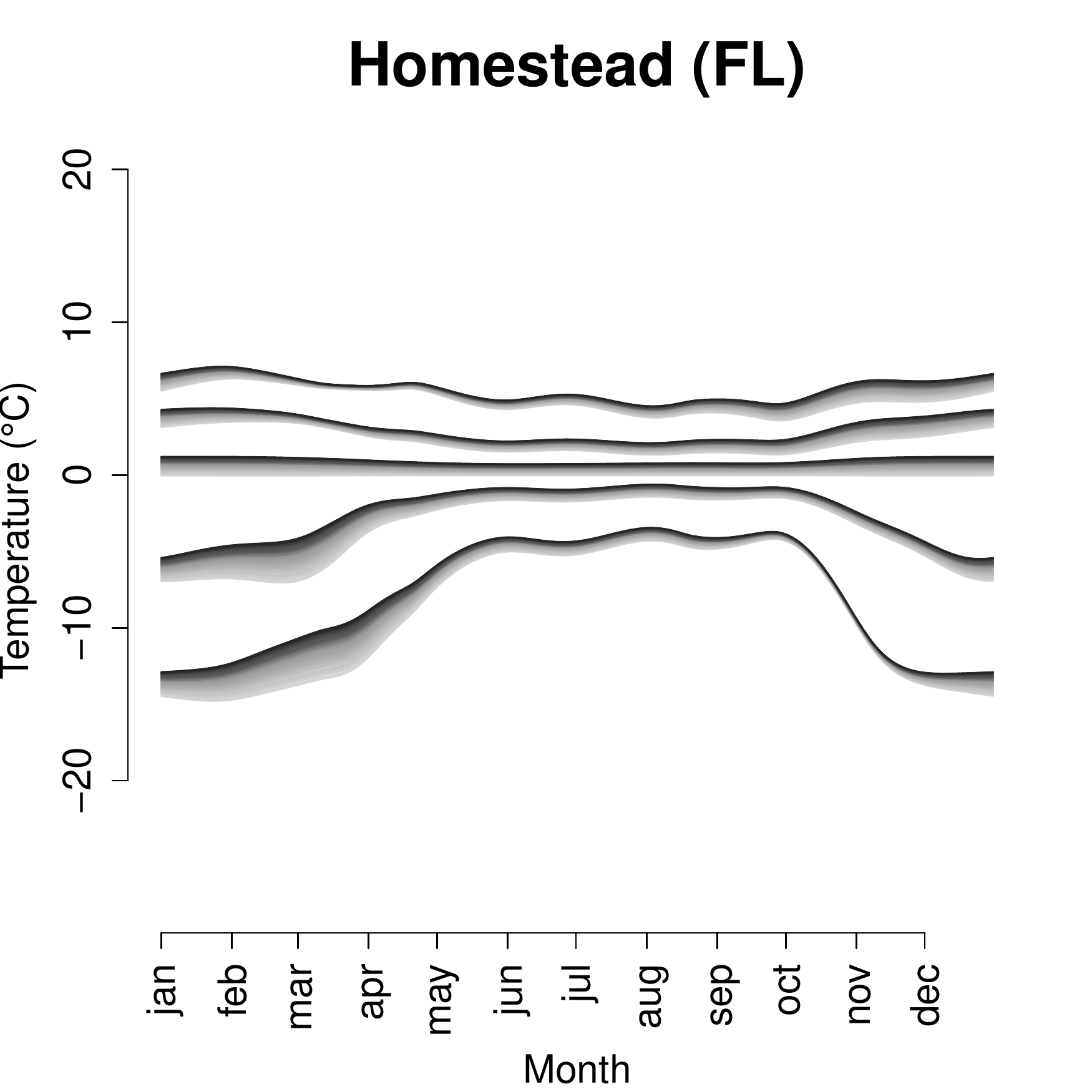}
\caption{Changes of SAT quantiles from the seasonal BATs model from the beginning of the observation period (light) to the end (dark). Curves are the 0.001, 0.1, 0.5, 0.9, and 0.999 quantiles minus the 0.5 quantile at the earliest year. To emphasize differences in magnitude between stations, all subplots have the same vertical axis.}
\label{fig:all_quantiles_differences}
\end{figure}

We conclude our analysis by illustrating the ability of the proposed nonstationary BATs model to capture changes in SAT distributions and their seasonal patterns between years in a nonstationary climate. 
Our model explicitly accounts for a proxy of climate change through the log CO$_2$ equivalent incorporated in the model parameters and through the interaction between seasonality and this long-term trend. 
First, in Figure \ref{fig:all_quantiles_differences}, we show how estimated seasonal quantiles from the  seasonal BATs model fits evolve over the period of the observational record. Specifically, the graphs use gradations of shading from grey to black to show time evolution of the differences of the estimated 0.001, 0.1, 0.5, 0.9, and 0.999 yearly quantiles from the 0.5 quantile at the starting year. Most combinations of quantiles and stations show a warming trend (the curves become darker with increasing time). %the $y$-axis value increases as the color changes from light to dark, indicating a warming trend. % and stretching trend of the distributions.  
%Warm and cold quantiles show stronger year-to-year changes than median quantiles, with the main commonality among all stations being a warming trend. 
However, the details of this trend vary substantially across locations, quantile levels, and time of the year. Bethel displays substantial warming, particularly in its lower quantiles at winter months. %, which goes along with scientific consensus that colder quantiles are  warming faster than warmer quantiles. 
%\julie{keep this? 
%\citet{huang2016} examined computer model outputs and pointed out that North-American warm extremes exhibit similar trends to temperature means, whereas cold extremes are warming faster than temperature means. 
Spring and summer months in San Diego also experience a prominent increase in the 0.999 quantile.
The long-term trend and its interaction with seasonal patterns can capture evolving seasonal patterns. 
Changes in seasonality over the years exhibit very different patterns depending upon station location and the quantile level considered.  
For instance, Hilo has the largest increases in median and hot temperatures over late summer and early fall, whereas cold quantiles at Bethel warm the most during winter. A notable departure from warming is seen at Hilo, Blythe, and Colorado Springs, where the 0.001 quantile shows a cooling trend in August through December. Many cold quantiles exhibit a less pronounced seasonality in the recent years compared with past years. In particular, the seasonality of the winter cold quantiles at warmer cities flattens over time; this phenomenon is less prominent in the warm quantiles.
%\julie{rephrase: Many cold quantiles exhibit a less pronounced seasonality in the recent years (dark shades) compared with past years (light shades), with in particular warmer winter cold quantiles flattening the seasonal pattern. This pattern is less prominent in the warm quantiles.}
%

\begin{figure}
    \centering
     \includegraphics[scale=.22]{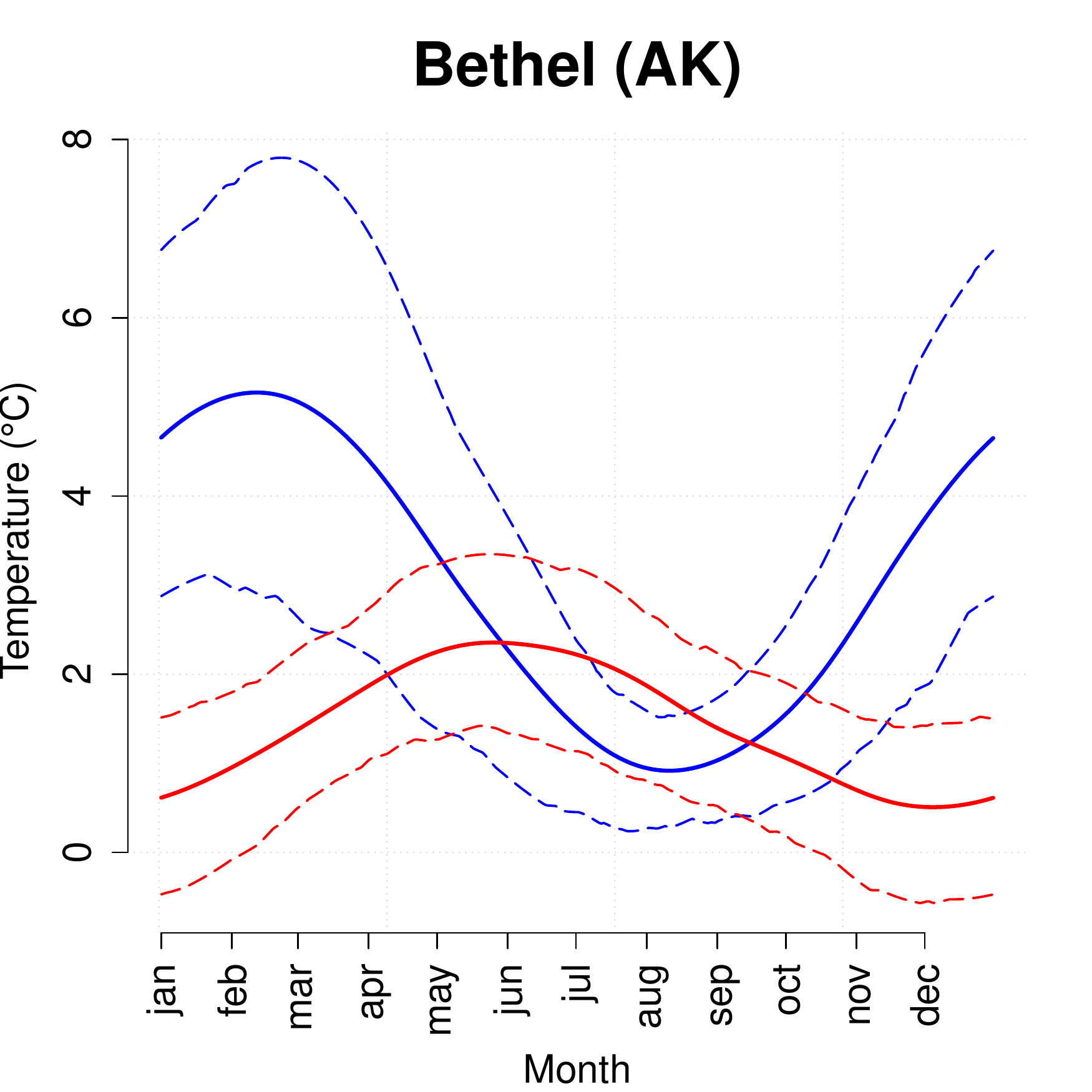}
    \includegraphics[scale=.22]{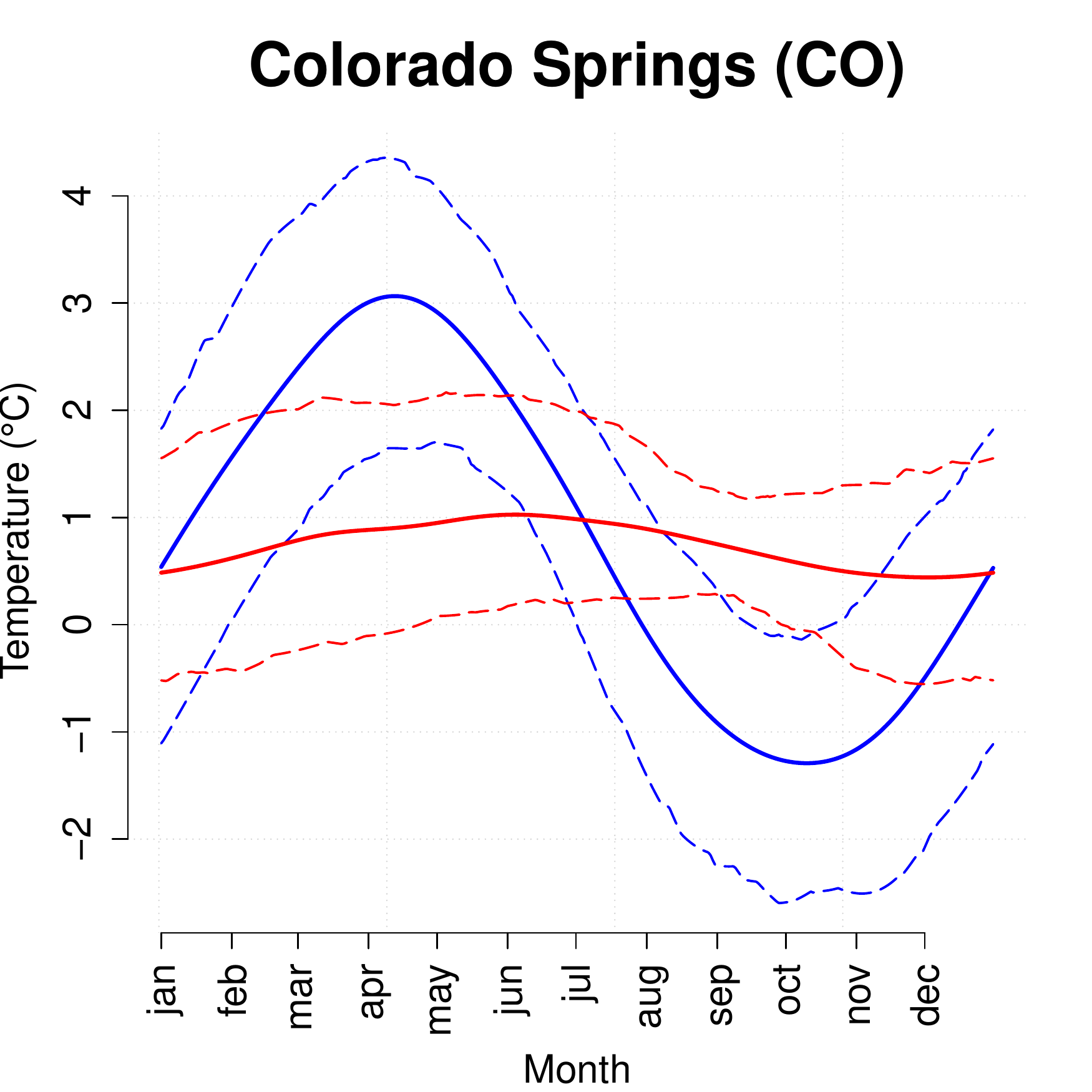}
    \includegraphics[scale=.22]{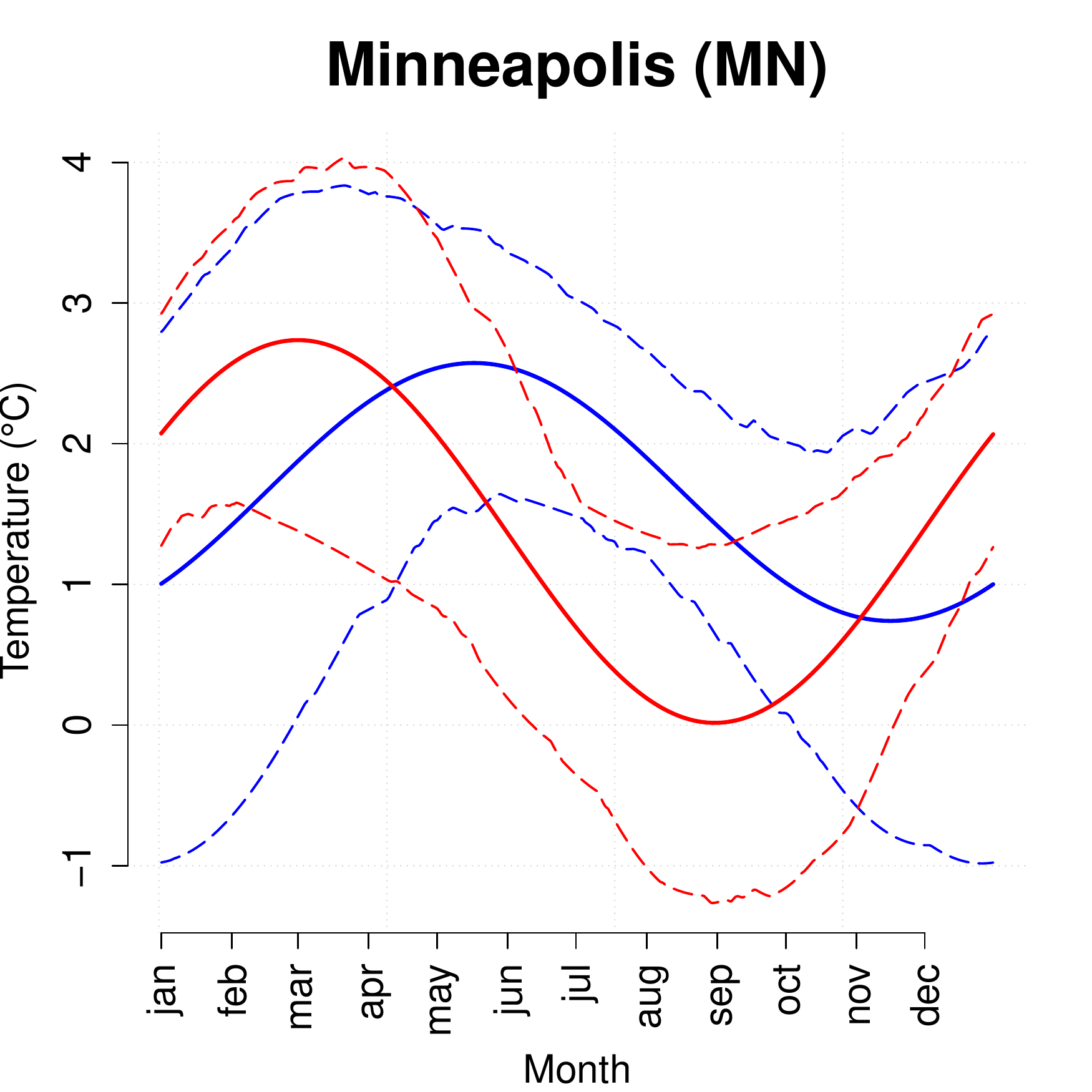}
    \includegraphics[scale=.22]{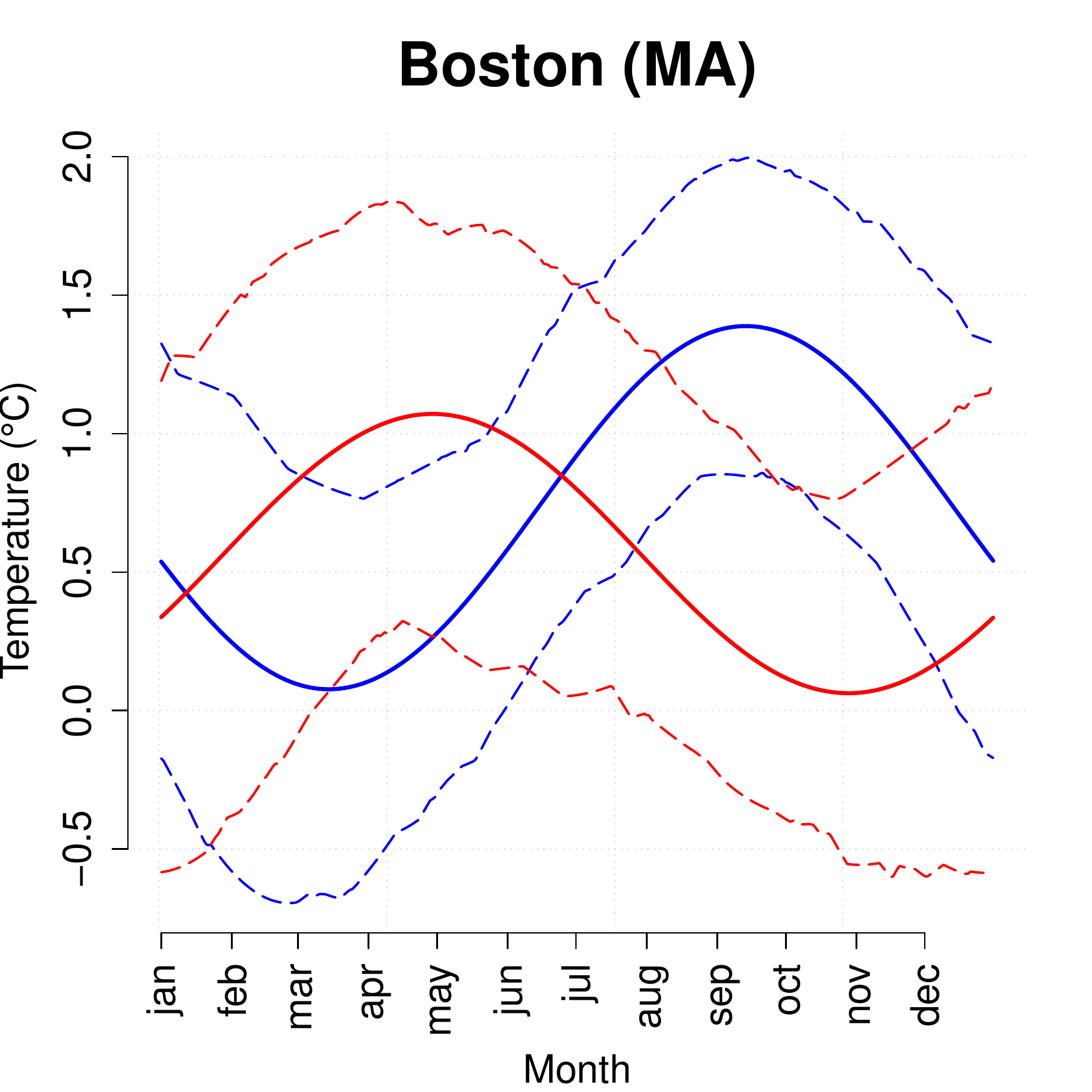}
    \includegraphics[scale=.22]{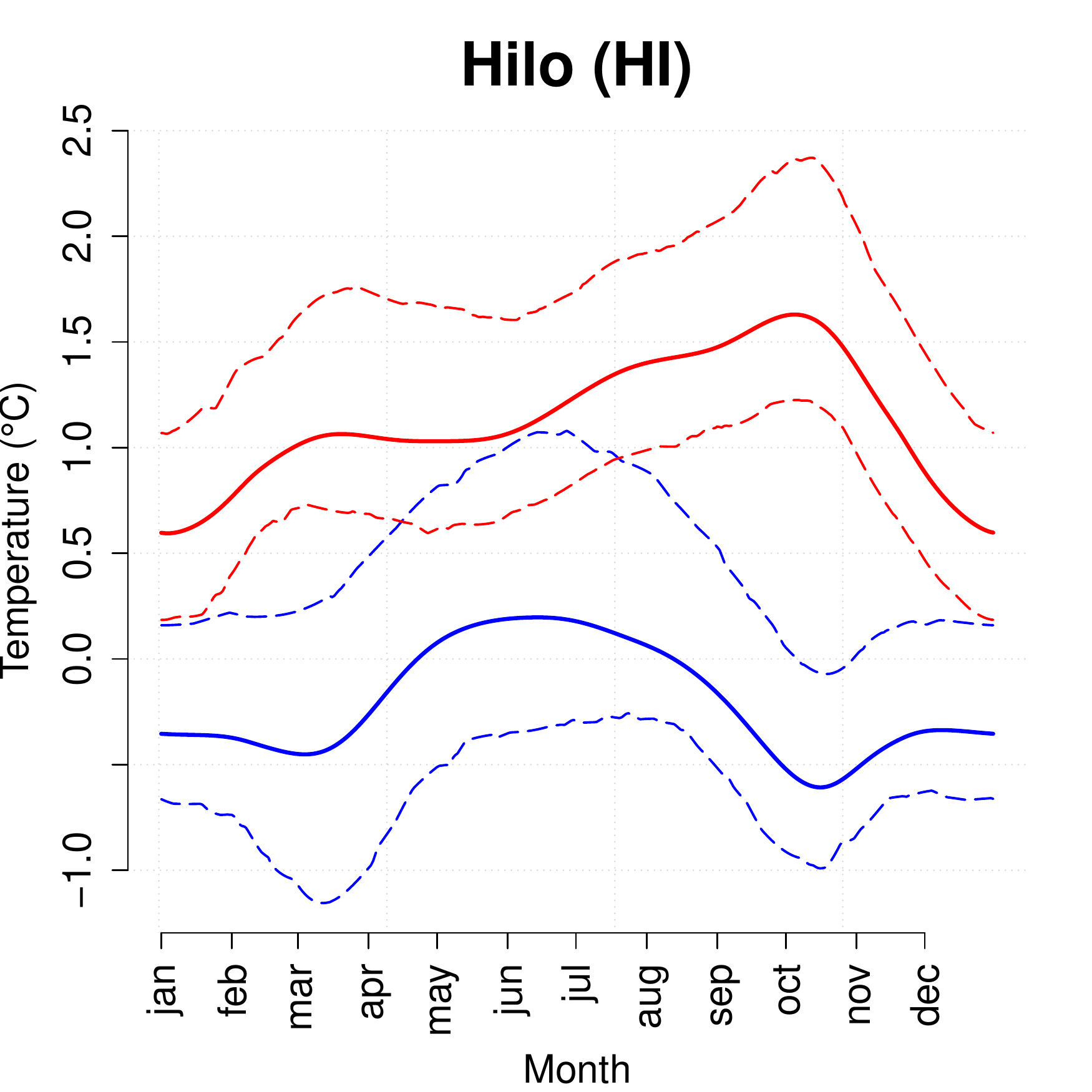}
    \includegraphics[scale=.22]{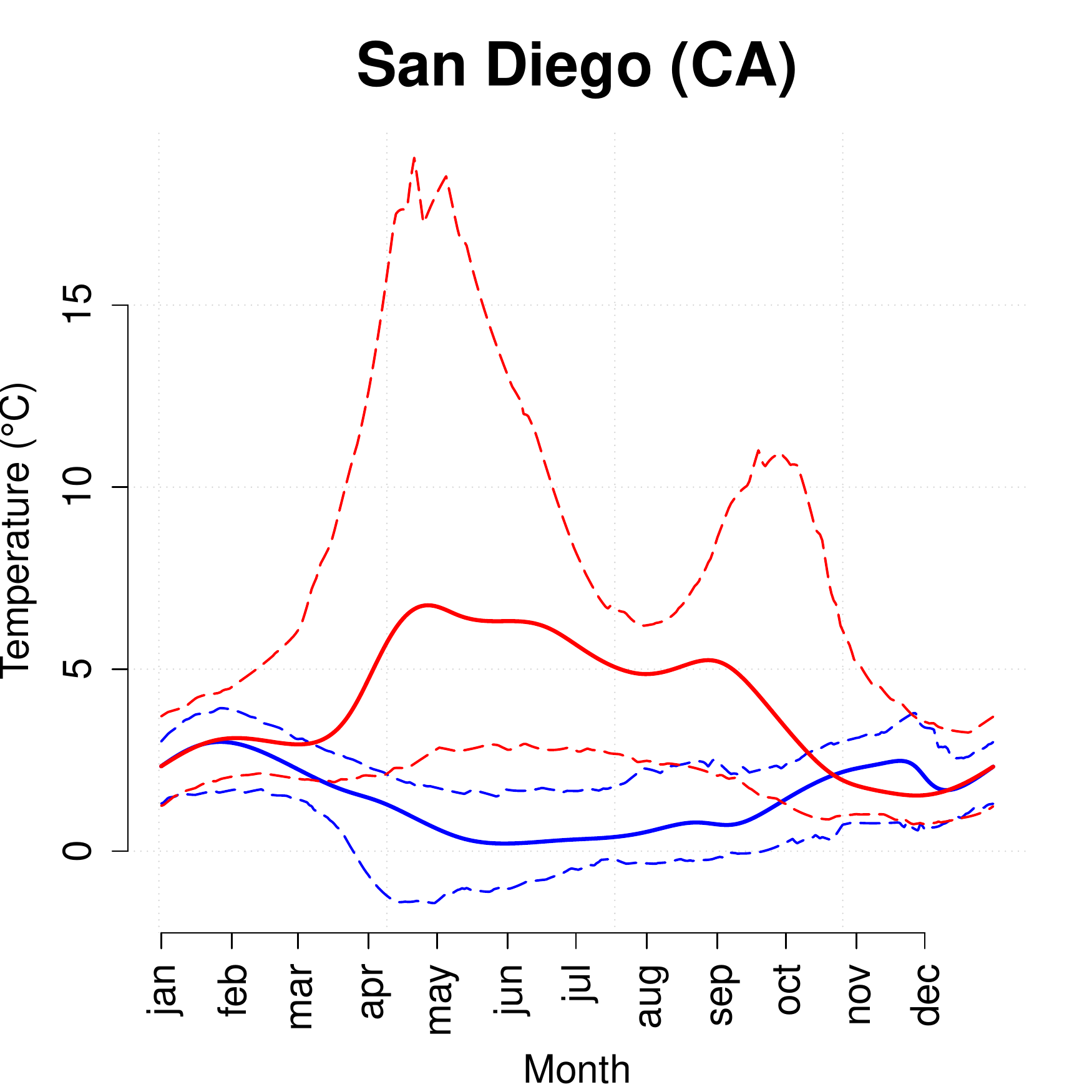}
    \includegraphics[scale=.22]{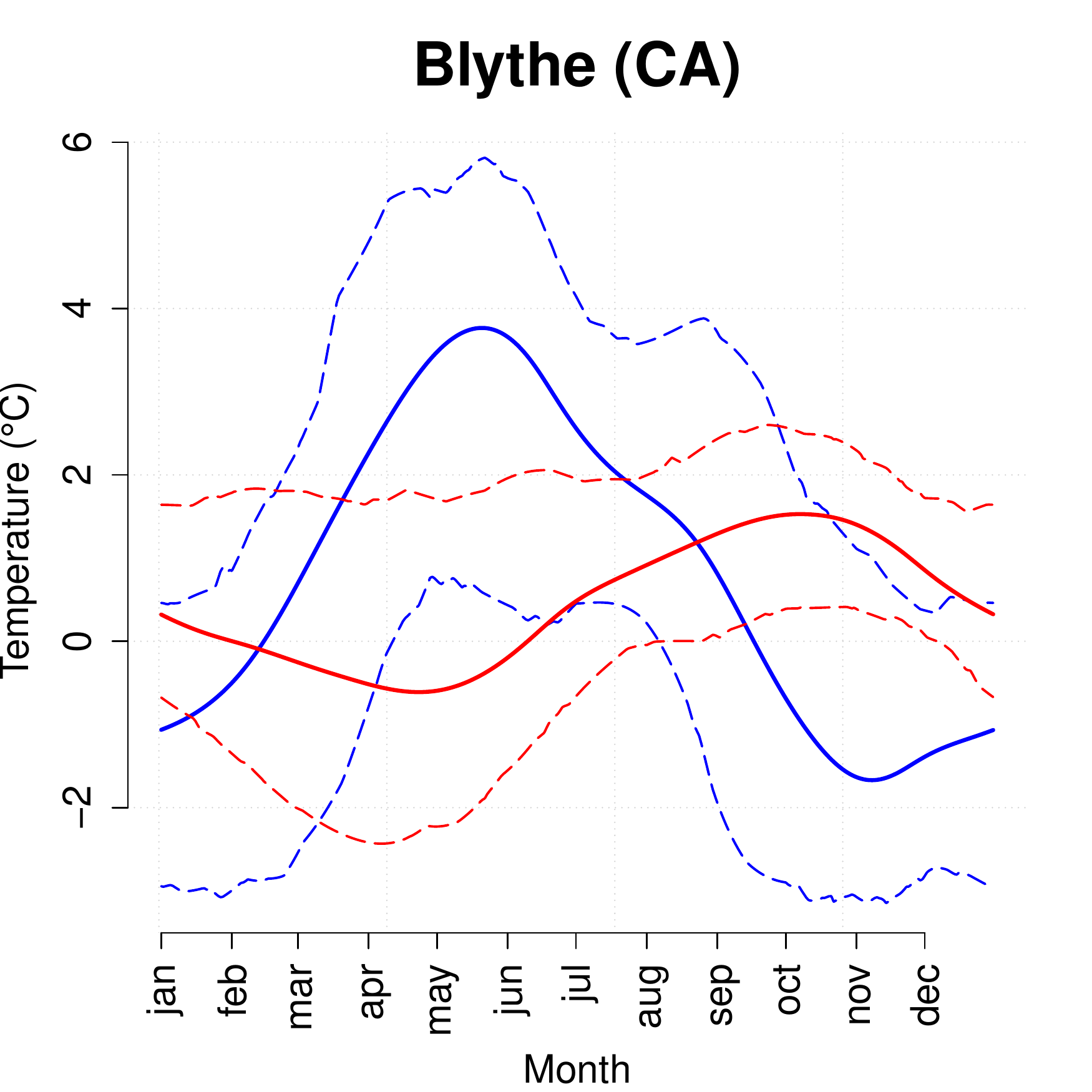}
    \includegraphics[scale=.22]{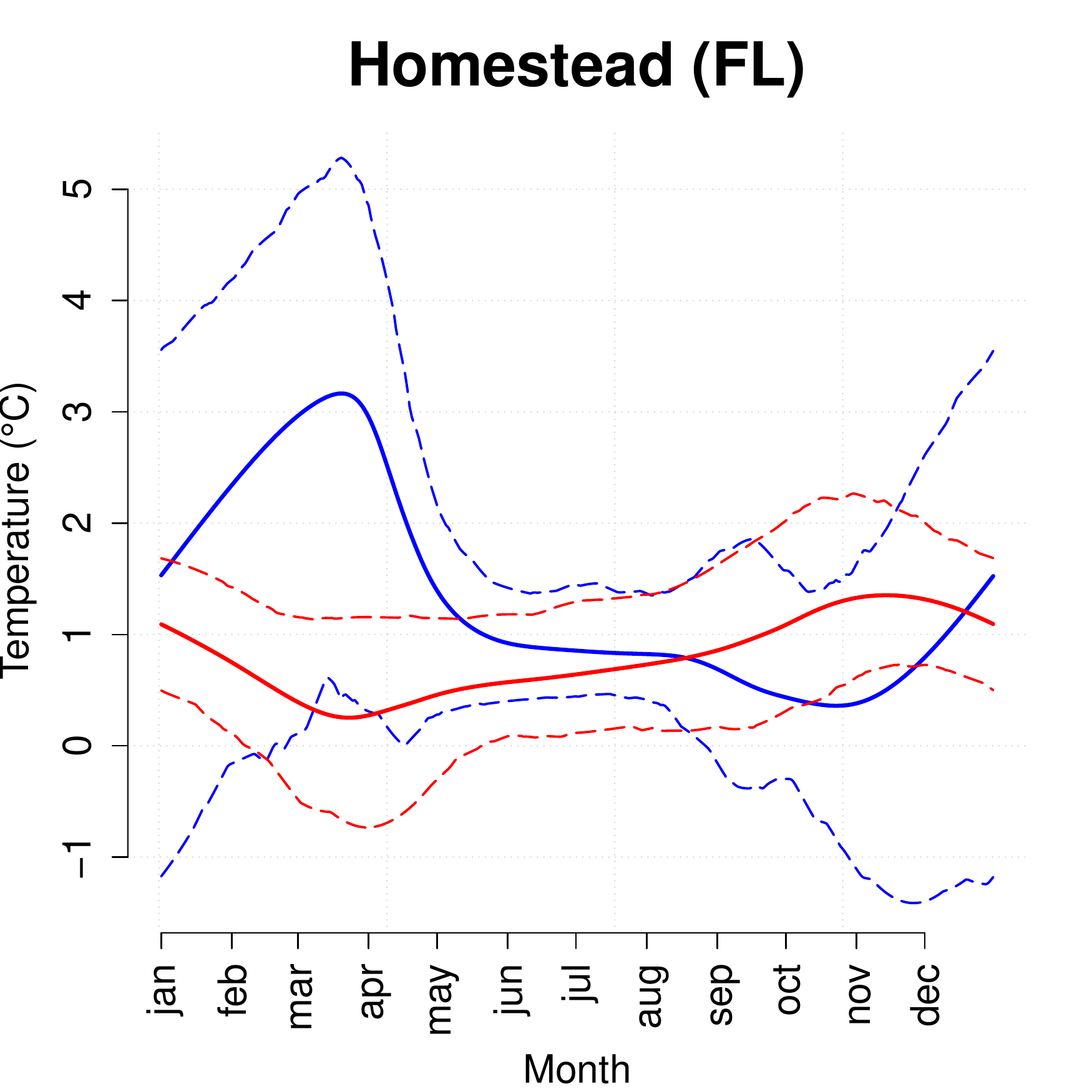}
\caption{Bootstrap uncertainty estimates for the change in 0.001 (blue) and 0.999 (red) SAT quantiles from the starting observation year to 2020. Solid line represents the estimated quantile change, and dashed lines represent pointwise 95\% bootstrap confidence intervals.}
\label{fig:bootstraps}
\end{figure}

Figure \ref{fig:bootstraps} provides uncertainty estimates for the changes in the 0.001 and 0.999 quantiles between the  first and last years of the studied period. These results are obtained from the stratified block bootstrap procedure accounting for temporal dependence discussed in Section \ref{sec:mle}. 
%Estimation uncertainty is larger for extreme quantiles than the median, partially due to the lower amount of data available, but also because the largest changes tend to occur in extreme quantiles and are naturally accompanied by the largest estimation uncertainty.
% The seasonal parameterization of the model enables to observe and quantify how changes in quantiles and their uncertainty vary across the year. 
% For instance, cold quantiles in Bethel (AK) are warming the most in winter whereas hot quantiles in San Diego (CA) are warming the most from April to September. 
% Some of the most changing quantiles over the years are also showing the most estimation uncertainty, and most changes are indicating some warming of the quantiles (except Winter cold quantiles in Hilo (HI) and Spring warm quantiles in Blythe (CA)). 
% The largest changes are observed Bethel (AK) and San Diego (CA) respectively on cold and warm quantiles. 
Note that, in general, the uncertainty bounds are not symmetric around the estimated quantile change.
In particular, San Diego exhibits a very large uncertainty associated with change in 0.999 quantile from April to June, consistent with the high variability in the upper tail observed in Figures \ref{fig:boxplot_season} and \ref{fig:quantiles} at that time of the year. Table \ref{tab:kappaboottable} shows that the BATs model in San Diego favors a heavy upper tail, and the confidence interval for the shape parameter of the upper GPD tail also includes positive values, indicating that a heavy tail is a possibility. Typical shape parameter estimates for temperature distributions range from $-0.3$ to 0 or at times are slightly positive \citep{nogaj:hal-03189825, toolkit, bommier2014}. Indeed, Figure \ref{fig:bootstraps} and  confidence intervals from Table \ref{tab:kappaboottable} suggest that the BATs estimate of the upper tail at San Diego may be too heavy.
This large upper tail uncertainty in San Diego must be interpreted with caution, but overall, the block bootstrap helps us quantify uncertainty without neglecting temporal dependence or climate change.

%To further investigate changes in SAT distributions, we consider the yearly evolution of quantile differences (Figure \ref{fig:quantiledifference_data_empirical}), which enables us to quantify changes in the spread of bulk and tails of the distribution. Quantile estimates are obtained from a quantile regression which is discussed Section  \ref{sec:seasonal_model}. %whose covariates are eight periodic splines to  quantify seasonal patterns, a long-term trend to quantify climate change, and an interaction term for long-term change in seasonal patterns. These covariates are discussed in more detail in Section  \ref{sec:seasonal_model}. 
%Figure \ref{fig:quantiledifference_seasonal_empirical} in the Appendix shows an additional interpretation of the quantile regression differences by plotting the yearly quantile difference curves instead of averaging over years. 

\begin{figure}
    \centering
    \includegraphics[scale=.22]{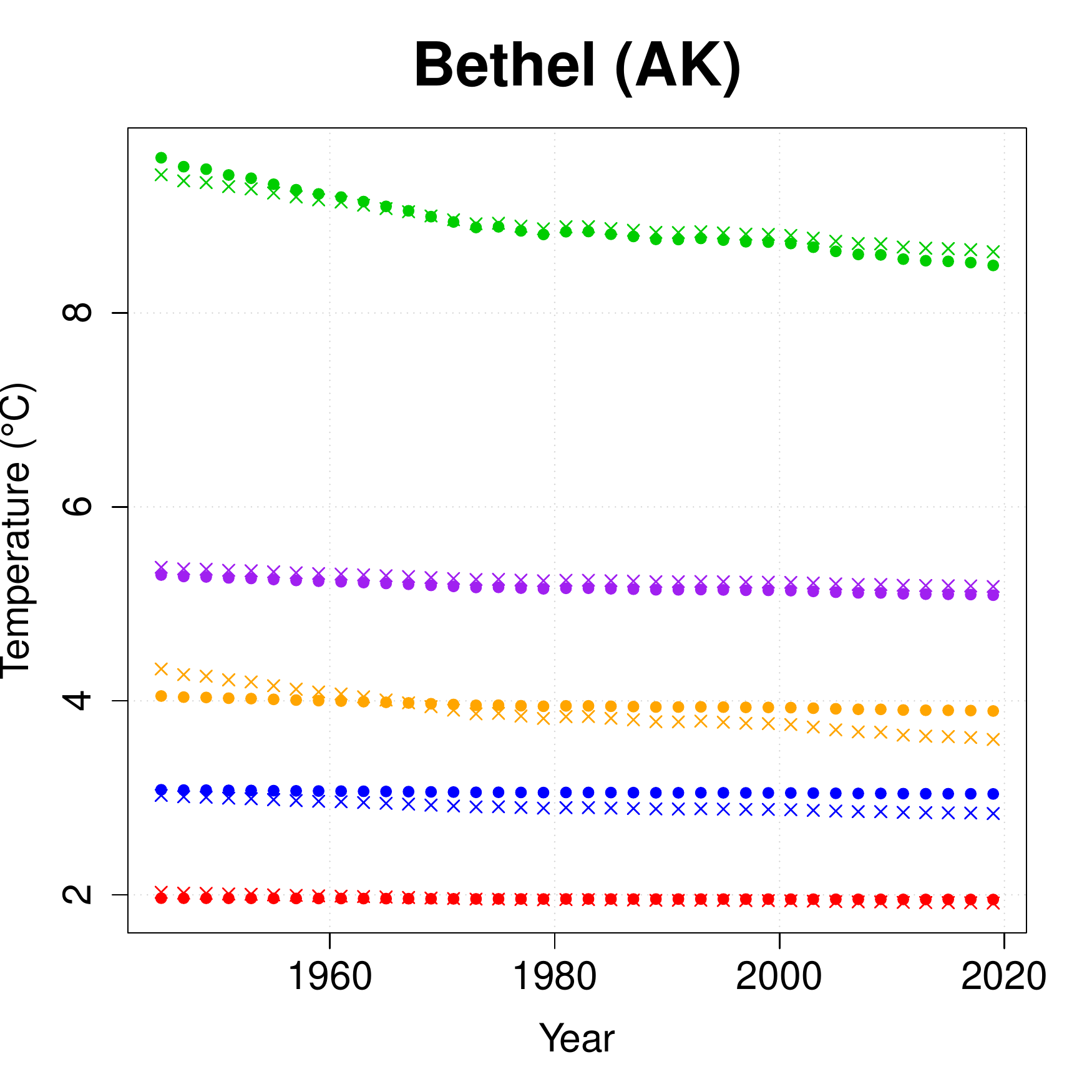} 
    \includegraphics[scale=.22]{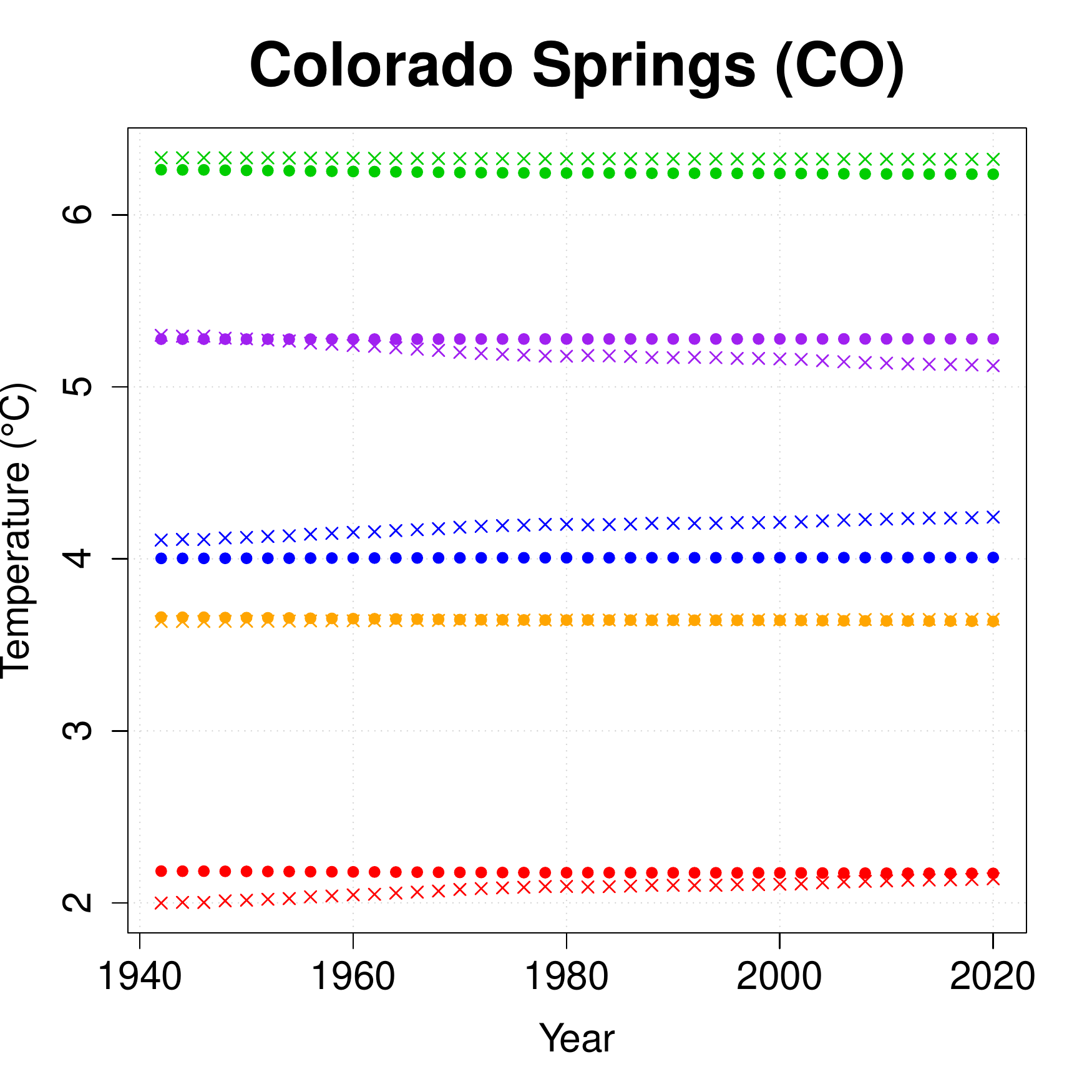}
    \includegraphics[scale=.22]{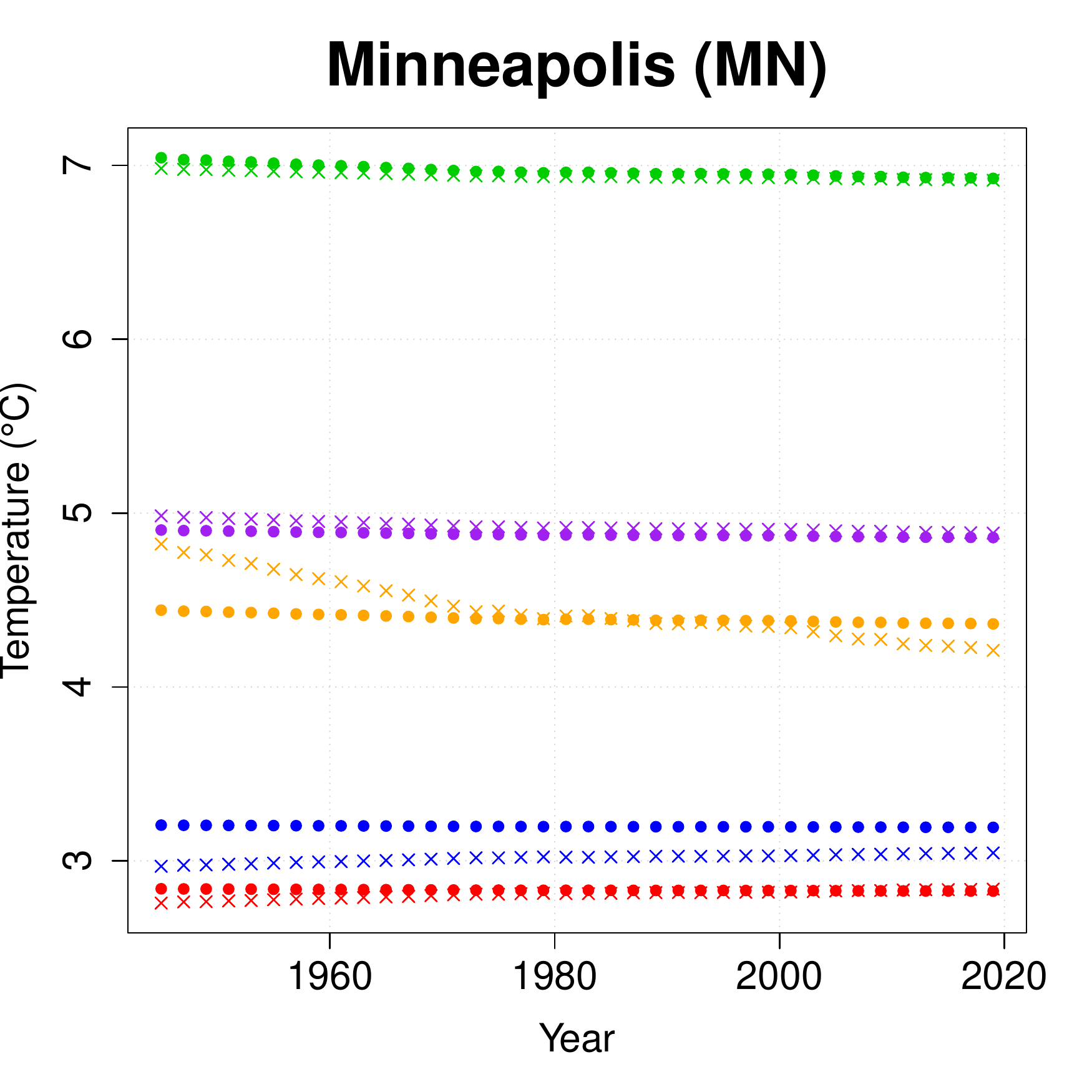}
     \includegraphics[scale=.22]{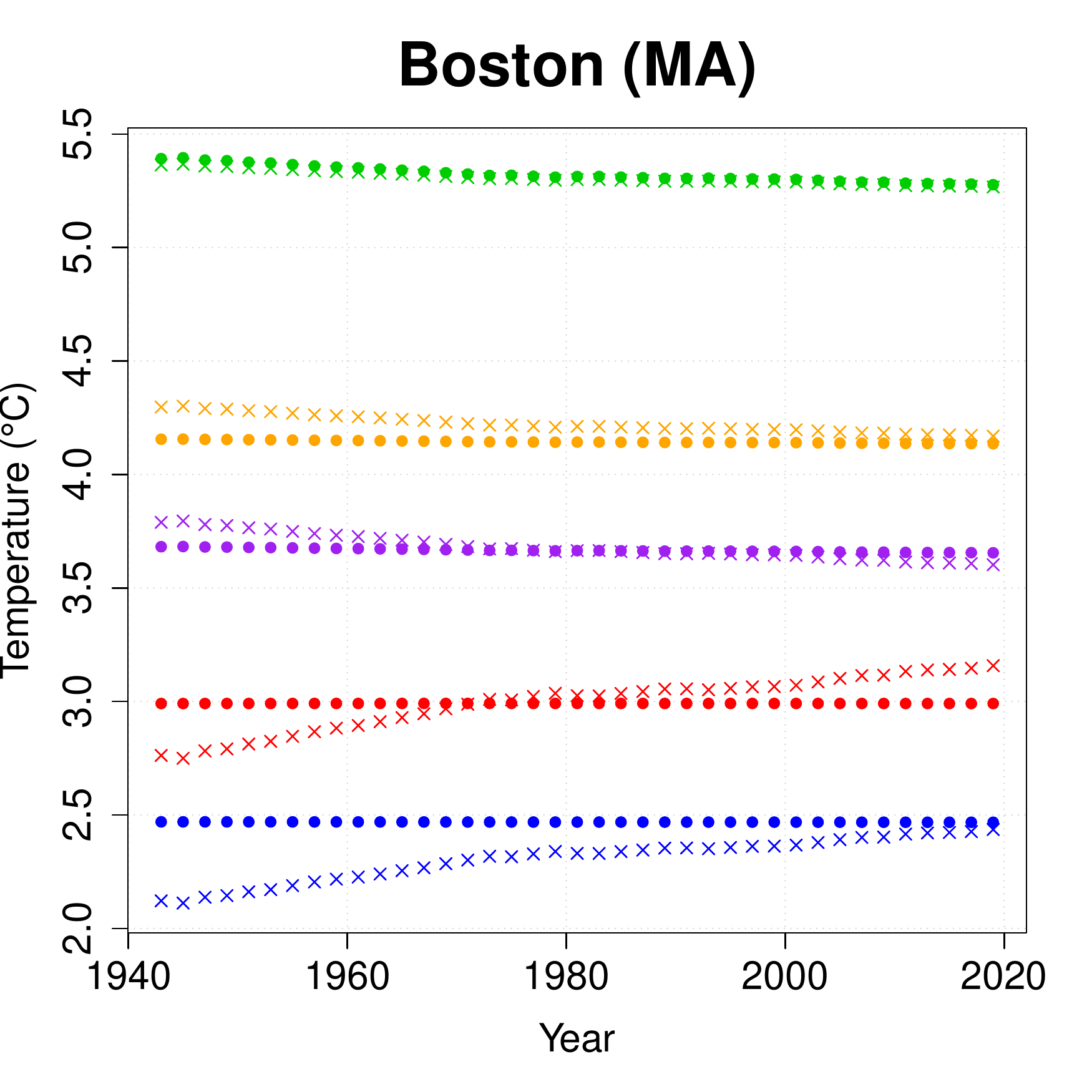}
    \includegraphics[scale=.22]{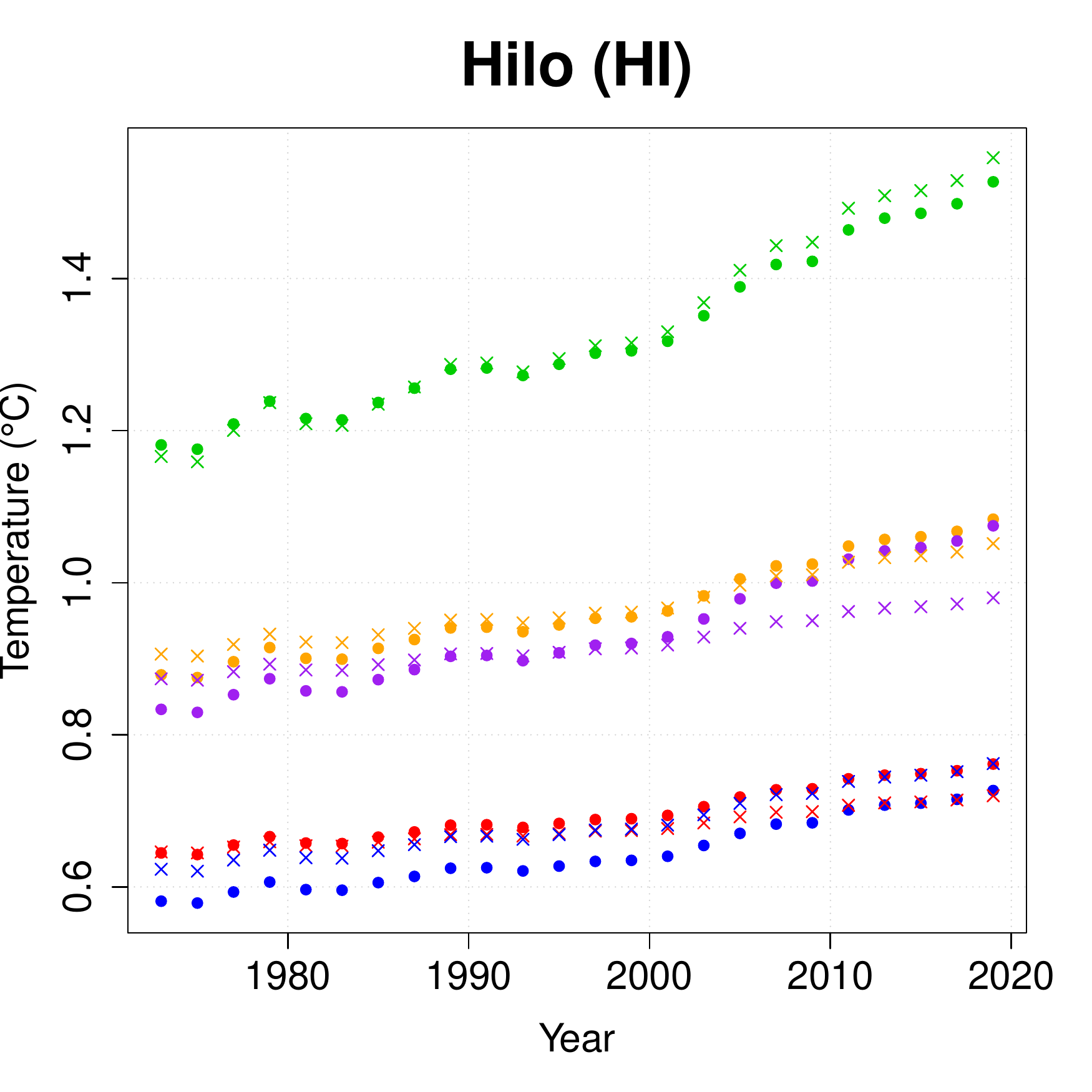} 
     \includegraphics[scale=.22]{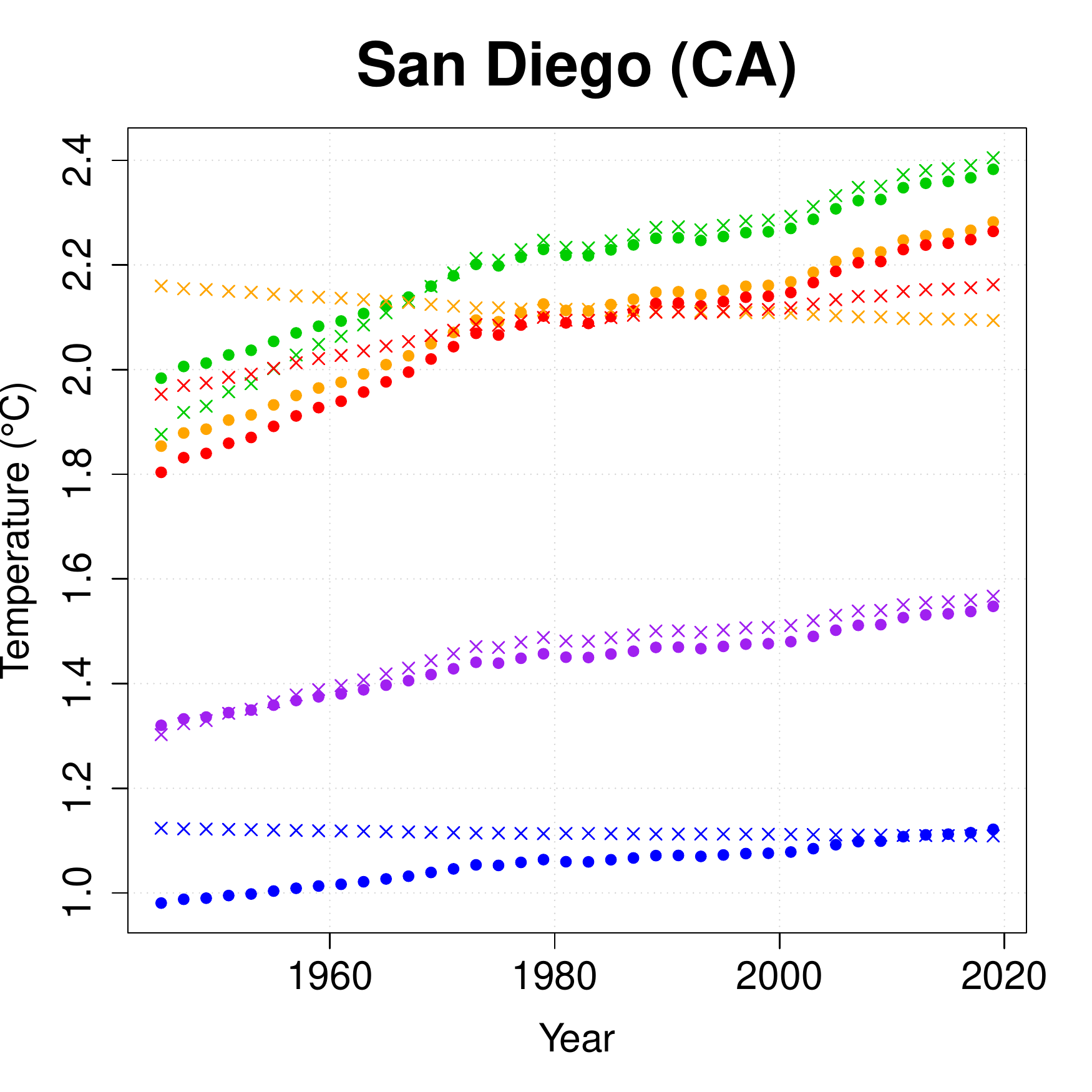} 
     \includegraphics[scale=.22]{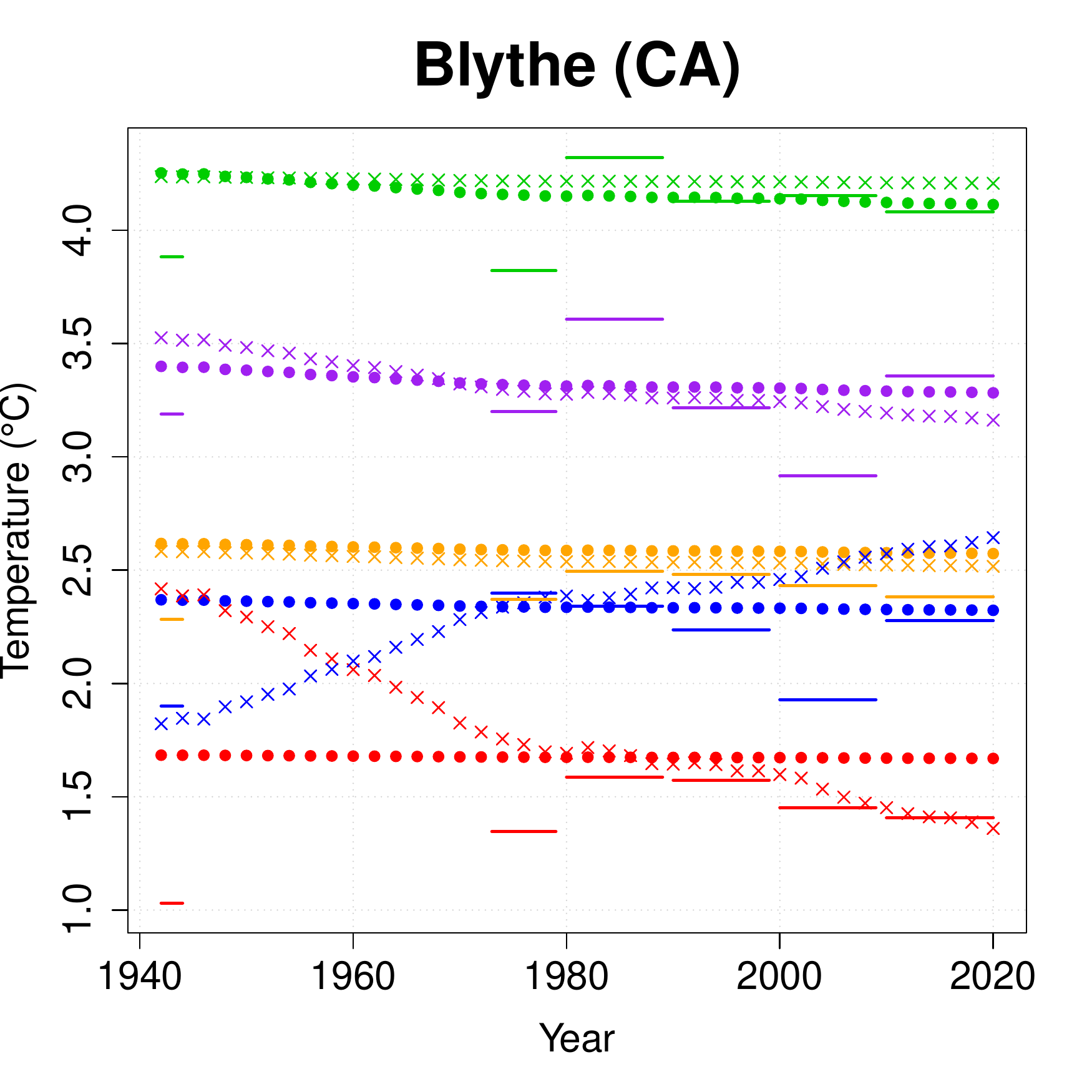}
     \includegraphics[scale=.22]{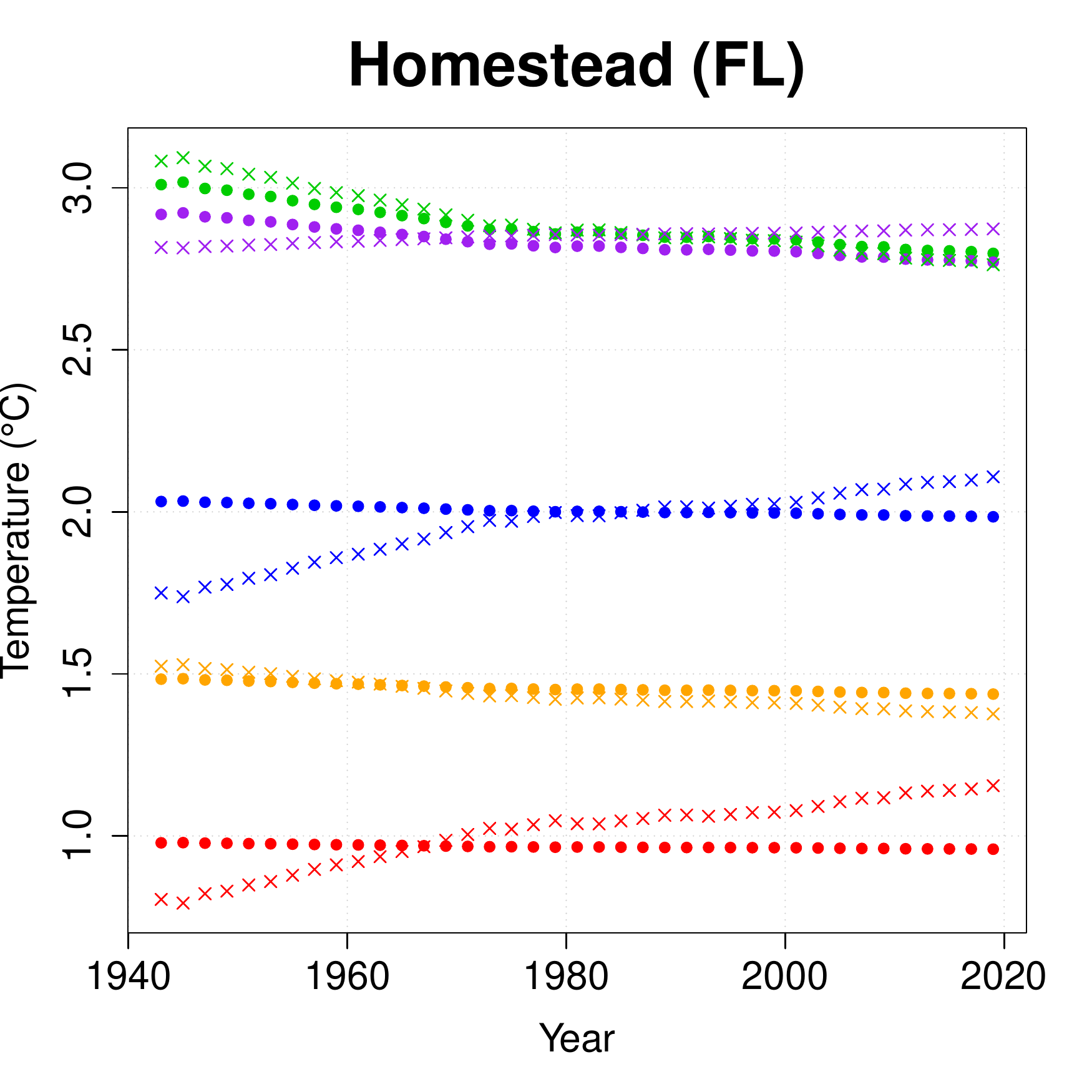}
     
     \caption{Average differences between quantiles for each year based on the BATs model ($\mathtt{o}$) and quantile regression ($\mathtt{x}$). Only every other year is shown. Red is $q_{0.99}-q_{0.95}$, orange is $q_{0.95}-q_{0.75}$, green is $q_{0.75}-q_{0.25}$, purple is $q_{0.25}-q_{0.05}$, and blue is $q_{0.05}-q_{0.01}$. Due to missing data at Blythe, we add lines which show stationary quantile regressions for each decade of data.}
    \label{fig:quantiledifference_data}
\end{figure}

Finally, we consider the yearly evolution of quantile differences across the study period (Figure \ref{fig:quantiledifference_data}), which enables us to quantify changes in the spread of bulk and tails of the distribution. Specifically, we examine the interquartile range and quantile differences in the tails to assess how the bulk spread and tail spreads change over time. For comparison with the BATs model fits, quantile estimates are also obtained from a quantile regression whose covariates are eight periodic splines to  quantify seasonal patterns, a long-term trend to quantify climate change, and an interaction term for long-term change in seasonal patterns (i.e.,\ same covariates as the location parameter in \eqref{steinparameterization}).  We observe a strong spatial variability in the behavior of quantile spreads. In general, the width of the bulk of the distribution decreases over time, with the IQR decreasing over time in most places (e.g.,\ Bethel) but increasing elsewhere (e.g.,\ Hilo).
Quantile differences in both the near (orange and purple) and far (red and blue) tails experience different rates of changes  for cold  and warm temperature. \cite{huang2016} and \cite{rhines2017}, respectively in climate projections and historical reanalysis, reported a decrease in variability of winter temperature in most North American regions. 
The  far  warm tails (red) often show  a stretching trend,  whereas the  near cold  tails (purple) tend to contract.
For most locations considered, the far tail spreads (red and blue) are smaller than the near tail spreads (orange and purple), with some few exceptions. In Colorado Springs and Homestead, both lower tail spreads are larger than the upper tail spreads. In San Diego, both upper tail spreads are larger than the lower tail spreads, and the far upper tail spread has changed at a pace that rivals the interquartile range. At Hilo, all measures of spread appear to be increasing over time. 
For the most part, results from BATs and quantile regression agree with one another in capturing these aspects of SAT. The largest discrepancy is in Blythe, where the BATs model lacks the increase in far lower tail spread and decrease in far upper tail spread found in results for the quantile regression. This difference seems to result from the many years of missing data at Blythe, as a stationary quantile regression for each decade (shown with horizontal lines) does not exhibit this crossing of far upper and lower tail spreads. Other stations also display a general characteristic of ``flatness'' in the BATs results when compared to the corresponding estimates from quantile regression.  Allowing other parameters in the BATs model to depend upon log CO$_2$ may help capture more subtle year-to-year changes in distribution.

%%---------------
\section{Discussion and Conclusions}

In this paper, we have applied the seasonal bulk-and-tails (BATs) model with a long-term trend to daily mean surface air temperature data from eight cities in the United States. The model has demonstrated its ability to capture different seasonality and climate change patterns as well as different behaviors in the upper and lower tails.  A cross-validation comparison shows superior fit of the BATs distribution to the skew-normal distribution on the entire distribution at all cities and to the generalized Pareto distribution in both tails at many cities. 

For practitioners, this model offers a great flexibility with a closed-form density that is relatively easy to work with. 
Only a few statistical models proposed in extremes literature possess a density describing the bulk and both tails \citep{naveau2016, tencaliec2020,stein2020, steinextremes}. Extreme value analysis is still  traditionally performed  with block-maxima or peak-over-threshold techniques, which limits analysis to a tail at a time and avoids an all-encompassing model that describes the entire distribution \citep{coles}. 
The proposed framework also allows parameters in the BATs distribution to change over time, both seasonally and long-term. %, resulting in a total of 45 parameters. Automatic differentiation software was used to help guide the maximum likelihood  estimation. 
A stratified block bootstrap procedure was conducted for uncertainty quantification. 

Of the seven parameters in the standard BATs model, we have taken the shape parameters $\kappa_0$  and $\kappa_1$ and the smoothness $\nu$ to be constant over time. This choice was made due to the difficulty of estimating the shape and smoothness even when these parameters are constant.
%but it may be appropriate to let a subset of these parameters also change over time.
To capture long-term variation of the nonstationary parameters, we included an annual log CO$_2$ equivalent covariate in the location parameters. Other parameters could depend upon this covariate, particularly the scale parameters, which would reflect a long-term change in the temperature variability, or the shape parameters, which would reflect long-term changes in tail behavior. Figure \ref{fig:quantiledifference_data} indicates that it may be worth allowing other parameters to depend on the long-term trend and its seasonal interaction. %However, constant shape parameters are already often challenging to numerically estimate, and this would necessitate further adaptation of the model.  
Although the logarithm of CO$_2$ equivalent serves as a proxy for climate change induced by greenhouse gases, it is important to note that not all long-term changes in this paper are due to global-scale anthropogenic climate change. Some sites may be affected by localized changes, particularly the larger cities and those with measurements taken at airports.
 
 The proposed model provides a flexible statistical tool to examine the bulk and tails of distributions changing over time.
We have quantified how different parts of the SAT distribution (i.e.,\ different quantiles) exhibit different seasonal and long-term changes at cities representative of different geographies and local climates.
 Hot and cold quantiles tend to experience larger long-term changes than median quantiles and also present different seasonal patterns.
 In particular, studied locations with warmer climates experience the most warming for hot quantiles in summer, and locations with colder climate experience the most warming for cold quantiles in winter. 
 
% An interesting idea we did not consider is studying changes in distribution on a year-to-year basis. Climate indices such as atmospheric circulation patterns or other large-scale drivers could also be included as covariates to account for interannual variability. \julie{I wonder if we want to discuss the interannual variability} 
Finally, the proposed model is fit to observations while neglecting spatial and temporal dependence. Modeling multidimensional extremes (as would be needed to explicitly account for spatial and/or temporal dependence) is an on-going challenge for the community \citep{huser2020}, where various techniques have been proposed, relying on hierarchical models \citep{gaetan2007}, copulas \citep{lee2018, krupskii2021}, and  conditional modeling \citep{wadsworth2019, simpson2021, huang2021}.  
Most of these methods ignore the bulk of the data or only focus on a single tail. Developing a bulk-and-tails model for multiple variables is a formidable task because of the wide range of behaviors that can occur in distributions for multivariate extremes \citep{huser2016, huser2017, huser2019, huang2019b, davison2013}.  

%Most of these models focus on one tail only, adding difficulties when pursuing a bulk-and-tails model. 

% 
% Another important aspect lacking in our analysis is dependence in space and time at the model level. Simultaneously modeling spatial and temporal dependence in the bulk and tails of a distribution is a challenging problem outside the scope of this paper.

% %%---------------
% \section{Things we want to mention at some points}
% - parameterization of the pdf in locations and scale ($\frac{\phi_{0}+\phi_{1}}{2}$ and $\phi_{0}-\phi_{1}$)
% - Literature of statistical models for temperature \cite{haugen2018}
% - One highlight of our model is the flexibility in modeling pdf across seasons. 
% - parameterization of the pdf in locations and scale ($\frac{\phi_{0}+\phi_{1}}{2}$ and $\phi_{0}-\phi_{1}$)
% - compute some scores between fitted distributions and data
% - Uncertainty on  estimated lower and upper bounds $L$ and $U$ (Delta method from uncertainty on estimated parameters?)
% - show some trust on the data
% - we want to check if climate change affects temperature variability and maybe tail heaviness. if yes, we might want to include a trend in the modeling of the scale and maybe shape parameters.

\section*{Acknowledgements}
Part of the effort of Mitchell Krock, Julie Bessac and Michael Stein is based in part on work supported by the U.S. Department of Energy, Office of Science, Office of Advanced Scientific Computing Research (ASCR) under Contract DE-AC02-06CH11347. Adam Monahan acknowledges the support of the Natural Sciences and Engineering Research Council of Canada (NSERC) (funding reference RGPIN-2019-204986).

\bibliographystyle{apalike}
\bibliography{biblio}

%\section{Appendix}

\appendix
\section{Complementary Data Information}\label{app:data}

\begin{table}[h!] 
\tabcolsep=0.1cm
\centering

\begin{tabular}{L{2cm}|C{.5cm}C{1cm}L{5cm}}
     & \% & Start & Missing \\ \hline
BET (AK) & 89 & 1945 &  \\
BLY (CA) & 86 & 1942 & 1945-1972 \\
BOS (MA) & 95 & 1943 &  \\
COL (CO) & 100 & 1942 & 1965-1972 \\
HIL (HI) & 99 & 1973 &   \\
HOM (FL) & 94 & 1943 & 1946-1955, 1971-1972, 2000-2004 \\
MIN (MN) & 100 & 1945 & 1965-1972 \\
SAN (CA) & 98 & 1945 &  \\
\end{tabular}
\caption{Summary description of the data. In particular, the percentage of  daily mean SAT measurements taken with an average of over 20 hours of data, the starting year of measurements, and any years which are entirely missing data. End year is 2020.}
\label{tab:stationinfo}
\end{table}

\section{Parameter Estimates}\label{app:param}

 \begin{table}[H] 
 \centering
\resizebox{\linewidth}{!}{%
\begin{tabular}{L{1.7cm}|R{1cm}L{2.4cm}|R{1cm}L{2.4cm}|R{1cm}L{2.4cm}|R{1cm}L{2.4cm}}
            & \multicolumn{2}{c}{$\kappa_0/\nu$} & \multicolumn{2}{c}{$\kappa_1/\nu$} & \multicolumn{2}{c}{$\xi_0$} & \multicolumn{2}{c}{$\xi_1$} \\ \hline
BET (AK) &         0.019 &   (-0.031, 0.027) &        -0.048 &  (-0.144, -0.008) & -0.234 &  (-0.328, -0.184) & -0.146 &   (-0.235, -0.130) \\
BLY (CA) &        -0.034 &   (-0.068, 0.013) &        -0.005 &   (-0.028, 0.016) & -0.222 &  (-0.326, -0.164) & -0.103 &  (-0.188, -0.049) \\
BOS (MA) &        -0.245 &  (-0.327, -0.215) &         -0.190 &  (-0.258, -0.149) & -0.188 &  (-0.285, -0.177) & -0.221 &  (-0.288, -0.177) \\
COL (CO) &        -0.088 &  (-0.134, -0.023) &        -0.156 &  (-0.268, -0.071) & -0.182 &  (-0.251, -0.139) & -0.165 &  (-0.208, -0.096) \\
HIL (HI) &        -0.051 &   (-0.084, 0.008) &        -0.008 &   (-0.036, 0.019) & -0.242 &  (-0.331, -0.176) & -0.093 &  (-0.169, -0.057) \\
HOM (FL) &        -0.002 &   (-0.052, 0.033) &        -0.009 &   (-0.047, 0.017) & -0.215 &  (-0.285, -0.167) & -0.104 &   (-0.243, -0.060) \\
MIN (MN) &         -0.220 &  (-0.273, -0.157) &        -0.251 &    (-0.300, -0.209) & -0.187 &  (-0.241, -0.143) & -0.235 &  (-0.303, -0.191) \\
SAN (CA) &         0.003 &   (-0.141, 0.034) &         0.119 &    (0.022, 0.239) & -0.122 &  (-0.217, -0.117) & -0.006 &   (-0.086, 0.013) \\
\end{tabular}}
\caption{Parameter estimates with 95\% confidence interval. Two left columns are tail behaviors of the BATs model, and two right columns are tail behaviors of the GPD models. Confidence intervals obtained from the percentile method with 200 block bootstrap samples.}
 \label{tab:kappaboottable}
 \end{table}

\section{Quantile Regression}\label{app:qr}
The $\tau$\textsuperscript{th} quantile $(0 < \tau < 1$) of a random variable $Y$ with cdf $F$ is defined as $Q_Y(\tau) = \inf \{y \mid y \ge F(y)  \}$. Quantile regression is used to estimate quantiles as a linear function of covariates $X$; that is, $X \beta \in \mathbb{R}^n$, where $X$ is a $n \times p$ matrix of covariates. Estimates of the coefficient vector are  given  by 
\begin{align*}
\hat \beta = \argmin \limits_{\beta \in {\mathbb{R}^p}} \sum_{i=1}^n \rho_\tau(y_i - x_i \beta)
\end{align*}
where $x_i$ is the $i$\textsuperscript{th} row of $X$ and $\rho_\tau(y) = y (\tau - \mathbf{1}[y < 0])$.
All quantile regressions were fit with the \texttt{R} package \texttt{quantreg} \citep{quantreg}.

\section{Kernel Density Estimate}\label{app:kde}
 Suppose we are interested in the KDE at day of year $d_0$. Let $A_{1943} = \{d_{-7},\dots,d_7\}$ be the window of days centered at $d_0$ for the first observation year, say 1943. With $A = A_{1943} \cup \dots \cup A_{2020}$, the KDE formula reads
\begin{equation*} \label{eq:KDE}
K_h(x) = \frac{1}{|A|} \sum_{x_i \in A} K \left( \frac{x-x_i}{h} \right)
\end{equation*}
where $K$ is a Gaussian kernel function and $h$ is a bandwidth. Estimates were obtained from the \texttt{density} \texttt{R} command with bandwidth selected according to \citet{sheather1991}.

%The temperature records present missing data that are not filled in this study. Except for two stations, Bethel and Blythe, more than $94\%$ of the daily statistics are computed on at least $20$ data-points, providing reliable daily summaries. Respectively, $89\%$ and $87\%$ of the daily statistics of Bethel and Blythe are computed on more than $20$ data-points. Blythe has $98\%$ of its daily summary computed on at least $10$ datapoints, whereas Bethel has almost $10\%$ of its daily summary computed with $8$ data-points.  %\\suggesting that 3-hourly data were used to compute daily summaries for periods of time in Bethel. %This provides an additional confidence in the studied data.

\end{document}